\documentclass[a4paper,12pt]{book}

\usepackage[croatian,english]{babel}
\usepackage[dvips]{graphicx}
\usepackage{makeidx}
\usepackage{amsmath}
\usepackage{amsfonts}
\usepackage{amssymb}
\usepackage[usenames,dvipsnames]{color}
\usepackage{tocbibind}
\usepackage{lscape}
\usepackage{rotating}

\usepackage{setspace}

\definecolor{grey}{rgb}{0.55,0.55,0.55}
\definecolor{red}{rgb}{0.8,0.0,0.0}

\makeindex

\begin{document}

\pagenumbering{roman}

\begin{onehalfspace}

\thispagestyle{empty}

\begin{center}

\vspace*{1.0cm} 

{\normalsize \textsc{University of Zagreb}}\\

\vspace{0.1cm}

{\normalsize \textsc{Faculty of Science}}\\

\vspace{0.2cm}

{\normalsize \textsc{Department of Physics}}

\vspace{2.0cm}

{\large Marinko Jablan}\\

\vspace{3.0cm}

{\Large \textbf{Electrodynamic properties of graphene and their technological applications}}\\


\vspace{4.5cm}

{Doctoral Thesis submitted to the Department of Physics\\
Faculty of Science, University of Zagreb\\
for the academic degree of\\
Doctor of Natural Sciences (Physics)\\

\vspace{2.5cm}

Zagreb, 2012.}

\vspace{1.0cm}


\end{center}

\newpage
\thispagestyle{empty}
\mbox{}

\newpage

\thispagestyle{empty}

\begin{center}

\vspace*{1.0cm} 

{\normalsize \textsc{Sveu\v{c}ili\v{s}te u Zagrebu}}\\

\vspace{0.1cm}

{\normalsize \textsc{Prirodoslovno-matemati\v{c}ki fakultet}}\\

\vspace{0.2cm}

{\normalsize \textsc{Fizi\v{c}ki odsjek}}

\vspace{2.0cm}

{\large Marinko Jablan}\\

\vspace{3.0cm}

{\Large \textbf{Elektrodinami\v{c}ka svojstva grafena i primjene u tehnologiji}}\\


\vspace{4.5cm}

{\normalsize Doktorska disertacija\\
predlo\v{z}ena Fizi\v{c}kom odsjeku\\
Prirodoslovno-matemati\v{c}kog fakulteta Sveu\v{c}ili\v{s}ta u Zagrebu\\
radi stjecanja akademskog stupnja\\
doktora prirodnih znanosti fizike\\

\vspace{2.5cm}

Zagreb, 2012.}

\vspace{1.0cm}


\end{center}

\newpage
\thispagestyle{empty}
\mbox{}

\newpage

\thispagestyle{empty}

\begin{center}
{\large \textsc{ }}\\
\end{center}


\noindent
{\normalsize University of Zagreb \hspace{5.0cm} Doctoral Thesis\\
Faculty of Science\\
Department of Physics}\\


\begin{center}
{\large \textbf{Electrodynamic properties of graphene 
and their technological applications}}\\

\vspace{0.33cm}

{\normalsize \textsc{Marinko Jablan}\\

\vspace{0.2cm}

Faculty of Science, University of Zagreb}\\
\end{center}


Graphene is a novel two-dimensional material with fascinating electrodynamic properties like the ability to support collective electron oscillations (plasmons) accompanied by tight confinement of electromagnetic fields. Our goal is to explore light-matter interaction in graphene in the context of plasmonics and other technological applications but also to use graphene as a platform for studying many body physics like the interaction between plasmons, phonons and other elementary excitations. Plasmons and plasmon-phonon interaction are analyzed within the self-consistent linear response approximation. We demonstrate that electron-phonon interaction leads to large plasmon damping when plasmon energy exceeds that of the optical phonon but also a peculiar mixing of plasmon and optical phonon polarizations. Plasmon-phonon coupling is strongest when these two excitations have similar energy and momentum. We also analyze properties of transverse electric plasmons in bilayer graphene. Finally we show that thermally excited plasmons strongly mediate and enhance the near field radiation transfer between two closely separated graphene sheets. We also demonstrate that graphene can be used as a thermal emitter in the near field thermophotovoltaics leading to large efficiencies and power densities. Near field heat transfer is analyzed withing the framework of fluctuational electrodynamics.

\begin{center}
(99 pages, 71 references, original in English)\\
\end{center}


\begin{tabular}{ll}
 & \\
Supervisor: & Prof. Dr. sc. H. Buljan\\
 & \\
Co-supervisor: & Prof. Dr. sc. M. Solja\v{c}i\'{c},
                 Massachusetts Institute of Technology, USA \\
 & \\
Committee: 
	& Prof. Dr. sc. A. Bjeli\v{s} \\
	& Prof. Dr. sc. H. Buljan \\
	& Prof. Dr. sc. M. Solja\v{c}i\'{c},  
     Massachusetts Institute of Technology, USA \\
	& Dr. sc. I. Kup\v{c}i\'{c}, v. zn. sur. \\
	& Dr. sc. M. Kralj, v. zn. sur., Institute of Physics \\
 & \\
Replacements: 
	& Prof. Dr. sc. S. Bari\v{s}i\'{c} \\
	& Dr. sc. I. Bogdanovi\'{c}-Radovi\'{c}, v. zn. sur., 
     Ru{\dj}er Bo\v{s}kovi\'{c} Institute\\
& \\
Thesis accepted: & 2012. 
\end{tabular}
\newpage
\thispagestyle{empty}
\mbox{}

\newpage


\thispagestyle{empty}

\begin{center}
{\large \textsc{Temeljna dokumentacijska kartica}}\\
\end{center}


\noindent
{\normalsize Sveu\v{c}ili\v{s}te u Zagrebu \hspace{5.0cm} Doktorska disertacija\\
Prirodoslovno-matemati\v{c}ki fakultet\\
Fizi\v{c}ki odsjek}\\


\begin{center}
{\large \textbf{Elektrodinami\v{c}ka svojstva grafena i primjene u tehnologiji}}\\

\vspace{0.33cm}

{\normalsize \textsc{Marinko Jablan}\\

\vspace{0.2cm}

Prirodoslovno-matemati\v{c}ki fakultet, Sveu\v{c}ili\v{s}te u Zagrebu}\\
\end{center}


Grafen je tek nedavno otkriveni dvo-dimenzionalan materijal s vrlo zanimljivim elektrodinami\v{c}kim svojstvima poput mogu\'{c}nosti podr\v{z}avanja kolektivnih oscilacija elektronskog plina (plazmona) pra\'{c}enih s jakom loklizacijom elektromagnetskog polja. Cilj ovog doktorata je prou\v{c}iti interakciju svjetlosti i materije u grafenu u konteksu plazmonike i drugih tehnolo\v{s}kih primjena ali tako\dj{}er upotrijebiti grafen kao platformu za ista\v{z}ivanje pojava fizike mno\v{s}tva \v{c}estica kao \v{s}to su interakcija izme\dj{}u plazmona, fonona i drugih elementarnih pobu\dj{}enja. Plazmone i plazmon-fonon interakciju analiziramo u kontekstu aproksimacije samo-konzistentnog linearnog odziva. Pokazujemo da elektron-fonon interakcija vodi k jakom gu\v{s}enju plazmona kada energija plazmona prije\dj{}e energiju opti\v{c}kog fonona ali tako\dj{}er neobi\v{c}no mije\v{s}anje polarizacija plazmona i opti\v{c}kog fonona. Plazmon-fonon vezanje je najja\v{c}e kad ta dva pobu\dj{}enja imaju usporedivu energiju i impuls. Tako\dj{}er analiziramo svojstva transverzalnog elektri\v{c}nog plazmona u dvo-sloju grafena. Kona\v{c}no pokazujemo da termalno pobu\dj{}eni plazmoni kanaliziraju i bitno pospje\v{s}uju radiativni transfer topline izm\dj{}u dvije bliske ravnine grafena. Tako\dj{}er pokazujemo da se grafen mo\v{z}e koristiti kao termalni emiter u termofotovoltaicima bliskog polja \v{s}to vodi k velikim efikasnostima i gusto\'{c}i snage. Prijenos topline u bliskom polju analiziramo u kontekstu fluktuacijske elektrodinamike.
\\
\\
Klju\v{c}ne rije\v{c}i: grafen / plazmonika / gu\v{s}enja / plazmon  / transverzalni elekri\v{c}ni mod/ plazmon-fonon vezanje / blisko-polje / prijenos topline  / termofotovoltaici

\begin{center}
(99 stranica, 71 literaturnih navoda, jezik izvornika engleski)\\
\end{center}


\begin{tabular}{ll}
 & \\
Mentor: & Prof. Dr. sc. H. Buljan\\
 & \\
Ko-mentor: & Prof. Dr. sc. M. Solja\v{c}i\'{c},
             Massachusetts Institute of Technology, SAD \\
 & \\
Komisija: 
	& Prof. Dr. sc. A. Bjeli\v{s} \\
	& Prof. Dr. sc. H. Buljan \\
	& Prof. Dr. sc. M. Solja\v{c}i\'{c},  
     Massachusetts Institute of Technology, SAD \\
	& Dr. sc. I. Kup\v{c}i\'{c}, v. zn. sur. \\
	& Dr. sc. M. Kralj, v. zn. sur., Institut za fiziku \\
& \\
Zamjene: 
	& Prof. Dr. sc. S. Bari\v{s}i\'{c} \\
	& Dr. sc. I. Bogdanovi\'{c}-Radovi\'{c}, v. zn. sur., 
     Institut Ru{\dj}er Bo\v{s}kovi\'{c}\\
& \\
Radnja prihva\'{c}ena: & 2012. 
\end{tabular}

\chapter*{Acknowledgements}
\addcontentsline{toc}{chapter}{Acknowledgements}  

The work in this thesis was completed under the supervision of Prof. Dr. Hrvoje Buljan at the University of Zagreb and Prof. Dr. Marin Solja\v{c}i\'{c} at the Massachusetts Institute of Technology.
I would like to thank both for their generous help and guidance 
during my research as a graduate student. I would like to thank Hrvoje for his tremendous patience but even more for opening my mind to discussion and showing me how powerful it can be when two very different minds cooperate. 
I would also like to thank Marin for giving me the opportunity to work in the stimulating environment of MIT but even more for his incredible insight into the current trends of modern science. At last I would like to thank Dr. Ivan Celanovi\'{c} and Ognjen Ili\'{c}  for the effort and time they invested in our papers. I would especially like to thank Ognjen for our stimulating discussions and for reminding me that in the end it is math that tells you how big things really are.

The work presented in this thesis has been published in several articles. The reference for each chapter is given below.\\

{\bf Chapter 3}: \\
M. Jablan, H. Buljan and M. Solja\v{c}i\'{c}\\
\emph{Plasmonics in graphene at infrared frequencies},\\
Phys. Rev. B {\bf 80}, 245435 (2009).\\

{\bf Chapter 4}: \\
M. Jablan, H. Buljan and M. Solja\v{c}i\'{c}\\
\emph{Transverse electric plasmons in bilayer graphene},\\
Optics Express {\bf 19}, 11236 (2011).\\

{\bf Chapter 5}: \\
M. Jablan, M. Solja\v{c}i\'{c} and H. Buljan,\\
\emph{Unconventional plasmon-phonon coupling in graphene},\\
Phys. Rev. B {\bf 83}, 161409(R) (2011).\\

{\bf Chapter 6}: \\
O. Ilic, M. Jablan, J. D. Joannopoulos, I. Celanovic, H. Buljan and M. Solja\v{c}i\'{c} \\
\emph{Near-field thermal radiation transfer controlled by plasmons in graphene},\\
arXiv:1201.1489, accepted for publication in Phys. Rev. B (2012). \\

\noindent
O. Ilic, M. Jablan, J. D. Joannopoulos, I. Celanovic and M. Solja\v{c}i\'{c} \\
\emph{Overcoming the black body limit in plasmonic and graphene near-field thermophotovoltaic systems},\\
Optics Express, {\bf 20}, A366 (2012).

\tableofcontents
\newpage

\pagenumbering{arabic} \setcounter{page}{1}

\chapter{Introduction} \label{chap:intro}

Carbon is a basic ingredient of life and all organic chemistry which is consequence of its abundance in nature and his chemical reactivity. With four valence electrons distributed to one 2$s$ and three 2$p$ orbitals, which can hybridize in many different ways, carbon is characterized by a large flexibility of chemical bonding. One particularly interesting case is $sp$2 hybridization which creates three strong $\sigma$-bonds in plane, while the remaining $p$ orbital is weakly bound with neighboring atoms creating $\pi$-bond. In this thesis we will be studying graphene: a two-dimensional (2D) crystal of carbon atoms assembled in a honeycomb structure. While $\sigma$-bond is responsible for the most of the structural integrity of graphene, $\pi$-bond determines low-energy electric and optical properties. Very peculiar property of graphene is that its low-energy electrons behave as massless Dirac particles \cite{Wallace1954, Semenoff1984} (near the corners of the Brillouin zone). Since graphene is essentially a 2D material, one can simply tune its Fermi level through an electrostatic gating which brings about large control over electrical and optical properties, important for various technological applications.

\section{Experimental realization}
\label{Sec:experiments}

Scientists were puzzled for long time whether nature allows existence of a two-dimensional crystal. In 1930's Peierls \cite{Peierls1935} and Landau \cite{Landau1937} showed that thermal fluctuations would destroy long range order and essentially melt 2D lattice at any finite temperature. Therefore it came as a surprise when Geim and Novoselov announced \cite{Novoselov2004, Novoselov2005a, Novoselov2005b} in 2004 a discovery of a first 2D crystal made of carbon atoms - graphene. Scientist were further astonished by a shear simplicity of the experimental method which essentially used a scotch tape to exfoliate graphite (graphite can be viewed as a simple stack of weakly bound graphene planes). The 2010 Nobel prize in physics came as a credit for this great discovery but it is interesting that even today in 2012 experimentalists still use this "scotch tape technique" since it offers exceptionally pure graphene samples on a small scale, important for fundamental research. Of course it is impractical on a large scale production which is required by various industrial applications, and soon after the discovery of graphene several other methods were developed for graphene production, most notably chemical vapor deposition (CVD) \cite{CVD}, segregation by heat treatment of carbon-containing substrates \cite{SiC} and liquid phase exfoliation \cite{LiquidExfol}. The most promising of these methods, for large scale graphene growth, is CVD which is also used \cite{Kralj} by the group of Dr. Marko Kralj from the Institute of Physics in Zagreb, Croatia. They heat ethylene ($C_2H_4$) gas, up to a temperature of $~1000^\circ$C, above the metal surface which serves both as a catalyst for ethylene decomposition and substrate for graphene growth.

It is interesting to note that various groups claim they have seen graphene in their experiments prior to 2004 but it wasn't until Geim and Novoselov groundbreaking experiments that the true potential and importance of graphene was recognized. 

While graphene's intriguing mechanical properties are still debated, this thesis concerns primarily with electrical and optical properties which are a subject of intense research and numerous practical applications. 

\section{Plasmonics}
\label{Sec:plasmonics}

Plasmonics studies collective electron surface charge oscillations (surface plasmons at surfaces of bulk materials or plasmons in a pure 2D materials like graphene) accompanied by tight confinement of electromagnetic (EM) fields. In recent years, an enormous interest has been surrounding the field of 
plasmonics, because of the variety of tremendously exciting and novel phenomena 
it could enable. On one hand, plasmonics seems to be the only viable path 
toward realization of nanophotonics: control of light at scales substantially 
smaller than the wavelength \cite{Barnes2003,Atwater2005,Yablonovitch2005,
Aristeidis2005}. On the other hand, plasmonics is a 
crucial ingredient for implementation of most metamaterials, and thereby 
all the exciting phenomena that they support \cite{Veselago1968,Shalaev2007,
Pendry2000,Smith2004}, including negative refraction, superlensing, and cloaking. 
However, there is one large and so far insurmountable obstacle towards 
achieving this great vision: plasmonic materials (most notably metals) have 
enormous losses in the frequency regimes of interest. 
This greatly motivates us to explore plasmons and their losses in 
a newly available material with unique properties: 
graphene \cite{Novoselov2004,Novoselov2005a,
Novoselov2005b}. 

Plasmons are also very interesting phenomenon from the point of view of many-body physics. Since losses are in a large manner determined by phonons we will encounter interactions between various elementary excitations and interesting many-body effects like plasmon-phonon coupling.

\section{Near field thermo-photo-voltaics}
\label{Sec:TPV}

\noindent Radiative heat transfer between two bodies can be greatly enhanced in
the \emph{near field}, i.e. by bringing the surfaces close together
to allow tunneling of evanescent photon modes \cite{Rytov,Polder,Pendry1999a}. This happens because near field radiation transfer involves thermal excitation of various surface modes which can have much greater wave vectors (and density of states) than the freely propagating modes (limited by the light line). Since each wave vector corresponds to a heat channel, vacuum becomes better heat conductor in the near field. However, due to their localization and evanescent nature, it is only at sub-wavelength separations that these modes become relevant. While measuring near field transfer has been experimentally difficult
\cite{Hargreaves1969,Narayanaswamy2008a,Shen2009a,Rousseau2009},
the promise of order-of-magnitude enhancements over the far field
Stefan-Boltzman black body limit has made transfer in the near field the
topic of much research.

With the current world energy demand and large environmental impact of fossil fuels there is a worldwide shift toward renewable energy sources. In that respect, thermo-photo-voltaics (TPVs) are a promising class of heat to electricity conversion devices \cite{TPV1,TPV2} where Sun can heat up an emitter that selectively re-radiates frequencies matched to the band gap of the photo-voltaic cell thus minimizing the thermalization losses. TPVs are not limited by the Sun source and can use any hot (terrestrial) object like a factory furnace or various hot car parts as a heat source. From the perspective of future energy crisis there is a large demand for more efficient energy management where TPVs can play important role by turning wasted heat into electricity.

Near field TPVs \cite{NFTPV1,NFTPV2,NFTPV3} further offer greater power densities since the near field heat transfer can be orders of magnitude larger than the far field limit. Finally, due to evanescent nature of EM modes, one does not need to worry about losing energy through modes with frequencies below the photo-voltaic band gap, resulting in even larger device efficiencies.

\section{Objectives and results}
\label{Sec:objectives}

The objective of this research is to study electrodynamic properties of graphene and especially high-frequency collective oscillations of electrons (plasmons). We will analyze plasmon excitations in the context of plasmonics and other technological applications, but we will also look at the same problem from the point of view of many-body physics as an interaction between various elementary excitations (plasmons, phonons, etc.). Finally we study near field heat transfer with graphene (mediated by thermally excited plasmons) in the context of TPVs.

We study plasmon excitations in graphene in the context of the Random Phase Approximation (RPA) \cite{PinesBook} and number-conserving relaxation-time approximation \cite{Mermin1970} and we show that plasmons in doped graphene can have both low losses and large localization for frequencies below optical phonon energy at 0.2 eV. Large plasmon damping occurs in the regime of interband single particle excitations which can be shifted towards larger energies for stronger doping values. We demonstrate that for sufficiently large doping there is a frequency interval from optical phonon frequency to boundary of interband regime, where the plasmon damping is dominated by emission of optical phonon and electron-hole pair. To describe impurity scattering we use DC relaxation time since we don't expect significant frequency dependance. The phonon contribution is estimated from the electron self-energy induced by electron-phonon interaction. 

We also explore electron-phonon interaction in graphene as an interesting problem from the aspect of many-body physics. By measuring Raman shift of optical phonon energy it was demonstrated that Born-Oppenheimer approximation (BOA) is not a valid approximation in graphene \cite{BOA}. The measured Raman shift is a consequence of the interaction with single particle excitations, however the breakdown of BOA means that electrons and phonons move on comparable energy scales which leads to a possibility of interaction between phonons and collective electron excitations (plasmons). We show that a peculiar type of hybridization of plasmon and optical phonon modes occurs around the point where the two modes cross in energy and momentum simultaneously since then the electron-phonon interaction will be drastically increased due to collective electron response. We demonstrate that the electron-phonon interaction leads to polarization mixing of the two modes so that longitudinal plasmon (LP) couples exclusively to the transverse optical phonon (TO) mode, while the tranverse electric mode, also referred to as the transverse plasmon (TP), couples exclusively to longitudinal optical phonon (LO) mode; thus there is no coupling between LPs and LO modes. Formally, we analyze plasmon-phonon coupling in the self-consistent linear-response formalism which describes interaction of phonons with both single particle and collective electronic excitations. We emphasize that the phonon interaction with collective excitations is much larger than the phonon interaction with single particle excitations (measured by Raman) which means that plasmon-phonon interaction can serve as a magnifier for exploring electron-phonon interaction in graphene. Further on, our calculations give a slight correction to the standard result of Raman shift of the optical phonon energy since the longwave phonons can interact also with radiative EM modes so that we predict increasing Raman linewidths for higher dopings. Finally we note that LO phonon decouples from all (single particle and collective) electronic excitations when its dispersion crosses the light line.

While longitudinal charge density oscillation can be referred to as longitudinal plasmon, which is also polarized like transverse-magnetic (TM) EM mode, we also analyze properties of the unusual transverse plasmon in 2D systems \cite{Mikhailov2007}, which is polarized like transverse-electric (TE) mode, and accompanied by transverse current density oscillation. These kind of modes are possible only if the imaginary part of 2D conductivity is negative which in principle requires interband transitions. From that perspective bilayer graphene is an interesting candidate for exploring these modes, because it has a rich band structure and particularly two perfectly nested bands with a gap of 0.4 eV which results in large joint density of states considering the vertical interband transitions. We show that plasmon properties (localization) of TE modes are much more pronounced in bilayer than in single layer graphene. 

We also show that thermally excited plasmons strongly mediate and enhance the near field radiation transfer between two closely separated graphene sheets. 
Near field heat transfer is analyzed within the framework of fluctuational electrodynamics and we predict several orders of magnitude larger values of heat transfer between two graphene sheets in the near field than the case of heat transfer between two black bodies, of the same temperatures, in the far field. Finally we demonstrate that graphene can be used as a thermal emitter in the near field thermophotovoltaics leading to large efficiencies and power densities.

The thesis is organized into chapters as follows. In Chapter \ref{chap:chap2} we present theoretical methods and tools that will be used throughout the text. We first calculate electron dispersion and electron-phonon interaction Hamiltonian in graphene within the tight binding approximation. Next we give the density-density and current-current response functions in the linear approximation and use fluctuation-dissipation theorem to calculate current-current correlation function due to thermal fluctuations in the system. Finally we use this to calculate the radiative heat transfer between two graphene sheets. In Chapter \ref{chap:chap3} we calculate plasmon dispersion and damping due to electron-impurity and electron-phonon scattering. In Chapter \ref{chap:chap4} we calculate dispersion od TE modes in single and bilayer graphene. In Chapter \ref{chap:chap5} we calculate plasmon-phonon interaction within the self-consistent linear response formalism. In Chapter \ref{chap:chap6} we calculate near field heat transfer between two graphene sheets and we analyze near field TPV device with graphene as a thermal emitter. Finally, in Chapter \ref{chap:summary} we summarize.

\chapter{Methods} \label{chap:chap2}

In this chapter, for the sake of the clarity of the presentation, we derive basic physical quantities used to describe graphene such as the low energy Dirac Hamiltonian and the electron-phonon interaction. We will also define standard response functions, like the conductivity, density-density, and current-current response functions, that will be used in later chapters. 
This chapter is intended to provide an introduction and overview of these concepts so the reader already familiar with them can skip the corresponding sections. 
Finally we will derive an expression for the radiative heat transfer between two graphene sheets at different temperatures by employing the fluctuation-dissipation theorem.

\section{Tight binding approximation in graphene}

In this section we use the tight-binding approximation to derive the electron band structure of graphene, Dirac equation valid at low energies and electron-phonon interaction.

\subsection{Electron band structure}
\label{Sec:Electron}

\begin{figure}[h!]
\begin{center}
\includegraphics[width=0.5 \textwidth ]{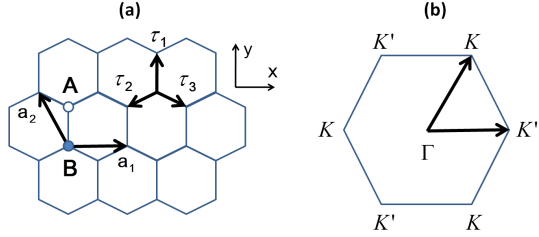}
\caption{
a) Graphene crystal structure. Unit cell vectors are $\bf{a}_1$ and $\bf{a}_2$ while A and B are atoms of the basis. b) Brillouin zone. We mark high symmetry points $\bf{K}$ and $\bf{K'}$ where the low-energy electron excitations are described by massless Dirac equation.
}
\label{Figure1_c2}
\end{center}
\end{figure}

Graphene crystal structure is determined by a Bravais lattice with two atoms in a basis (see figure \ref{Figure1_c2}). We can choose unit cell vectors as ${\bf a}_1=a(1,0)$ and ${\bf a}_2=a(-1/2,\sqrt{3}/2)$, while the vectors connecting first neighbors are given by $\mbox{\boldmath $\tau$}_1=a(0,1/\sqrt{3})$, $\mbox{\boldmath $\tau$}_2=a(-1/2,-1/2\sqrt{3})$, and 
$\mbox{\boldmath $\tau$}_3=a(1/2,-1/2\sqrt{3})$. Here $a=0.25$ nm is a lattice constant while the nearest neighbor carbon-carbon distance is 
$|\mbox{\boldmath $\tau$}_l|=b=a/\sqrt{3}=0.14$ nm.

Unit cell vectors of reciprocal lattice are given by ${\bf c}_1=(2\pi/a)(1,1/\sqrt{3})$ and 
${\bf c}_2=(2\pi/a)(0,2/\sqrt{3})$, while we are primarily interested in the vertex points of the Brillouin zone i.e. vectors ${\bf K}=(2\pi/a)(1/3,1/\sqrt{3})$ and 
${\bf K'}=(2\pi/a)(2/3,0)$. The remaining four vertex points are equivalent to the points ${\bf K}$ i ${\bf K'}$ since they are connected to them by a simple translation with the reciprocal vector $n_1 {\bf c}_1+n_2 {\bf c}_2$, where $n_1$ and $n_2$ are integers.

As we already pointed out in the introduction, $sp$2 hybridization is responsible for the mechanical stability of graphene by creating three strong $\sigma$ bonds in $xy$ plane while the remaining $p_z$ orbital weakly interacts with the neighborings $p_z$ orbitals creating the $\pi$ bond. Since we are particularly interested in $\pi$ bond, the entire problem is very well described with the tight binding approximation \cite{Wallace1954}.

Let us define now an operator $c_{\bf R}^\dag$ that creates a free $p_z$ orbital at the lattice point $\bf R$, i.e. $|p_z({\bf R})\rangle=c_{\bf R}^\dag|0\rangle$. Let us further denote by $-\gamma_0$ the hopping integral between nearest neighbor $p_z$ orbitals (next-nearest neighbor interaction is negligible and $\gamma_0\approx 2.8$ eV \cite{Wallace1954}). Since we are only interested in the behavior of the electron energies near the $p_z$ orbital energy, our system is well described by a tight binding Hamiltonian
\begin{equation}
H = -\gamma_0\sum_{{\bf R_A},\mbox{\boldmath $\tau$}_l} 
c_{{\bf R_A}-\mbox{\boldmath $\tau$}_l}^\dag  c_{{\bf R_A}}^{} 
    -\gamma_0\sum_{{\bf R_B},\mbox{\boldmath $\tau$}_l} 
c_{{\bf R_B}+\mbox{\boldmath $\tau$}_l}^\dag  c_{{\bf R_B}}^{},
\label{TBA}
\end{equation}
where the sum over lattice points is divided into two parts that contain different basis atoms i.e. 
${\bf R_A}=n_1 {\bf a}_1+n_2 {\bf a}_2+\mbox{\boldmath $\tau$}_1$ and 
${\bf R_B}=n_1 {\bf a}_1+n_2 {\bf a}_2$ ($n_1$ and $n_2$ are integers). In equation (\ref{TBA}), we have assumed that zero energy corresponds to $p_z$ orbital energy (i.e. $E(p_z)=0$) and we have neglected overlapping of the two neighboring orbitals. We have also omitted the notion of electron spin since it only plays the role of additional degree of freedom. Eigenstates of the Hamiltonian (\ref{TBA}) must take the form of the linear combination of $p_z$ orbitals that satisfy the Bloch condition
\begin{equation}
c_{\bf k}^\dag = 
\frac{1}{\sqrt N}\sum_{\bf R_A}e^{i{\bf k \cdot R_A}} c_{\bf R_A}^\dag f_A({\bf k}) + 
\frac{1}{\sqrt N}\sum_{\bf R_B}e^{i{\bf k \cdot R_B}} c_{\bf R_B}^\dag f_B({\bf k}) Z,
\label{Bloch}
\end{equation}
where we have explicitly separated the phase $Z$ ($Z^*Z=1$) which will be defined later so that analytical expressions would look as simple as possible. Let us define now Fourier transform of operators $c_{\bf R_A}^\dag$ and $c_{\bf R_B}^\dag$ as
\begin{equation}
A_{\bf k}^\dag = \frac{1}{\sqrt N}\sum_{\bf R_A}e^{i{\bf k \cdot R_A}} c_{\bf R_A}^\dag , \mbox{and}
\label{Fourier-A}
\end{equation}
\begin{equation}
B_{\bf k}^\dag = \frac{Z}{\sqrt N}\sum_{\bf R_B}e^{i{\bf k \cdot R_B}} c_{\bf R_B}^\dag .
\label{Fourier-B}
\end{equation}
Then, the Bloch eigenstate (\ref{Bloch}) is $c_{\bf k}^\dag =f_A({\bf k})A_{\bf k}^\dag + f_B({\bf k})B_{\bf k}^\dag$, and we can also write the inverse Fourier transforms (since $Z^*Z=1$) as
\begin{equation}
c_{\bf R_A}^\dag = \frac{1}{\sqrt N}\sum_{\bf k}e^{-i{\bf k \cdot R_A}} A_{\bf k}^\dag , \mbox{and}
\label{A:Inv-Fourier-A}
\end{equation}
\begin{equation}
c_{\bf R_B}^\dag = \frac{Z^*}{\sqrt N}\sum_{\bf k}e^{-i{\bf k \cdot R_B}} B_{\bf k}^\dag .
\label{A:Inv-Fourier-B}
\end{equation}
Let us look now at the first sum from (\ref{TBA}) and notice that every vector ${\bf R_A}-\mbox{\boldmath $\tau$}_l$ is in fact one of the $\bf R_B$ vectors, so we have
\begin{align}
\sum_{{\bf R_A},\mbox{\boldmath $\tau$}_l} 
c_{{\bf R_A}-\mbox{\boldmath $\tau$}_l}^\dag  c_{{\bf R_A}}^{} & =
\sum_{{\bf R_A},\mbox{\boldmath $\tau$}_l} 
\left(
\frac{Z^*}{\sqrt N}\sum_{\bf k'}e^{-i{\bf k'} \cdot 
           ({\bf R_A}-\mbox{\boldmath $\tau$}_l)} B_{\bf k'}^\dag
\right)
\left(
\frac{1}{\sqrt N}\sum_{\bf k}e^{i{\bf k} \cdot {\bf R_A}} A_{\bf k}
\right)
\nonumber \\
& = \sum_{{\bf k},{\bf k'},\mbox{\boldmath $\tau$}_l} Z^* 
e^{i{\bf k'} \cdot \mbox{\boldmath $\tau$}_l}
B_{\bf k'}^\dag A_{\bf k}
\sum_{\bf R_A} \frac{1}{N}e^{i({\bf k}-{\bf k'}) \cdot {\bf R_A}}.
\label{A:sum1}
\end{align}
However, since
\begin{equation}
\sum_{\bf R_A} \frac{1}{N}e^{i({\bf k}-{\bf k'}) \cdot {\bf R_A}} = \delta_{\bf k,k'} ,
\label{A:delta}
\end{equation}
we obtain in the first sum
\begin{equation}
\sum_{{\bf R_A},\mbox{\boldmath $\tau$}_l} 
c_{{\bf R_A}-\mbox{\boldmath $\tau$}_l}^\dag c_{{\bf R_A}}^{} 
=
\sum_{{\bf k},\mbox{\boldmath $\tau$}_l} Z^* 
e^{i{\bf k} \cdot \mbox{\boldmath $\tau$}_l} B_{\bf k}^\dag A_{\bf k}.
\label{A:sum11}
\end{equation}
In a similar manner we get the second sum
\begin{align}
\sum_{{\bf R_B},\mbox{\boldmath $\tau$}_l} 
c_{{\bf R_B}+\mbox{\boldmath $\tau$}_l}^\dag  c_{{\bf R_B}}^{} & =
\sum_{{\bf R_B},\mbox{\boldmath $\tau$}_l} 
\left(
\frac{1}{\sqrt N}\sum_{\bf k'}e^{-i{\bf k'} \cdot 
         ({\bf R_B}+\mbox{\boldmath $\tau$}_l)} A_{\bf k'}^\dag
\right)
\left(
\frac{Z}{\sqrt N}\sum_{\bf k}e^{i{\bf k} \cdot {\bf R_B}} B_{\bf k}
\right)
\nonumber \\
& = \sum_{{\bf k},\mbox{\boldmath $\tau$}_l} Z 
e^{-i{\bf k} \cdot \mbox{\boldmath $\tau$}_l}
A_{\bf k}^\dag B_{\bf k}.
\label{A:sum2}
\end{align}
Finally, the Hamiltonian (\ref{TBA}) becomes
\begin{equation}
H = -\gamma_0 \sum_{\bf k} 
\left( 
     \sum_l Z^* e^{i{\bf k}\cdot\mbox{\boldmath $\tau$}_l} 
     \cdot B_{\bf k}^\dag A_{\bf k}^{} + 
     \sum_l Z e^{-i{\bf k}\cdot\mbox{\boldmath $\tau$}_l} 
     \cdot A_{\bf k}^\dag B_{\bf k}^{} 
\right).
\label{Fourier-H}
\end{equation}
Relation (\ref{Fourier-H}) contains a specially important function
\begin{equation}
T({\bf k})=-\gamma_0\sum_l Z e^{-i{\bf k}\cdot\mbox{\boldmath $\tau$}_l},
\label{T}
\end{equation}
so finally we can write equation (\ref{Fourier-H}) in a matrix form
\begin{equation}
H = \sum_{\bf k}
\left( {\begin{array}{cc}
A_{\bf k}^\dag & B_{\bf k}^\dag
\end{array}} \right)
\left( {\begin{array}{cc}
0 & T({\bf k}) \\
T^*({\bf k}) & 0 \\
\end{array}} \right)
\left( {\begin{array}{c}
A_{\bf k} \\
B_{\bf k} \\
\end{array}} \right).
\label{H-matrix}
\end{equation}
Now, since the Bloch state 
$c_{\bf k}^\dag =f_A({\bf k})A_{\bf k}^\dag + f_B({\bf k})B_{\bf k}^\dag$
has to diagonalize this Hamiltonian, we can also write
\begin{equation}
H = \sum_{\bf k} E({\bf k}) c_{\bf k}^\dag c_{\bf k}^{} =
\sum_{\bf k} E({\bf k})
\left( {\begin{array}{cc}
A_{\bf k}^\dag & B_{\bf k}^\dag
\end{array}} \right)
\left( {\begin{array}{c}
f_A({\bf k}) \\
f_B({\bf k}) \\
\end{array}} \right)
\left( {\begin{array}{cc}
f_A^*({\bf k}) & f_B^*({\bf k})
\end{array}} \right)
\left( {\begin{array}{c}
A_{\bf k} \\
B_{\bf k} \\
\end{array}} \right).
\label{H-dijag}
\end{equation}
By comparing equations (\ref{H-matrix}) and (\ref{H-dijag}) we need to have:
\begin{equation}
\left( {\begin{array}{cc}
0 & T({\bf k}) \\
T^*({\bf k}) & 0 \\
\end{array}} \right)
\left( {\begin{array}{c}
f_A({\bf k}) \\
f_B({\bf k}) \\
\end{array}} \right)
=
E({\bf k})
\left( {\begin{array}{c}
f_A({\bf k}) \\
f_B({\bf k}) \\
\end{array}} \right).
\label{eigenvalue}
\end{equation}
So we have reduced entire problem to the matrix diagonalization, while the eigenvalues (i.e. energies) are given by:
\begin{equation}
\left| {\begin{array}{cc}
-E({\bf k}) & T({\bf k}) \\
T^*({\bf k}) & -E({\bf k}) \\
\end{array}} \right|
=0.
\label{determinanta}
\end{equation}
Solution of the determinant equation (\ref{determinanta}) determines the electron band structure in graphene as  \cite{Wallace1954}:
\begin{equation}
E_{\pm}({\bf k}) = \pm \sqrt{T({\bf k})T^*({\bf k})}=
\pm \gamma_0 \sqrt{1 + 4\cos{\frac{a k_x}{2}}\cos{\frac{a k_y\sqrt3}{2}} + 
4\cos^2{\frac{a k_x}{2}}}.
\label{e-vrpce}
\end{equation}

\begin{figure}
\centerline{
\mbox{\includegraphics[width=0.5\textwidth]{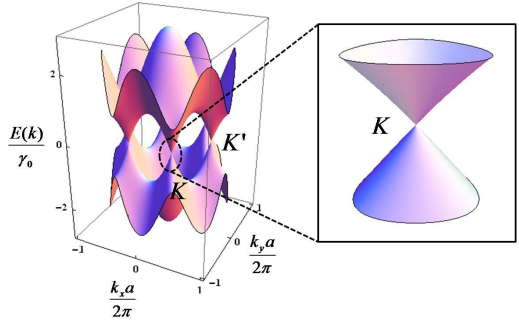}}
}
\caption{
Graphene electron band structure with Dirac cones around $\bf K$ point (magnified). Intrinsic graphene has Fermi level $E_F=E_{\bf K}=0$.
}
\label{Figure2_c2}
\end{figure}

Figure \ref{Figure2_c2} shows the function $E({\bf k})$ and we can notice the peculiar behavior of the bands at the Brillouin zone vertex points ${\bf K}$ and ${\bf K'}$. Further on, since each graphene unit cell contains two atoms in basis and each atom donates one free electron into the band, Fermi energy is defined such that there are enough electrons to fill precisely one Brillouin zone in the reciprocal space.  
Relation (\ref{e-vrpce}) tells us that electron bands are divided into positive and negative states that touch precisely at the the vertex point of the Brillouin zone (see also figure \ref{Figure2_c2}), such that we have $E_F=E_{\bf K}=0$. Because of the fact that electron states around the Fermi energy determines the low-energy properties, we will focus precisely on the area around the ${\bf K}$ and the ${\bf K'}$ points.
Finally we note that, since valence (negative) and conduction (positive) band touch at only 6 points (${\bf K}$, ${\bf K'}$ and the remaining four equivalent vertex points), that are located precisely at the Fermi level, the intrinsic graphene is an unusual zero gap semiconductor. 

\subsection{Dirac electron dispersion in graphene }
\label{Sec:Dirac}

We can write equation (\ref{eigenvalue}) as an eigenvalue equation: $H_{\bf k}\psi_{\bf k}=E_{\bf k}\psi_{\bf k}$, where the Hamiltonian and the wave function (eigenfunction) are given by
\begin{equation}
H_{\bf k}=
\left( {\begin{array}{cc}
0 & T({\bf k}) \\
T^*({\bf k}) & 0 \\
\end{array}} \right), \mbox{and}
\label{Hamilton1}
\end{equation}
\begin{equation}
\psi_{\bf k}=
\left( {\begin{array}{c}
f_A({\bf k}) \\
f_B({\bf k}) \\
\end{array}} \right).
\label{wavef}
\end{equation}
Let us focus now on the area around the $\bf K$ point and change the origin of our wave vector as ${\bf k} \rightarrow {\bf k+K}$, so that we have $|{\bf k}|<<|{\bf K}|$. Now we can make a Taylor expansion of the function $T({\bf k})$ as follows:
\begin{equation}
T({\bf k}) = -\gamma_0\sum_l Z e^{-i({\bf k+K})\cdot\mbox{\boldmath $\tau$}_l}
\approx -\gamma_0\sum_l Z e^{-i{\bf K}\cdot\mbox{\boldmath $\tau$}_l} 
(1-i{\bf k}\cdot\mbox{\boldmath $\tau$}_l).
\label{T-sum}
\end{equation}
Next we calculate the following sums:
\begin{equation}
\sum_l  e^{-i{\bf K}\cdot\mbox{\boldmath $\tau$}_l} = 0, \mbox{and}
\label{sum1}
\end{equation}
\begin{equation}
\sum_l \mbox{\boldmath $\tau$}_l e^{-i{\bf K}\cdot\mbox{\boldmath $\tau$}_l} = 
e^{-i2\pi/3} \frac{\sqrt3}{2} a (i \hat{x} + \hat{y}).
\label{sum2}
\end{equation}
Now we will choose the phase $Z=e^{-i\pi/3}$ so that we have
\begin{equation}
\sum_l \mbox{\boldmath $\tau$}_l Z e^{-i{\bf K}\cdot\mbox{\boldmath $\tau$}_l} = 
e^{-i\pi} \frac{\sqrt3}{2} a (i \hat{x} + \hat{y}) = 
\frac{\sqrt3}{2} a (-i \hat{x} - \hat{y}).
\label{sum22}
\end{equation}
Finally we get expressions for the function $T({\bf k})$ and the effective Hamiltonian $H_{\bf k}$ in the vicinity of point $\bf K$:
\begin{equation}
T({\bf k}) 
= \gamma_0 i {\bf k} \cdot \sum_l \mbox{\boldmath $\tau$}_l Z 
e^{-i{\bf K}\cdot\mbox{\boldmath $\tau$}_l}
= \frac{\sqrt 3}{2} a \gamma_0 (k_x - ik_y), \mbox{and}
\label{T-sumzum}
\end{equation}
\begin{equation}
H_{\bf k}= \frac{\sqrt3}{2} a \gamma_0
\left( {\begin{array}{cc}
0 & k_x - ik_y \\
k_x + ik_y & 0 \\
\end{array}} \right).
\label{Hamilton2}
\end{equation}
It is now convenient to introduce new variable: $\hbar v_F \equiv \frac{\sqrt3}{2} a \gamma_0$, where $v_F\approx 10^6$ m/s since $\gamma_0\approx 2.8$ eV \cite{Wallace1954}. Hamiltonian (\ref{Hamilton2}) now becomes
\begin{equation}
H_{\bf k}= 
\hbar v_F
\left[
\left( {\begin{array}{cc}
0 & 1 \\
1 & 0 \\
\end{array}} \right) k_x
+
\left( {\begin{array}{cc}
0 & -i \\
i & 0 \\
\end{array}} \right) k_y 
\right],
\label{Hamilton-Dirac}
\end{equation}
that is,
\begin{equation}
H_{\bf k}= \hbar v_F \mbox{\boldmath $\sigma$}\cdot{\bf k},
\label{Hamilton-Dirac2}
\end{equation}
where $\mbox{\boldmath $\sigma$}=\sigma_x\hat{\bf x}+\sigma_y\hat{\bf y}$, while $\sigma_x=
\left( {\begin{array}{cc}
0 & 1 \\
1 & 0 \\
\end{array}} \right)$
and 
$\sigma_y=
\left( {\begin{array}{cc}
0 & -i \\
i & 0 \\
\end{array}} \right)$
are the Pauli spin matrices. Here we note the remarkable property of graphene around $\bf K$ point where electrons behave precisely like massless Dirac particles of spin 1/2 \cite{Schiff}! We can also find energies (eigenvalues) and wave functions (eigenvectors) from the equation $H_{\bf k}\psi_{\bf k}=E_{\bf k}\psi_{\bf k}$:
\begin{equation}
E_{n,\bf k}= n \cdot \hbar v_F |{\bf k}| = 
n \cdot \hbar v_F  \sqrt{k_x^2+k_y^2} , \mbox{and}
\label{Energy-Dirac}
\end{equation}
\begin{equation}
\psi_{n,\bf k}({\bf r})= 
\langle {\bf r} | n{\bf k} \rangle=
\frac{1}{L\sqrt2}
\left( {\begin{array}{c}
n \\
e^{i \theta({\bf k})} \\
\end{array}} \right)
e^{i {\bf k}\cdot{\bf r}}.
\label{wavef-Dirac}
\end{equation}
Here $L^2$ is the area of graphene, $n=1$ ($n=-1$) denotes the conduction (valence) band, respectively, and the angle  $\theta({\bf k})=\tan^{-1}(k_y/k_x)$. 
Further on, we note that behavior around $K'$ point is easily found if we move the wave vector origin so that ${\bf k} \rightarrow {\bf k+K'}$. In that case it is more convenient to choose the phase $Z=1$ and the Hamiltonian (\ref{Hamilton1}) turns into $H'_{\bf k}=\hbar v_F \mbox{\boldmath $\sigma$}^*\cdot{\bf k}$.
Hamiltonian $H'_{\bf k}$ has eigenvalues:
$E'_{n,\bf k}= n \cdot \hbar v_F |{\bf k}|=E_{n,\bf k}$ that are degenerate with eigenvalues of Hamiltonian $H_{\bf k}$ so that $\bf K'$ point represents only an additional degree of freedom like electron spin. In other words we can limit ourself to the behavior around $\bf K$ point if we note that each state is four fold degenerate i.e. two spin and two valley (${\bf K}-{\bf K'}$) degenerate. 

Finally, let us find the electron density and electron current density operators for Dirac electrons in graphene. To start, note that the electron momentum is ${\bf p}=\hbar{\bf k}$, which can be written as an operator in the coordinate representation ${\bf p}=-i\hbar\mbox{\boldmath $\nabla$}$, so the Dirac Hamiltonian (\ref{Hamilton-Dirac2}) can be written as: 
$H=-i\mbox{\boldmath $\sigma$}\cdot\mbox{\boldmath $\nabla$}$. If we now describe this Dirac electron by a wave function 
$\psi_{\alpha}({\bf r})= \langle {\bf r}|\alpha\rangle$,
then the electron particle density is simply $\rho_\alpha=|\psi_{\alpha}({\bf r})|^2$ so the density operator is
\begin{equation}
\rho_{op}({\bf r})=\delta({\bf r}-{\bf r}_{op}).
\label{density_op_r}
\end{equation} 
To find the electron current density we can apply the equation of continuity: 
$-e\frac{\partial \rho}{\partial t}
+\mbox{\boldmath $\nabla$}\cdot {\bf j}=0$ (here we take $e>0$ so that $-e$ denotes the electron charge), with an equation of motion $H\psi=i\hbar\frac{\partial\psi}{\partial t}$. This yields the electron current density: 
${\bf j}_\alpha=-ev_F\psi_{\alpha}({\bf r})^*
\mbox{\boldmath $\sigma$}
\psi_{\alpha}({\bf r})$, i.e. the current density operator:
\begin{equation}
{\bf j}_{op}({\bf r})=
-e v_F \mbox{\boldmath $\sigma$} \delta({\bf r}-{\bf r}_{op}).
\label{current_density_op_r}
\end{equation} 
At last, the Fourier transforms of these quantities are given by
\begin{equation}
\rho_{op}({\bf q})=\frac{1}{L^2} e^{-i{\bf q}\cdot{\bf r}_{op}}, \mbox{and}
\label{density_op_q}
\end{equation}
\begin{equation}
{\bf j}_{op}({\bf q})=
-\frac{e v_F}{L^2} \mbox{\boldmath $\sigma$} e^{-i{\bf q}\cdot{\bf r}_{op}}.
\label{current_density_op_q}
\end{equation}


\subsection{Electron-phonon interaction}
\label{Sec:Electron-phonon}

\begin{figure}
\centerline{
\mbox{\includegraphics[width=0.5\textwidth]{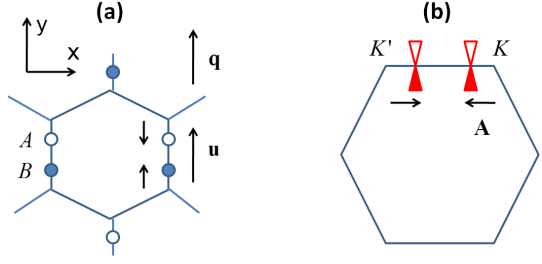}}
}
\caption{
a) Atom motion in graphene during the longitudinal optical phonon oscillations.
b) Motion of basis atoms described by a vector $\bf u$ in the real space induces vector potential  $\bf A$ that moves Dirac points in the reciprocal space. Dirac point symmetry causes unusual polarization of this vector potential: ${\bf A} \perp {\bf u}$.
}
\label{Figure3_c2}
\end{figure}

Since graphene is a 2D crystal with two atoms per basis, there are also two optical phonon branches (transverse and longitudinal) that are degenerate at energy $\hbar\omega_0=0.196$ eV and mostly independent of wave vector $q$ (for long wave modes $q<<2\pi/a$).
Let us denote by 
${\bf{u}}({\bf R})=[{\bf u}_A({\bf R})-{\bf u}_B({\bf R})]/\sqrt{2}$ 
motion of the basis atom A relative to the atom B in the unit cell at the position $\bf R$ (see figure \ref{Figure3_c2}).
If A and B where oppositely charged ions like in polar crystals, then their motion would result in the electric dipole moment i.e. electric field in the direction of the vector ${\bf{u}}$ and strong electron-phonon interaction. However, since A and B are completely equivalent carbon atoms, graphene belongs to the class of covalent crystals, and electron-phonon interaction is considerably reduced compared to the case of polar crystals. 
We will also see that electron-phonon interaction in graphene acquires unusual form in the vicinity of the Dirac ($\bf K$ and $\bf K'$) point and we will demonstrate that optical phonon oscillation creates effective electric field that is perpendicular to the vector ${\bf{u}}$. That fact will lead to peculiar mixing of plasmon and optical phonon polarizations.

The rigorous calculation of electron-phonon interaction in graphene is given in references \cite{Ando_opt_ph, Ando_el_ph}, while we only sketch here the main steps. Let us start with the tight binding Hamiltonian (\ref{Hamilton1})
\begin{equation}
H_{\bf k}=
\left( {\begin{array}{cc}
0 & T({\bf k}) \\
T^*({\bf k}) & 0 \\
\end{array}} \right).
\label{Hamilton11}
\end{equation}
The effect of phonon on the electron motion can be simply found by considering the change in the hopping integral ($-\gamma_0$) with the change in the nearest neighbor distance. 
Let us now observe atom A at a position $\bf R_A$ and a neighboring atom B at a position 
${\bf R_B}={\bf R_A}-\mbox{\boldmath $\tau$}_l$ whose equilibrium relative distance is simply $|\mbox{\boldmath $\tau$}_l|=b$. If we move these two atoms out of equilibrium positions, new distance is: 
$|\mbox{\boldmath $\tau$}_l+{\bf u_A}({\bf R_A})-
{\bf u_B}({\bf R_A}-\mbox{\boldmath $\tau$}_l)|$,
and the leading order change in the hopping integral is
\begin{align}
-\gamma & = - \gamma_0 - \frac{\partial \gamma_0(b)}{\partial b} 
\left[
|\mbox{\boldmath $\tau$}_l+{\bf u_A}({\bf R_A})-{\bf u_B}({\bf R_A} -\mbox{\boldmath $\tau$}_l)|-b
\right] 
\nonumber \\
& \approx - \gamma_0 - \frac{\partial \gamma_0(b)}{\partial b} 
\frac{1}{b}\mbox{\boldmath $\tau$}_l \cdot
\left[
{\bf u_A}({\bf R_A})-{\bf u_B}({\bf R_A}-\mbox{\boldmath $\tau$}_l)
\right] .
\label{delta-gamma}
\end{align}

Since we are interested in long wavelength optical phonons ($q<<2\pi/a$), then instead of discrete vector $\bf R$, we can write a continuous coordinate $\bf r$ in the expression ${\bf{u}}({\bf R})=[{\bf u}_A({\bf R})-{\bf u}_B({\bf R})]/\sqrt{2}$, i.e. we write
\begin{equation}
{\bf u_A}({\bf R_A})-{\bf u_B}({\bf R_A}-\mbox{\boldmath $\tau$}_l)
\approx
{\bf u_A}({\bf r})-{\bf u_B}({\bf r}-\mbox{\boldmath $\tau$}_l)
\approx
{\bf u}({\bf r})\sqrt 2 .
\label{longwave}
\end{equation}
Finally the change in the hopping integral (\ref{delta-gamma}), in the long wavelength limit, is given by:
\begin{equation}
-\gamma \approx - \gamma_0 - \frac{\partial \gamma_0(b)}{\partial b} 
\frac{\sqrt 2}{b}\mbox{\boldmath $\tau$}_l \cdot {\bf u}({\bf r}).
\label{delta-gamma2}
\end{equation}
Note here that all three neighboring carbon atoms ($l=1,2,3$) see the same phonon amplitude ${\bf u}({\bf r})$, which will not be true in the case of finite wavevector $q$. However this change in amplitude will come with an extra factor $q\cdot a$, so unless we are working with phonon wavevectors on the order of Brillouin zone, the long wavelength limit is a great approximation concerning the interaction between electrons and optical phonons.  

We can write phonon motion ${\bf u}({\bf r})$ as a sum over normal modes
\begin{equation}
{\bf u}({\bf r})= \sum_{{\bf q},\mu} \frac{1}{\sqrt{NM}}Q_{{\bf q}\mu}
{\bf e}_{{\bf q}\mu} e^{i{\bf q}\cdot{\bf r}}.
\label{normal-mode}
\end{equation}
Here $M$ is the mass of a carbon atom, $\mu=L,T$ denotes longitudinal i.e. transverse polarization, and if we define an angle $\varphi({\bf q})=\tan^{-1}(q_y/q_x)$, then polarization vectors are given by
\begin{equation}
{\bf e}_{{\bf q}L}=
i(\cos\varphi({\bf q})\hat{x}+\sin\varphi({\bf q})\hat{y}) ,\mbox{and}
\label{polari-L}
\end{equation}
\begin{equation}
{\bf e}_{{\bf q}T}=
i(-\sin\varphi({\bf q})\hat{x}+\cos\varphi({\bf q})\hat{y}).
\label{polari-T}
\end{equation}
Finally we can write the phonon amplitude $Q_{{\bf q}\mu}$ through the creation ($b_{{\bf q}\mu}^\dag$) and annihilation operators ($b_{{\bf q}\mu}$) as
\begin{equation}
Q_{{\bf q}\mu}=\sqrt{\frac{\hbar}{2\omega_0}}
(b_{{\bf q}\mu} + b_{-{\bf q}\mu}^\dag).
\label{phonon-ampl}
\end{equation}
To find how the phonon motion ${\bf u}({\bf r})$ influences the electrons around the $\bf K$ point let us change the origin of wave vector as before: ${\bf k} \rightarrow {\bf k+K}$. Now the function $T({\bf k})$ becomes
\begin{align}
T({\bf k},{\bf r}) & = \sum_l (-\gamma) Z 
e^{-i({\bf k}+{\bf K})\cdot\mbox{\boldmath $\tau$}_l}
\nonumber \\
& = \sum_l 
\left( 
-\gamma_0 - \frac{\partial \gamma_0(b)}{\partial b} 
\frac{\sqrt2}{b}\mbox{\boldmath $\tau$}_l \cdot {\bf u}({\bf r}) 
\right)
Z e^{-i({\bf k}+{\bf K})\cdot\mbox{\boldmath $\tau$}_l}
\nonumber \\
& \approx \sum_l 
\left( 
-\gamma_0 - \frac{\partial \gamma_0(b)}{\partial b} 
\frac{\sqrt2}{b}\mbox{\boldmath $\tau$}_l \cdot {\bf u}({\bf r}) 
\right)
Z e^{-i{\bf K}\cdot\mbox{\boldmath $\tau$}_l}
(1-i{\bf k} \cdot \mbox{\boldmath $\tau$}_l).
\label{phonon-T}
\end{align}
By looking at the leading order expansion in the phonon motion ${\bf u}({\bf r})$ and electron wave vector ${\bf k}$ we have
\begin{equation}
T({\bf k},{\bf r}) =  
\left( 
\gamma_0 i {\bf k} - \frac{\partial \gamma_0(b)}{\partial b} 
\frac{\sqrt2}{b} {\bf u({\bf r})}
\right)
\cdot
\sum_l \mbox{\boldmath $\tau$}_l Z e^{-i{\bf K}\cdot\mbox{\boldmath $\tau$}_l}.
\label{phonon-T2}
\end{equation}
We recognize the first part of the expression (\ref{phonon-T2}) from the equation (\ref{T-sumzum}) for bare Dirac electrons
\begin{equation}
T_0({\bf k}) 
= \gamma_0 i {\bf k} \cdot \sum_l \mbox{\boldmath $\tau$}_l Z 
e^{-i{\bf K}\cdot\mbox{\boldmath $\tau$}_l}
= \frac{\sqrt3}{2} a \gamma_0 (k_x - ik_y) ,
\label{T-sumzum2}
\end{equation}
while the other part of the sum (\ref{phonon-T2}) gives
\begin{align}
T_{e-ph}({\bf r}) & =  
- \frac{\partial \gamma_0(b)}{\partial b} 
\frac{\sqrt2}{b} {\bf u({\bf r})}
\cdot
\sum_l \mbox{\boldmath $\tau$}_l Z e^{-i{\bf K}\cdot\mbox{\boldmath $\tau$}_l} 
\nonumber \\
& =
- \frac{\partial \gamma_0(b)}{\partial b} 
\frac{\sqrt2}{b} \frac{\sqrt3}{2} a (-iu_x-u_y).
\label{T-e-ph}
\end{align}
With the substitution: $\hbar v_F =\frac{\sqrt3}{2} a \gamma_0$, expressions above transform into a simpler form
\begin{equation}
T_0({\bf k}) = \hbar v_F (k_x - ik_y) , \mbox{and}
\label{T-sumzum3}
\end{equation}
\begin{equation}
T_{e-ph}({\bf r}) =  
\hbar v_F\frac{\partial \gamma_0(b)}{\partial b} 
\frac{\sqrt 2}{b\gamma_0}  (iu_x+u_y).
\label{T-e-ph2}
\end{equation}
Finally since $T({\bf k},{\bf r})=T_0({\bf k})+T_{e-ph}({\bf r})$, we can also write for the total Hamiltonian $H_{\bf k}=H_{\bf k}^0+H_{e-ph}$ where
\begin{equation}
H_{\bf k}^0=
\left( {\begin{array}{cc}
0 & T_0({\bf k}) \\
T_0^*({\bf k}) & 0 \\
\end{array}} \right)
=
\hbar v_F
\left( {\begin{array}{cc}
0 & k_x - ik_y \\
k_x + ik_y & 0 \\
\end{array}} \right) , \mbox{and}
\label{Hamilton0}
\end{equation}
\begin{equation}
H_{e-ph}=
\left( {\begin{array}{cc}
0 & T_{e-ph} \\
T_{e-ph} & 0 \\
\end{array}} \right)
=
\hbar v_F\frac{\partial \gamma_0(b)}{\partial b} 
\frac{\sqrt2}{b\gamma_0}
\left( {\begin{array}{cc}
0 & u_y + iu_x \\
u_y - iu_x & 0 \\
\end{array}} \right).
\label{Hamilton-e-ph}
\end{equation}
If we introduce here the notation 
$\mbox{\boldmath $\sigma$} \times {\bf{u}}=\sigma_x u_y-\sigma_y u_x$
, then we can write equation (\ref{Hamilton-e-ph}) in a convenient form \cite{Ando_anomaly, Ando_opt_ph, Ando_el_ph}:
\begin{equation}
H_{e-ph} = -\hbar v_F \frac{\partial \gamma_0(b)}{\partial b} 
\frac{\sqrt2}{b\gamma_0} \mbox{\boldmath $\sigma$} \times {\bf{u(r)}}.
\label{Hamilton-e-ph-convenient}
\end{equation}
From this expression we can immediately see the unusual property of the electron-phonon interaction in the vicinity of Dirac point. Namely the total Hamiltonian $H_{\bf k}$ in the presence of the phonons, can be obtained from the bare Hamiltonian 
$H_{\bf k}^0=\hbar v_F \mbox{\boldmath $\sigma$}\cdot{\bf k}$, by a simple substitution:  
\begin{equation}
k_x \rightarrow k_x + Ku_y ,
\label{kx}
\end{equation}
\begin{equation}
k_y \rightarrow k_y - Ku_x ,
\label{ky}
\end{equation}
where $K=\frac{\partial \gamma_0(b)}{\partial b}\frac{\sqrt2}{b\gamma_0}$. But this is precisely equivalent to the action of the vector potential $\bf A$:
\begin{equation}
\hbar{\bf k} \rightarrow \hbar{\bf k} + e{\bf A}.
\label{vectorA}
\end{equation}
In other words influence of phonons on the electron motion is equivalent to the presence of vector potential with components: $A_x \propto u_y$ and $A_y \propto -u_x$. This will in turn lead to the unusual mixing of plasmon and optical phonon polarizations. To understand this let us assume that the phonon wave vector is oriented in the $y$ direction (${\bf q}=q\hat{\bf y}$) and let us look at the longitudinal optical phonon motion that have $u_x=0$ and $u_y\neq0$ (see figure  \ref{Figure3_c2}).
Then the phonon influence is given by transverse vector potential since $A_x\neq0$ and $A_y=0$. In other words longitudinal phonon oscillation is equivalent to the transverse vector potential oscillation i.e. transverse electric field. On the other hand, since plasmons are collective charge density oscillations, accompanied by a longitudinal electric field, there won't be any interaction between plasmon and longitudinal optical phonon. We will further show that there is a strong interaction of plasmon and transverse optical phonon which is a very counter intuitive result from the perspective of polar crystals. 

The simplest way to analyze the electron-phonon interaction in graphene is to show how the phonon amplitude couples to the electron current density. In that regards let us take electron-phonon Hamiltonian (\ref{Hamilton-e-ph-convenient}) and write expansion of the phonon motion ${\bf u}({\bf r})$ over the normal modes from equation (\ref{normal-mode}) to obtain:
\begin{equation}
H_{e-ph} = -\hbar v_F \frac{\partial \gamma_0(b)}{\partial b} 
\frac{\sqrt2}{b\gamma_0} \frac{1}{\sqrt{NM}}
\sum_{{\bf q},\mu}  e^{i{\bf q}\cdot{\bf r}}
\mbox{\boldmath $\sigma$} \times {\bf e}_{{\bf q}\mu}
Q_{{\bf q}\mu}.
\label{Hamilton-e-ph-convenient_expand}
\end{equation}
Here we recognize the current density operator
${\bf j}_{\bf q}^\dag=
-\frac{e v_F}{L^2} \mbox{\boldmath $\sigma$} e^{i{\bf q}\cdot{\bf r}}$
from equation (\ref{current_density_op_q}), and if we introduce the factor 
$F=\frac{\hbar}{e} \frac{\partial \gamma_0(b)}{\partial b} 
\frac{\sqrt2}{b\gamma_0} \frac{1}{\sqrt{NM}}$, we can finally write for the electron-phonon interaction Hamiltonian:
\begin{equation}
H_{e-ph} = L^2 F \sum_{{\bf q},\mu}  
{\bf j}_{\bf q}^\dag \times {\bf e}_{{\bf q}\mu}
Q_{{\bf q}\mu}.
\label{Hamilton-e-ph-current}
\end{equation}

A more convenient way to write electron-phonon interaction is to show how the phonon amplitude couples to the electron density. In that respect, let us define the quantities: 
\begin{equation}
E_{{\bf q}L} \equiv
e_{{\bf q}L}\cdot\hat{y}+ie_{{\bf q}L}\cdot\hat{x} =
i \sin\varphi({\bf q}) - \cos\varphi({\bf q})=
-e^{-i\varphi({\bf q})} , \mbox{and}
\label{eL}
\end{equation}
\begin{equation}
E_{{\bf q}T} \equiv
e_{{\bf q}T}\cdot\hat{y}+ie_{{\bf q}T}\cdot\hat{x} =
i \cos\varphi({\bf q}) + \sin\varphi({\bf q})=ie^{-i\varphi({\bf q})} .
\label{eT}
\end{equation}
Then by using the normal mode expansion (\ref{normal-mode}) we obtain
\begin{equation}
u_y+iu_x=
\sum_{{\bf q},\mu} \frac{1}{\sqrt{NM}}Q_{{\bf q}\mu}
(e_{{\bf q}\mu}\cdot\hat{y}+ie_{{\bf q}\mu}\cdot\hat{x})
e^{i{\bf q}\cdot{\bf r}}
=
\sum_{{\bf q},\mu} \frac{1}{\sqrt{NM}}Q_{{\bf q}\mu}
E_{{\bf q}\mu} e^{i{\bf q}\cdot{\bf r}}.
\label{ef1}
\end{equation}
Finally the electron-phonon interaction Hamiltonian (\ref{Hamilton-e-ph}) can be written as:
\begin{equation}
H_{e-ph}=
\hbar v_F\frac{\partial \gamma_0(b)}{\partial b} 
\frac{\sqrt2}{b\gamma_0}
\sum_{{\bf q},\mu} \frac{1}{\sqrt{NM}}Q_{{\bf q}\mu}
\left( {\begin{array}{cc}
0 &  E_{{\bf q}\mu}\\
E_{{\bf q}\mu}^* & 0 \\
\end{array}} \right) 
e^{i{\bf q}\cdot{\bf r}} .
\label{Hamilton-e-ph2}
\end{equation}
If we now define
\begin{equation}
g \equiv
\hbar v_F
\frac{\partial \gamma_0(b)}{\partial b} 
\frac{\sqrt2}{b\gamma_0}
\frac{1}{\sqrt{NM}} , \mbox{and}
\label{g}
\end{equation}
\begin{equation}
M_{{\bf q}\mu} \equiv
\left( {\begin{array}{cc}
0 &  E_{{\bf q}\mu}\\
E_{{\bf q}\mu}^* & 0 \\
\end{array}} \right) ,
\label{M}
\end{equation}
then we can write electron-phonon interaction as a coupling between phonon amplitude 
$Q_{{\bf q}\mu}$ and electron density operator $\rho_{\bf q}^\dag$ from equation (\ref{density_op_q}) as:
\begin{equation}
H_{e-ph}=
L^2 \sum_{{\bf q},\mu} g M_{{\bf q}\mu} \rho_{\bf q}^\dag Q_{{\bf q}\mu}.
\label{Hamilton-e-ph-density}
\end{equation}

Formula (\ref{Hamilton-e-ph-current}) and (\ref{Hamilton-e-ph-density}) are equivalent. However, the response of the system to the interaction Hamiltonian (\ref{Hamilton-e-ph-current}) is most easily described by utilizing the current-current response function while the response to the interaction Hamiltonian  (\ref{Hamilton-e-ph-density}) is most easily described by the density-density response function. 


\section{Response functions}

In this section we present response functions which describe response of our system to an external perturbation. Specifically we calculate graphene's conductivity, density-density, and current-current response functions in the weak coupling approximation i.e. linear response theory. These three functions are all connected by simple relations, however it will be more convenient to use one or another depending on the specific nature of the problem being studied. 

\subsection{Conductivity}

\subsubsection{Semiclassical model - Drude conductivity}

If we are only interested in the response of the graphene under the influence of external electromagnetic field, we can simply calculate the conductivity function $\sigma(\omega)$. The semiclassical model gives a simple relation for the Drude conductivity \cite{AM}
\begin{equation}
\mbox{\boldmath $\sigma$}_D(\omega)=e^2\int 4\frac{d{\bf k}}{4\pi^2}
\frac{{\bf v}({\bf k}){\bf v}({\bf k})}{1/\tau-i\omega}
\left( -\frac{\partial f}{\partial E} \right)_{E=E({\bf k})} ,
\label{semiclassical_conductivity_AM}
\end{equation}
where $\tau$ is the relaxation time, 
${\bf v}({\bf k})=\frac{1}{\hbar}\frac{\partial E({\bf k})}{\partial{\bf k}}$ is the electron velocity, $f(E)=\frac{1}{e^{(E-\mu)/kT}+1}$ is the Fermi-Dirac distribution function, and factor $4$ stands for two spin and two valley degeneracy.
Semiclassical model is simply a generalization of the Drude model for free electrons to the case of an arbitrary band structure $E({\bf k})$, however we will see that it can describe lot of interesting phenomena in a qualitatively correct way. At zero temperature one has $-\partial f/\partial E=\delta(E-\mu)$ and it is straightforward to show that for the case of Dirac electrons in graphene
$H_{\bf k}= \hbar v_F \mbox{\boldmath $\sigma$}\cdot{\bf k}$, 
the Drude conductivity is given by
\begin{equation}
\sigma_D(\omega)=e^2\frac{\mu}{\pi\hbar^2}\frac{1}{1/\tau-i\omega}.
\label{semiclasical_conductivity_AM_0T}
\end{equation}
It is a slightly more tedious task to show that at finite temperature $T$ one has
\begin{equation}
\sigma_D(\omega)=e^2\frac{2 k T}{\pi\hbar^2}
\ln\left( 2 \cosh \frac{\mu}{2 k T} \right)
\frac{1}{1/\tau-i\omega} .
\label{semiclasical_conductivity_AM_finiteT}
\end{equation}

\subsubsection{Fermi's golden rule - interband conductivity}

The semiclassical model has a serious limitation since it cannot describe transitions between different bands \cite{AM}, which is particularly important in graphene that has zero band-gap between valence and conduction bands. 
To take into account these interband transitions we will calculate the response of graphene to an external electric field, in the first order perturbation theory (using the Fermi's golden rule). 

Let us imagine that an electromagnetic plane wave of frequency $\omega$ is incident under the normal angle onto the graphene sheet. We can choose the gauge so that the scalar potential $\varphi=0$, while the vector potential 
${\bf A}={\bf A}_0e^{-i\omega t}$, so that the electric field is given by 
${\bf E}=-\frac{\partial {\bf A}}{\partial t}=i\omega{\bf A}=
{\bf E}_0e^{-i\omega t}$
, and ${\bf E}_0=i\omega{\bf A}_0$. Electrons in graphene are described by a Dirac Hamiltonian (\ref{Hamilton-Dirac2})
$H_0= \hbar v_F \mbox{\boldmath $\sigma$}\cdot{\bf k}=
v_F \mbox{\boldmath $\sigma$}\cdot{\bf p}$, where ${\bf p}$ is the graphene's electron momentum so that the interaction with the vector potential is simply described by a substitution ${\bf p} \rightarrow {\bf p}+e{\bf A}$. In other words, the total Hamiltonian in the presence of electromagnetic field can be written as 
$H=v_F \mbox{\boldmath $\sigma$}\cdot({\bf p}+e{\bf A})=H_0+H_{int}$, where the interaction part of the Hamiltonian is given by
\begin{equation}
H_{int}=e v_F \mbox{\boldmath $\sigma$}\cdot{\bf A} =
\frac{e v_F}{i\omega} \mbox{\boldmath $\sigma$}\cdot{\bf E}_0 e^{-i\omega t}.
\label{H_int_A}
\end{equation}
Here we have kept only the time dependent part ($e^{-i\omega t}$) responsible for the absorption process. Then the Fermi's golden rule \cite{Sakurai} gives the probability for a transition from an initial state $i$ to the final state $f$, with an absorption of a photon:
\begin{equation}
\frac{d w_{i\rightarrow f}}{dt}=
\frac{2\pi}{\hbar}|\langle i|H_{int}|f \rangle|^2 
\delta(\hbar\omega_{if}-\hbar\omega) f_i (1-f_f).
\label{Fermi_golden_rule}
\end{equation}
The total power absorbed from the incident wave can be written in two ways. First, one can write
\begin{equation}
P_a=\sum_{i,f}\hbar\omega\frac{d w_{i\rightarrow f}}{dt} .
\label{P_fermi_golden_rule}
\end{equation}
On the other hand, since ${\bf j}(\omega)=\sigma(\omega) {\bf E}(\omega)$, one can write (for harmonic fields) \cite{Jackson} 
\begin{equation}
P_a=2\Re\int {\bf j}(\omega)\cdot{\bf E}^*(\omega) d{\bf r} = 
2\Re\sigma(\omega) |{\bf E}_0|^2 L^2 ,
\label{P_conductivity}
\end{equation}
where $L^2$ is the area of graphene sheet, and we used the fact that 
${\bf E}(\omega)={\bf E}_0$ is uniform along the graphene plane for the case of normal incident wave. Finally we have
\begin{equation}
\Re\sigma(\omega)=\frac{\hbar\omega}{2|{\bf E}_0|^2 L^2} \sum_{i,f}
\frac{2\pi}{\hbar}|\langle i|H_{int}|f \rangle|^2 
\delta(\hbar\omega_{if}-\hbar\omega) f_i (1-f_f) .
\label{Re_conductivity_fermi_golden_rule}
\end{equation}
Now, let us denote the initial (final) state of the electron by a band index $n$ ($n'$) and a wave vector $\bf k$ ($\bf k'$) i.e. $|i\rangle=|n {\bf k}\rangle$ ($|f\rangle=|n' {\bf k'}\rangle$). Without loss of generality we can assume that the electric field is polarized along the $x$ direction: 
${\bf E}_0=E_0\hat{\bf x}$. Then we can write for the matrix element:
\begin{equation}
\langle i|H_{int}|f \rangle=\frac{e v_F}{i\omega} E_0 
\langle n' {\bf k'}|\sigma_x|n {\bf k} \rangle .
\label{H_if}
\end{equation}
Further on, by using explicit form (\ref{wavef-Dirac}) for the Dirac electron wave function $\psi_{n,\bf k}$, it is simple to show that
\begin{equation}
\langle n' {\bf k'}|\sigma_x|n {\bf k} \rangle=
\int \psi_{n',\bf k'}^*({\bf r}) \sigma_x \psi_{n,\bf k}({\bf r}) d{\bf r} =
\frac{1}{2}\delta_{{\bf k},{\bf k'}}
(n e^{-i\theta_{\bf k}}+n' e^{i\theta_{\bf k}}) ,
\label{matric_element_sigma_x}
\end{equation}
so we obtain expressions for the matrix element
\begin{equation}
|\langle i|H_{int}|f \rangle|^2=\frac{e^2 v_F^2}{\omega^2} |E_0|^2
\delta_{{\bf k},{\bf k'}}\frac{1}{2}(1+n n'\cos 2\theta_{\bf k}) ,
\label{matrix_element}
\end{equation}
and conductivity
\begin{align}
\Re\sigma(\omega)=\frac{\hbar\omega}{2}4\sum_{n,n'} &
\frac{1}{4\pi^2}\int kdk \int d\theta_{\bf k}
\frac{2\pi}{\hbar}\frac{e^2 v_F^2}{\omega^2}
\frac{1}{2}(1+n n'\cos 2\theta_{\bf k}) \times
\nonumber \\
& \delta( n'\hbar v_F k-n\hbar v_F k-\hbar\omega)
f_{n k}(1-f_{n' k}) ,
\label{Re_conductivity_fermi_golden_rule_long} 
\end{align}
where we also took into account 2 spin and 2 valley degeneracy. 
We now take into account only (interband) transitions between conduction and valence bands, because the intraband transitions are already taken into account by the Drude conductivity.  
After lengthy but straightforward calculation, one obtains simple expression for the real part of the conductivity
\begin{equation}
\Re\sigma(\omega)=\frac{e^2}{4\hbar}
f(-\hbar\omega/2)[1-f(\hbar\omega/2)] .
\label{Re_conductivity_fermi_golden_rule_final}
\end{equation}
It is instructive to look at this result at zero temperature
\begin{equation}
\Re\sigma(\omega)=\frac{e^2}{4\hbar} \theta(\hbar\omega-2\mu) .
\label{Re_conductivity_fermi_golden_rule_zeroT}
\end{equation}
Here $\theta(x)$ is a simple step function [$\theta(x<0)=0$ and $\theta(x>0)=1$]. 
The real part of the conductivity provides us with absorption of the electromagnetic field incident on a graphene sheet. We see that there is no absorption for $\hbar\omega<2\mu$ which is result of the Pauli exclusion principle. On the other hand above this threshold, when $\hbar\omega>2\mu$ one will have uniform absorption. Since the incident energy flux is given by $W_i=2|E_0|^2/\mu_0 c$ (see reference \cite{Jackson}), and the absorbed energy per unit time per unit area is given by 
$W_a=P_a/L^2=2\Re\sigma(\omega)|E_0|^2$ (see equation (\ref{P_conductivity})),
the absorption coefficient can be written as
\begin{equation}
|a|^2=\frac{W_a}{W_i}=
\frac{\mu_0 c e^2}{4\hbar} = 2.3\% .
\label{absorption_coefficient}
\end{equation}
This result has been confirmed by experiment \cite{Nair2008}.
Further on, note that if we include the emission process, then we obtain the following expression for the conductivity:
\begin{align}
\Re\sigma(\omega) & = \frac{e^2}{4\hbar} 
[f(-\hbar\omega/2)[1-f(\hbar\omega/2)]-
[1-f(-\hbar\omega/2)]f(\hbar\omega/2)]
\nonumber \\
& = \frac{e^2}{4\hbar} 
[f(-\hbar\omega/2)-f(\hbar\omega/2)] .
\label{Re_conductivity_fermi_golden_rule_all}
\end{align}
Finally we can obtain the imaginary part of the conductivity by using the Kramers-Kronig relations \cite{PinesBook}:
\begin{align}
\Im\sigma(\omega) & = -\frac{2\omega}{\pi} 
{\cal P}\int_{0}^{\infty}
\frac{\Re\sigma(\omega')}{\omega'^2-\omega^2}d\omega'
\nonumber \\
& = -\frac{e^2}{4\hbar} \frac{4\hbar\omega}{\pi}
{\cal P}\int_{0}^{\infty}
\frac{f(-\epsilon)-f(\epsilon)}{(2\epsilon)^2-(\hbar\omega)^2}d\epsilon .
\label{Kramers_Kronig}
\end{align}
It is convenient to introduce the following function:
\begin{equation}
G(\epsilon)\equiv f(-\epsilon)-f(\epsilon)=
\frac{\sinh \frac{\epsilon}{kT}}
{\cosh \frac{\mu}{kT}+\cosh \frac{\epsilon}{kT}} .
\label{function_G}
\end{equation}
Then, we can simply write for the total interband conductivity 
(see also \cite{Falkovsky2008}):
\begin{equation}
\sigma_I(\omega) = \frac{e^2}{4\hbar} 
\left( 
G(\omega/2) + i \frac{4\hbar\omega}{\pi} \int_{0}^{\infty}
\frac{G(\epsilon)-G(\hbar\omega/2)}{(2\epsilon)^2-(\hbar\omega)^2}d\epsilon
\right) .
\label{Interband_conductivity}
\end{equation}
In the last expression, we took into account that principal value of the integral with $G(\hbar\omega/2)$ equals to zero, which removes singularities from the integral in the imaginary part of the conductivity.


\subsection{Density-density response function}

We now proceed to a more formal, but powerful, aspect of linear response theory by looking into the density-density response function. In the last section we assumed that there is no spatial dependence of external perturbation and calculated only frequency dependence of the conductivity. Let us assume that graphene is placed in an external scalar potential of arbitrary spatial and time dependence
\begin{equation}
\varphi_{ext}({\bf r}, t)= 
\int e^{-i\omega t} d\omega 
\sum_{\bf q} e^{i{\bf q}\cdot {\bf r}}
\varphi_{ext}({\bf q},\omega) .
\label{electrostatic_potential}
\end{equation}
Now, scalar potential simply couples to the electron charge density so one can write the interaction Hamiltonian \cite{PinesBook}
\begin{equation}
H_{int}= \int e^{-i\omega t} d\omega L^2 \sum_{\bf q} 
(-e \varphi_{ext}({\bf q},\omega))\rho_{\bf q}^\dag .
\label{Hamilt_poten_density}
\end{equation}
We now assume the weak coupling between the system (electron density) and a probe (external potential) so that we can focus on a single $({\bf q},\omega)$ component. The induced electron particle density is then given by
\begin{equation}
\langle \rho_{ind}({\bf q},\omega) \rangle = 
\chi ({\bf q},\omega) (-e \varphi_{ext}({\bf q},\omega)) ,
\label{density_response}
\end{equation}
where the density-density response function is given by \cite{PinesBook}
\begin{equation}
\chi({\bf q},\omega)=L^2\sum_{a,b}\frac{e^{-\beta E_b}}{Z}
|\langle a | \rho_{\bf q}^\dag | b \rangle|^2 
\left( \frac{1}{\hbar\omega-\hbar\omega_{ab}+i\eta} 
-\frac{1}{\hbar\omega+\hbar\omega_{ab}+i\eta} \right).
\label{d_d_response}
\end{equation}
Here $Z=\sum_b e^{-\beta E_b}$ is the partition function, and $\hbar\omega_{ab}=E_a-E_b$. Further on $|a\rangle$, and $E_a$ are exact many body state, and energy of the system in the presence of the perturbation. In other words we can write $H |a\rangle = E_a |a\rangle$ where $H=H_0+H_{int}$ is the total system Hamiltonian given by the sum of the Hamiltonian in the absence of perturbation ($H_0$) and the interaction term ($H_{int}$). Equation (\ref{d_d_response}) is exact in the limit of weak coupling (i.e. linear response), however one first needs to find the exact eigenstates of the total Hamiltonian $H$ which is not an easy task. We shall deal with this issue by working in the self-consistent approximation i.e. by introducing simple, yet powerful, concept of screening. In that regard let us note that the induced charge density $\langle\rho_{ind}({\bf q},\omega)\rangle$ will be accompanied by an scalar potential $\varphi_{ind}({\bf q},\omega)$ which can act back on the electrons through the interaction Hamiltonian (\ref{Hamilt_poten_density}). In other words, instead of equation (\ref{density_response}) we should write the self-consistent equation for the total induced particle density
\begin{equation}
\langle \rho_{ind}({\bf q},\omega) \rangle = 
\chi ({\bf q},\omega) 
(-e \varphi_{ext}({\bf q},\omega)-e \varphi_{ind}({\bf q},\omega)) .
\label{density_response_self_consistent}
\end{equation}
However, $\chi ({\bf q},\omega)$ is now the screened density-density response function which is again given by the equation (\ref{d_d_response}), only $|a,b \rangle$ are now simply the eigenstates of the noninteracting Hamiltonian $H_0$. This is the lowest order approximation which can also be traced down to the random phase approximation. For a system of Dirac electrons, described by a wave functions $\psi_{n,\bf k}$ given by equation (\ref{wavef-Dirac}), one then obtains for the screened response function \cite{PinesBook}:
\begin{align}
\chi({\bf q},\omega) & =\frac{1}{L^2}4\sum_{n n'{\bf k}}
|\langle n' {\bf k}+{\bf q} | e^{i{\bf q}\cdot{\bf r}} | n {\bf k} \rangle|^2 
\frac{f_{n {\bf k}}-f_{n' {\bf k}+{\bf q}}}
{\hbar\omega-E_{n' {\bf k}+{\bf q}}+E_{n {\bf k}}+i\eta} 
\nonumber \\
& = \frac{1}{L^2}4\sum_{n n'{\bf k}}
\frac{1}{2}[1+n n' \cos(\theta_{{\bf k}+{\bf q}}-\theta_{\bf k})]
\frac{f_{n {\bf k}}-f_{n' {\bf k}+{\bf q}}}
{\hbar\omega-E_{n' {\bf k}+{\bf q}}+E_{n {\bf k}}+i\eta} .
\label{d_d_response_0}
\end{align}

We will be particularly interested in the dielectric function of this system so we need to find the relation between the scalar potential $\varphi_{ind}$ and the induced surface charge density 
$-e\langle\rho_{ind}({\bf r},t)\rangle =
-e\langle\rho_{ind}({\bf q},\omega)\rangle 
e^{i{\bf q}\cdot{\bf r}}e^{-i\omega t}$. Let us define here the vector 
${\bf r}=x\hat{\bf x}+y\hat{\bf y}$ which lies in the graphene plane (located at $z=0$) while $z$ axis is perpendicular to graphene plane. Further on we assume graphene is sitting in between two dielectrics of permittivities $\epsilon_{r1}$ ($z<0$) and $\epsilon_{r2}$ ($z>0$). If we work in the electrostatic approximation ($q>>\omega/c$) then the scalar potential induced by the surface charge density located at the plane $z=0$ is simply given by
\begin{equation}
\varphi_{ind}({\bf r},z,t)=\varphi_{ind}({\bf q},\omega)
e^{i{\bf q}\cdot{\bf r}-q|z|}e^{-i\omega t}.
\label{electrostatic_potential_space}
\end{equation}
The electric field is given by 
${\bf E}=-\mbox{\boldmath $\nabla$}\varphi$. We can now separate the electric field ${\bf E}={\bf E}_{\bf r}+E_z\hat{\bf z}$ 
into component along the graphene plane 
${\bf E}_{\bf r}=-\mbox{\boldmath $\nabla$}_{\bf r}\varphi$
which is given by expression
\begin{equation}
{\bf E}_{\bf r}^{ind}({\bf r},z,t)=
-\varphi_{ind}({\bf q},\omega)i{\bf q}
e^{i{\bf q}\cdot{\bf r}-q|z|}e^{-i\omega t} ,
\label{E_r}
\end{equation}
and component perpendicular to the graphene plane 
$E_z=-\partial\varphi/\partial z$ which is given by expressions
\begin{equation}
{\bf E}_z^{ind}({\bf r},z>0,t)=
\varphi_{ind}({\bf q},\omega)iq
e^{i{\bf q}\cdot{\bf r}-qz}e^{-i\omega t} ,
\label{E_z_plus}
\end{equation}
\begin{equation}
{\bf E}_z^{ind}({\bf r},z<0,t)=
-\varphi_{ind}({\bf q},\omega)iq
e^{i{\bf q}\cdot{\bf r}+qz}e^{-i\omega t} .
\label{E_z_minus}
\end{equation}
Further on, the Gauss law can be written as a boundary condition across the graphene plane as \cite{Jackson}
\begin{align}
-e\langle\rho_{ind}({\bf r},t)\rangle & =
[{\bf D}^{ind}({\bf r},z=0^+,t)-{\bf D}^{ind}({\bf r},z=0-,t)]\cdot\hat{\bf z}
\nonumber \\
& = \epsilon_0\epsilon_{r1}E_z^{ind}({\bf r},z=0^+,t) -
\epsilon_0\epsilon_{r2}E_z^{ind}({\bf r},z=0^-,t) .
\label{boundary_condition}
\end{align}
Then by using the decomposition into Fourier components and equations (\ref{E_z_plus}) and (\ref{E_z_minus}) we obtain desired relation between the induced charge density and corresponding induced scalar potential:
\begin{equation}
-e\langle\rho_{ind}({\bf q},\omega)\rangle =
q \varphi_{ind}({\bf q},\omega) 2\bar{\epsilon}_r \epsilon_0 .
\label{density_potenital_ind}
\end{equation}
Here $\bar{\epsilon}_r=(\epsilon_{r1}+\epsilon_{r1})/2$, and we can introduce the external charge density corresponding to the external potential by the same relation
\begin{equation}
-e\rho_{ext}({\bf q},\omega) =
q \varphi_{ext}({\bf q},\omega) 2\bar{\epsilon}_r \epsilon_0 .
\label{density_potenital_ext}
\end{equation}
Let us now define the graphene dielectric function 
$\epsilon({\bf q},\omega)$ as \cite{PinesBook}:
\begin{equation}
\frac{\epsilon({\bf q},\omega)}{\bar{\epsilon_r}} = 
\frac{\rho_{ext}({\bf q},\omega)}
{\rho_{ext}({\bf q},\omega)+\langle\rho_{ind}({\bf q},\omega)\rangle}.
\label{dielectric_function_definition}
\end{equation}
Then from equations (\ref{density_response_self_consistent}), (\ref{density_potenital_ind}) and (\ref{density_potenital_ext}) we obtain
\begin{equation}
\frac{\epsilon({\bf q},\omega)}{\bar{\epsilon_r}} = 
1-\frac{e^2}{2 \bar{\epsilon_r}\epsilon_0 q} \chi({\bf q},\omega) .
\label{dielectric_function}
\end{equation}
Note that the zero of dielectric function ($\epsilon({\bf q},\omega)=0$) defines the collective electron oscillation (plasmon) which is the core subject of this thesis. 

Finally let us find the relation between density-density response function $\chi({\bf q},\omega)$ and conductivity $\sigma({\bf q},\omega)$. If we introduce the total scalar potential 
$\varphi_{tot}=\varphi_{ext}+\varphi_{ind}$, then by using equation (\ref{density_response_self_consistent}), we can write the induced surface charge density as
$-e\langle \rho_{ind}({\bf q},\omega) \rangle = 
\chi ({\bf q},\omega) e^2 \varphi_{tot}({\bf q},\omega)$.
On the other hand Ohm's law gives the induced surface current density 
$\langle {\bf j}_{ind}({\bf q},\omega) \rangle = 
\sigma({\bf q},\omega) {\bf E}_{\bf r}^{tot}({\bf q},\omega)$
, while the electric field can be found from equation (\ref{E_r}): 
${\bf E}_{\bf r}^{tot}({\bf q},\omega)=
-\varphi_{tot}({\bf q},\omega)i{\bf q}$. Finally, equation of continuity can be written with Fourier components as 
$-e\langle \rho_{ind}({\bf q},\omega) \rangle = 
{\bf q}\cdot\langle {\bf j}_{ind}({\bf q},\omega) \rangle /\omega$, so we obtain desired relation:
\begin{equation}
\sigma({\bf q},\omega)=i\frac{\omega e^2}{q^2}\chi ({\bf q},\omega) .
\label{conductivity_density_density}
\end{equation}
Note however that $\sigma({\bf q},\omega)$ refers only to the longitudinal conductivity since the scalar potential alone is not enough to decribe the transverse fields.


\subsection{Current-current response function}

In the last section we described response to the external scalar potential which we now supplement by calculating response to the external vector potential. Let us then start with the Hamiltonian (\ref{Hamilton-Dirac2}) describing free Dirac particles: $H_0=v_F \mbox{\boldmath $\sigma$}\cdot{\bf p}$, where $\bf p$ is the electron momentum. In the presence of external vector potential ${\bf A}_{ext}({\bf r},t)$, one can write for the total Hamiltonian 
$H=v_F\mbox{\boldmath$\sigma$}\cdot
({\bf p}+e{\bf A}_{ext}({\bf r},t))=H_0+H_{int}$, where the interaction part of the Hamiltonian is given by: 
$H_{int}=e v_F\mbox{\boldmath$\sigma$}\cdot{\bf A}_{ext}({\bf r},t)$. We can now decompose vector potential into Fourier components to obtain:
\begin{equation}
H_{int}=\sum_{\bf q} e^{i{\bf q}\cdot{\bf r}}
e v_F\mbox{\boldmath$\sigma$}\cdot{\bf A}_{ext}({\bf q},t) ,
\label{H_vector_potential}
\end{equation}
then by using the current density operator from equation (\ref{current_density_op_q}) we can write
\begin{equation}
H_{int}=-L^2\sum_{\bf q} {\bf j}_{\bf q}^\dag\cdot{\bf A}_{ext}({\bf q},t) .
\label{Hamilt_vector_potential_current}
\end{equation}
It is now convenient to introduce the longitudinal 
($V_{L}={\bf V}\cdot{\bf e}_{{\bf q}L}^*$) and transverse 
($V_{T}={\bf V}\cdot{\bf e}_{{\bf q}T}^*$) vector components by using the polarization vectors from equations (\ref{polari-L}) and (\ref{polari-T}). We can now write the interaction Hamiltonian
\begin{align}
H_{int} & = -L^2\sum_{{\bf q},\mu} 
j_{{\bf q},\mu}^\dag\cdot A_{ext,\mu}({\bf q},t)
\nonumber \\
& = \int e^{-i\omega t} d\omega
(-L^2)\sum_{{\bf q},\mu} 
j_{{\bf q},\mu}^\dag\cdot A_{ext,\mu}({\bf q},\omega) .
\label{Hamilt_vector_potential_current_L_T}
\end{align}
Finally, by assuming the weak coupling between the external probe and our system, precisely like in the last section, we obtain the induced current density:
\begin{equation}
\langle j_{ind,\mu}({\bf q},\omega) \rangle = 
\chi_\mu ({\bf q},\omega) 
(- A_{ext,\mu}({\bf q},\omega)) .
\label{current_response}
\end{equation}
Here the current-current response function is given by \cite{PinesBook}
\begin{equation}
\chi_\mu({\bf q},\omega)=L^2\sum_{a,b}\frac{e^{-\beta E_b}}{Z}
|\langle a | j_{{\bf q},\mu}^\dag | b \rangle|^2 
\left( \frac{1}{\hbar\omega-\hbar\omega_{ab}+i\eta} 
-\frac{1}{\hbar\omega+\hbar\omega_{ab}+i\eta} \right).
\label{j_j_response}
\end{equation}
We can now use the free electron states to write the screened function:
\begin{equation}
\chi_\mu({\bf q},\omega) =L^2 4\sum_{n n'{\bf k}}
|\langle n' {\bf k}+{\bf q} | j_{{\bf q},\mu}^\dag | n {\bf k} \rangle|^2 
\frac{f_{n {\bf k}}-f_{n' {\bf k}+{\bf q}}}
{\hbar\omega-E_{n' {\bf k}+{\bf q}}+E_{n {\bf k}}+i\eta} .
\label{j_j_response_0}
\end{equation}
At last, by using the exact form of the electron wave function from equation (\ref{wavef-Dirac}), we obtain different expressions for the longitudinal and transverse current-current response functions:
\begin{equation}
\chi_L({\bf q},\omega) =\frac{e^2 v_F^2}{L^2}4\sum_{n n'{\bf k}}
\frac{1}{2}[1+n n' \cos(\theta_{\bf k}+\theta_{{\bf k}+{\bf q}})]
\frac{f_{n {\bf k}}-f_{n' {\bf k}+{\bf q}}}
{\hbar\omega-E_{n' {\bf k}+{\bf q}}+E_{n {\bf k}}+i\eta} ,
\label{j_j_response_L}
\end{equation}
\begin{equation}
\chi_T({\bf q},\omega) =\frac{e^2 v_F^2}{L^2}4\sum_{n n'{\bf k}}
\frac{1}{2}[1-n n' \cos(\theta_{\bf k}+\theta_{{\bf k}+{\bf q}})]
\frac{f_{n {\bf k}}-f_{n' {\bf k}+{\bf q}}}
{\hbar\omega-E_{n' {\bf k}+{\bf q}}+E_{n {\bf k}}+i\eta} .
\label{j_j_response_T}
\end{equation}

Note here that expressions (\ref{j_j_response_L}) and (\ref{j_j_response_L}) actually diverge if we use Dirac states (\ref{wavef-Dirac}) instead of actual electron states in graphene limited by some band cut-off. However, his subtlety can be easily solved by subtracting from $\chi_{L}({\bf q},\omega)$ 
[$\chi_{T}({\bf q},\omega)$] the value $\chi_{L}({\bf q},\omega=0)$ 
[$\chi_{T}({\bf q} \to 0,\omega=0)$]
to take into account that there is no current response 
to the longitudinal [transverse] time [time and space] independent vector potential, 
see \cite{Principi2009, Falkovsky2007} for details.

Let us also find relation between the conductivity and the current-current response function. Note that the electric field is given by 
${\bf E}=-\partial{\bf A}/\partial t$ so that 
${\bf E}({\bf q},\omega)=i\omega {\bf A}({\bf q},\omega)$. Then we can write equation (\ref{current_response}) as 
$\langle j_{ind,\mu}({\bf q},\omega) \rangle= 
\chi_\mu ({\bf q},\omega) \frac{i}{\omega} E_{ext,\mu}({\bf q},\omega)$. In other words desired relation is simply: 
\begin{equation}
\sigma_\mu({\bf q},\omega)=\frac{i}{\omega}\chi_\mu({\bf q},\omega) .
\label{conductivity_current_current}
\end{equation}
Note here that longitudinal conductivity $\sigma_L({\bf q},\omega)$ (describing response of a system to the longitudinal field) is generally different from the transverse conductivity $\sigma_T({\bf q},\omega)$ (describing response of a system to the transverse field), unless we are working in the limit of small wave vectors ($q\rightarrow 0$).


\subsection{Fluctuation-dissipation theorem}

In this section we derive relation between current-current correlation function and the current-current response function at finite temperature, which is given by the fluctuation-dissipation theorem. We will use this result later to calculate the radiative heat transfer between two graphene sheets.

We start with the current-current correlation function:
\begin{equation}
K_\mu({\bf r},t,{\bf r'},t') =
\langle j_\mu({\bf r},t) j_\mu^\dag({\bf r'},t') \rangle .
\label{correlation_function}
\end{equation}
Due to translational invariance in space and time we can write 
$K_\mu({\bf r},t,{\bf r'},t')=
K_\mu({\bf r}-{\bf r'},t-t')=
K_\mu({\bf d},\tau)$
, where we have denoted by: ${\bf d}={\bf r}-{\bf r'}$ and $\tau=t-t'$. 
Then the Fourier transforms from the space and time domains are respectively given by
\begin{equation}
K_\mu({\bf q},\tau)=\frac{1}{L^2} \int 
K_\mu({\bf d},\tau) e^{-i{\bf q}\cdot{\bf d}} d {\bf d} ,
\label{K_q_tau}
\end{equation}
\begin{equation}
K_\mu({\bf q},\omega)=\frac{1}{2\pi} \int 
K_\mu({\bf q},\tau) e^{i\omega\tau} d \tau .
\label{K_q_omega}
\end{equation}
It will be more convenient for us to use these relations in a slightly different form. In that regards let us use relations (\ref{K_q_tau}) and (\ref{K_q_omega}) with translational invariance in space and time, respectively, to show that
\begin{align}
\langle j_\mu({\bf q}) j_\mu^\dag({\bf q'}) \rangle & =
\frac{1}{L^4} \int \int 
\langle j_\mu({\bf r}) j_\mu^\dag({\bf r'}) \rangle
e^{-i{\bf q}\cdot{\bf r}} e^{i{\bf q'}\cdot{\bf r'}}
d{\bf r} d{\bf r'} 
\nonumber \\
& =
\frac{1}{L^2} \int  e^{i({\bf q'}-{\bf q})\cdot{\bf r'}} d{\bf r'} 
\frac{1}{L^2} \int  K_{\mu} ({\bf r}-{\bf r'}) 
e^{-i{\bf q}\cdot({\bf r}-{\bf r'})} d{\bf r}
\nonumber \\
& =
\delta_{{\bf q},{\bf q'}} K_\mu({\bf q}) ,
\label{jj_q}
\end{align}
%
%
\begin{align}
\langle j_\mu(\omega) j_\mu^\dag({\omega}') \rangle & =
\frac{1}{4\pi^2} \int \int 
\langle j_\mu(t) j_\mu^\dag(t') \rangle
e^{i\omega t} e^{-i\omega ' t'}
dt dt' 
\nonumber \\
& =
\frac{1}{2\pi} \int  e^{i(\omega-\omega ')t'} dt' 
\frac{1}{2\pi} \int  K_{\mu} (t-t') 
e^{-i\omega(t-t')} dt
\nonumber \\
& =
\delta(\omega-\omega ') K_\mu(\omega) .
\label{jj_omega}
\end{align}
Relations (\ref{jj_q}) and (\ref{jj_omega}) simply state that there is no correlation between different $\bf q$ or different $\omega$ components. We can join these two relations in a single one
\begin{equation}
\langle j_\mu({\bf q},\omega) j_\mu^\dag({\bf q'},\omega ') \rangle =
\delta_{{\bf q},{\bf q'}} \delta(\omega-\omega ') K_\mu({\bf q},\omega) .
\label{jj_q_omega}
\end{equation}
To find the $K_\mu({\bf q},\omega)$ let us note that evolution of current operator, in the Heisenberg picture, is given by
$j_\mu ({\bf q},\tau)=e^{iH\tau/\hbar} j_\mu^\dag({\bf q},0) e^{-iH\tau/\hbar}$. Now we can write
\begin{align}
K_\mu({\bf q},\tau) & = 
\langle j_\mu({\bf q},\tau) j_\mu^\dag({\bf q},0) \rangle
\nonumber \\
& =
\langle 
e^{iH\tau/\hbar} j_\mu^\dag({\bf q},0) e^{-iH\tau/\hbar} j_\mu^\dag({\bf q},0) 
\rangle
\nonumber \\
& =
\sum_{a,b}\frac{e^{-\beta E_b}}{Z}
\langle b | e^{iH\tau/\hbar} j_\mu ({\bf q}) | a \rangle
\langle a | e^{-iH\tau/\hbar} j_\mu^\dag ({\bf q}) | b \rangle
\nonumber \\
& = \sum_{a,b}\frac{e^{-\beta E_b}}{Z} 
|\langle a | j_\mu^\dag ({\bf q}) | b \rangle|^2 e^{i\omega_{ab}\tau} .
\label{correlation_function_expanded}
\end{align}
Finally, the Fourier transform of this expression is given by
\begin{equation}
K_\mu({\bf q},\omega)=\sum_{a,b}\frac{e^{-\beta E_b}}{Z}
|\langle a | j_\mu^\dag ({\bf q}) | b \rangle|^2 \delta(\omega-\omega_{ab}).
\label{correlation_function_fourier_time}
\end{equation}
Note however that the imaginary part of the response function, calculated in equation (\ref{j_j_response}), is given by 
\begin{equation}
\Im \chi_\mu({\bf q},\omega)=L^2\sum_{a,b}\frac{e^{-\beta E_b}}{Z}
|\langle a | j_\mu^\dag ({\bf q}) | b \rangle|^2 \frac{-\pi}{\hbar}
[ \delta(\omega-\omega_{ab})-\delta(\omega+\omega_{ab}) ] .
\label{j_j_response_imag}
\end{equation}
We immediately see that correlation function is related to a response function in a simple manner:
\begin{equation}
\Im \chi_\mu({\bf q},\omega)=-\frac{\pi}{\hbar}L^2 
\left[ K_\mu({\bf q},\omega) - K_\mu({\bf q},-\omega) \right] .
\label{correlation_dissipation1}
\end{equation}
By applying the detail balancing condition here, we can write
\begin{equation}
\Im \chi_\mu({\bf q},\omega)=-\frac{\pi}{\hbar}L^2 
\left[ 1 - e^{-\beta\hbar\omega} \right] K_\mu({\bf q},\omega) .
\label{correlation_dissipation2}
\end{equation}
Finally we have
\begin{equation}
K_\mu({\bf q},\omega) = -\frac{\hbar}{\pi} 
\frac{1}{1 - e^{-\beta\hbar\omega}}
\frac{1}{L^2} \Im \chi_\mu({\bf q},\omega),
\label{FDT}
\end{equation}
or if we use the relation $\chi_\mu({\bf q},\omega)=-i\omega\sigma_\mu({\bf q},\omega)$ we can write this in a more convenient form as
\begin{equation}
K_\mu({\bf q},\omega) = \frac{1}{\pi} 
\frac{\hbar\omega}{1 - e^{-\beta\hbar\omega}}
\frac{1}{L^2} \Re \sigma_\mu({\bf q},\omega).
\label{FDT_conductivity}
\end{equation}
This is in fact the well know fluctuation-dissipation theorem stating that the correlation function ($K_\mu$) due to thermal fluctuations is directly related to the dissipation in the system ($\Re \sigma_\mu$ or $\Im \chi_\mu$). This result will be of use in the following section.


\section{Radiative heat transfer}

\begin{figure}
\centerline{
\mbox{\includegraphics[width=0.8\textwidth]{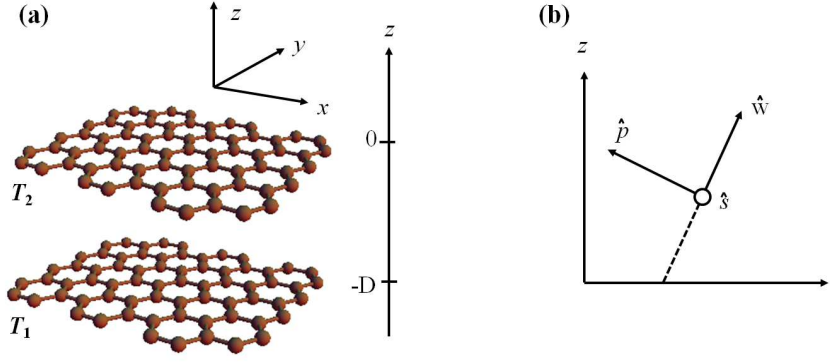}}
}
\caption{
(a) Schematic diagram of the radiation transfer problem: a free standing sheet of graphene at temperature $T_1$ is radiating to another free standing graphene sheet at temperature $T_2$ and distance $D$ away. (b) Polarization vectors defined in the text.
}
\label{Figure4_c2}
\end{figure}

In this section we analyze the radiative heat transfer between two graphene sheets separated by a distance $D$ and held at temperatures $T_1$ and $T_2$ (see figure \ref{Figure4_c2}). To calculate the heat transfer we shall start by looking into correlations between electric currents induced by the thermal fluctuations in the first graphene sheet. Following that we shall use Green function technique to find the electromagnetic fields in the second graphene sheet, induced by the fluctuating currents from the first sheet. Finally heat transfer can be found by calculating Ohmic losses, induced by this electromagnetic field, within the second graphene sheet. 

In the last section we calculated current-current correlation function due to thermal fluctuations. Fluctuation-dissipation theorem (\ref{FDT_conductivity}) and equation (\ref{jj_q_omega}) give the correlation function of the fluctuating currents in the first graphene sheet:
\begin{equation}
\langle j_{1\mu}({\bf q},\omega) j_{1\mu}^\dag({\bf q},\omega ') \rangle =
\delta(\omega-\omega ') \frac{1}{\pi} 
\frac{\hbar\omega}{1 - e^{-\beta_1\hbar\omega}}
\frac{1}{L^2} \Re \sigma_{1\mu}({\bf q},\omega) .
\label{FDT_q_omega_conductivity}
\end{equation}

To find the electromagnetic fields induced by these fluctuating currents we can use classical electrodynamics so we shall start with classical quantities and return to the quantum values only later when necessary. Since the system is translational invariant we can focus on a single ${\bf q},\omega$-component and write the Fourier transform of the surface current density from the first graphene sheet as
\begin{equation}
{\bf j}_1({\bf r}, t)= 
\int e^{-i\omega t} d\omega 
\sum_{\bf q} e^{i{\bf q}\cdot {\bf r}}
{\bf j}_1({\bf q},\omega) .
\label{current_fourier}
\end{equation}
Further on, let us assume the most simple case where there is only vacuum in between and around graphene sheets. Then the electric field satisfies a simple wave equation
\begin{equation}
(\nabla^2 + \omega^2/c^2){\bf E}({\bf r},z)=0 .
\label{wave_eq}
\end{equation}
This equation has a plane wave solution 
${\bf E}({\bf r},z)={\bf E}({\bf q},\omega) e^{i{\bf q}\cdot{\bf r}+i\gamma z}$, where we took into consideration that the periodicity in the $xy$ direction is determined by the wave vector $\bf q$. In other words we can write for the total wave vector: ${\bf w}={\bf q}+\gamma\hat{\bf z}$,
while the equation (\ref{wave_eq}) requires: 
$w^2=|{\bf w}|^2=q^2+\gamma^2=\omega^2/c^2$ i.e. the $z$ component of the wave vector $\bf w$ is given by:
\begin{equation}
\gamma=\sqrt{\omega^2/c^2-q^2}.
\label{w}
\end{equation}
Further on, since there is no free charge around graphene sheets, the Gauss law states that $\nabla\cdot {\bf E}({\bf r},z)=0$. This means that 
${\bf E}({\bf q},\omega)\cdot{\bf w}=0$ i.e. electric field is transversely polarized so it is convenient to introduce unit vectors $\hat{\bf s}$ and $\hat{\bf p}$ that are perpendicular to wave vector ${\bf w}$:
\begin{equation}
\hat{\bf s}=\hat{\bf q}\times\hat{\bf z}, \mbox{and}
\label{s_hat}
\end{equation}
\begin{equation}
\hat{\bf p}=w^{-1}(-\gamma\hat{\bf q}+q\hat{\bf z}).
\label{p_hat}
\end{equation}
In this way $(\hat{\bf s},\hat{\bf w},\hat{\bf p})$ is a set of right-handed orthonormal triad (see figure \ref{Figure4_c2}) where
\begin{equation}
\hat{\bf w}=\frac{\bf w}{w}=w^{-1}(q\hat{\bf q}+\gamma\hat{\bf z}).
\label{nu_hat}
\end{equation}
We also note that there is a simple connection with the longitudinal and transverse wave vectors introduced before in this chapter:
${\bf e}_{{\bf q}L}=i\hat{\bf q}$, ${\bf e}_{{\bf q}T}=-i\hat{\bf s}$.

To match the boundary conditions given by the surface current density ${\bf j}_1({\bf q},\omega)$ from the first graphene plane ($z=-D$) in the presence of the second graphene sheet ($z=0$), we use the Green function technique from reference \cite{Sipe1987} which is particularly convenient for the layered structures like ours. In that manner one obtains different electric field component ${\bf E}({\bf q},\omega)$ depending whether we are located below the first graphene sheet ($z<-D$), in between the sheets ($-D<z<0$), or above the second sheet ($z>0$). Since we are interested in the field in the second sheet ($z=0$), it is easiest to look into the expression for the field above the second graphene sheet ($z>0$) where one obtains
\begin{equation}
{\bf E}({\bf q},\omega)=
-\frac{\omega}{2\epsilon_0 c^2 \gamma}
\left( \hat{\bf s}T_{12}^s\hat{\bf s} + \hat{\bf p}T_{12}^p\hat{\bf p} \right)
\cdot {\bf j}_1({\bf q},\omega) .
\label{field_current}
\end{equation} 
Here we have explicitly separated $s$ and $p$ polarizations which have very different behavior, and $T_{12}$ is a transmission coefficient for a system of two parallel graphene sheets given by \cite{Sipe1987}
\begin{equation}
T_{12}=\frac{t_1 t_2 e^{i\gamma D}}{1 - r_1 r_2 e^{2i\gamma D}} .
\label{transmission}
\end{equation}
Note that the same expression is valid for $s$ and $p$ polarization, but reflection $r$ and transmission $t$ coefficients are different for different polarizations. It is a simple manner of elementary electrodynamics to demonstrate that these are
\begin{equation}
r^s=\frac{ -\frac{\omega\sigma_T}{2\gamma \epsilon_0 c^2} }
         { 1 + \frac{\omega\sigma_T}{2\gamma \epsilon_0 c^2}} ,
\label{r_s}
\end{equation}
\begin{equation}
t^s=\frac{1}{ 1 + \frac{\omega\sigma_T}{2\gamma \epsilon_0 c^2}} ,
\label{t_s}
\end{equation}
\begin{equation}
r^p=\frac{ \frac{\gamma \sigma_L}{2\epsilon_0\omega} }
         { 1+ \frac{\gamma \sigma_L}{2\epsilon_0\omega}} , \mbox{and}
\label{r_p}
\end{equation}
\begin{equation}
t^p=\frac{1}{ 1+ \frac{\gamma \sigma_L}{2\epsilon_0\omega}} .
\label{t_p}
\end{equation}
Note also that transverse conductivity ($\sigma_T$) determines the $s$-polarization, and longitudinal conductivity ($\sigma_L$) determines the $p$-polarization. Finally the total electric field in the space above the second graphene sheet ($z>0$) is given by
\begin{equation}
{\bf E}({\bf r}, z, t) = 
\int e^{-i\omega t} d\omega 
\sum_{\bf q} e^{i{\bf q}\cdot {\bf r} +iwz} {\bf E}({\bf q},\omega) .
\label{field_fourier_space}
\end{equation}
So the field, precisely at the second sheet ($z=0$) is
\begin{equation}
{\bf E}({\bf r}, t)=
{\bf E}({\bf r}, z=0^+, t) = 
\int e^{-i\omega t} d\omega 
\sum_{\bf q} e^{i{\bf q}\cdot {\bf r}} {\bf E}({\bf q},\omega) .
\label{electric_field}
\end{equation}

At last, the heat transfer from the first graphene sheet to the second graphene sheet is simply given by Ohmic losses induced by this electric field. The power dissipated per unit area is given by \cite{Jackson}
\begin{align}
H_{1\rightarrow 2}=\frac{1}{L^2}\frac{dE_{mech}}{dt} & = 
\frac{1}{L^2} \int {\bf j}_2 ({\bf r}, t) 
\cdot {\bf E}({\bf r}, t) d{\bf r}
\nonumber \\
& = 
\int \int d\omega d{\omega}' e^{-i(\omega-{\omega}')t} 
\sum_{\bf q} {\bf j}_2 ({\bf q}, \omega) 
\cdot {\bf E}^*({\bf q}, {\omega}') .
\label{Ohmic_loss}
\end{align}

Let us now take into account that current densities ${\bf j}_{1,2}({\bf q},\omega)$ have only vector components along the graphene ($xy$) plane. Then due to equation (\ref{p_hat}) one has 
$\hat{\bf p}\cdot{\bf j}_1({\bf q},\omega)=
-\frac{\gamma}{w}\hat{\bf q}\cdot{\bf j}_1({\bf q},\omega)$, 
and we can write equation (\ref{field_current}) again as
\begin{equation}
{\bf E}({\bf q},\omega)=
-\frac{\omega}{2\epsilon_0 c^2 \gamma} 
\left( \hat{\bf s}T_{12}^s\hat{\bf s} + 
\left(
       \frac{\gamma^2}{w^2} \hat{\bf q} - \frac{q\gamma}{w^2}\hat{\bf z}
\right)
T_{12}^p\hat{\bf q} \right)
\cdot {\bf j}_1({\bf q},\omega) .
\label{field_current2}
\end{equation}
Further on, the scalar product in the equation (\ref{Ohmic_loss}) can be written as ${\bf j}_2\cdot{\bf E}^*={\bf j}_2\cdot{\bf E}_{\bf r}^*$, where 
${\bf E}_{\bf r}={\bf E}-({\bf E}\cdot\hat{\bf z})\hat{\bf z}$ is the projection of the electric field vector to the graphene ($xy$) plane. In that way we can write equation (\ref{field_current2}) as
\begin{equation}
{\bf E}_{\bf r}({\bf q},\omega)=
-\frac{\omega}{2\epsilon_0 c^2 \gamma}
\left( \hat{\bf s}T_{12}^s\hat{\bf s} + 
       \frac{\gamma^2}{w^2} \hat{\bf q} 
T_{12}^p\hat{\bf q} \right)
\cdot {\bf j}_1({\bf q},\omega).
\label{field_current_r}
\end{equation}
Let us note here again that $\hat{\bf q}=i{\bf e}_{{\bf q}L}^*$ and $\hat{\bf s}=-i{\bf e}_{{\bf q}T}^*$ while the longitudinal ($\mu=L$) and transverse ($\mu=T$) components of the current density are defined as: 
$j_\mu({\bf q},\omega)={\bf j}({\bf q},\omega)\cdot{\bf e}_{{\bf q}\mu}^*$. 
Then we can write equation (\ref{field_current_r}) as
\begin{equation}
{\bf E}_{\bf r}({\bf q},\omega) =
-\frac{\omega}{2\epsilon_0 c^2 \gamma} 
\left( -{\bf e}_{{\bf q}T}^* T_{12}^s j_{1 T}({\bf q},\omega) +
-\frac{\gamma^2}{w^2} {\bf e}_{{\bf q}L}^* T_{12}^p j_{1 L}({\bf q},\omega)
\right) .
\label{field_current_LT}
\end{equation}
Finally due to Ohm's law 
${j}_{2\mu}({\bf q},\omega)=\sigma_{2\mu}({\bf q},\omega){E}_{\bf r,\mu}({\bf q},\omega)$ we have
\begin{align}
\langle 
{\bf j}_{2}({\bf q},\omega) \cdot {\bf E}_{\bf r}^*({\bf q},{\omega}') 
\rangle =
\frac{\omega^2}{4\epsilon_0^2 c^4 |\gamma|^2}
& \sigma_{2T}({\bf q},\omega) 
\langle j_{1T}({\bf q},\omega) j_{1T}({\bf q},{\omega}')^* \rangle
|T_{12}^s|^2
\nonumber \\
+ 
\frac{|\gamma|^2}{4\epsilon_0^2 \omega^2} 
& \sigma_{2L}({\bf q},\omega) 
\langle j_{1L}({\bf q},\omega) j_{1L}({\bf q},{\omega}')^* \rangle
|T_{12}^p|^2 .
\label{jdotE}
\end{align}
Here we have explicitly written the ensemble average which requires us to calculate precise quantum correlations of the current density operator. We have also used relation (\ref{jj_q_omega}) which states that there is no correlation between different $\omega$ components due to translational invariance in the time domain. In fact the current-current correlation function (\ref{FDT_q_omega_conductivity}) is given by
\begin{equation}
\langle j_{1\mu}({\bf q},\omega) j_{1\mu}^\dag({\bf q},\omega ') \rangle =
\delta(\omega-\omega ') \frac{1}{\pi} 
\frac{\hbar\omega}{1 - e^{-\beta_1\hbar\omega}}
\frac{1}{L^2} \Re \sigma_{1\mu}({\bf q},\omega) .
\label{FDT_q_omega_conductivity2}
\end{equation}
Since the final result has to be a real quantity, we can simply look into real part of the expression (\ref{jdotE}) 
\begin{align}
\Re \langle 
{\bf j}_{2}({\bf q},\omega) \cdot {\bf E}_{\bf r}^*({\bf q},{\omega}') 
\rangle = &
\delta(\omega-\omega ') \frac{1}{\pi} 
\frac{\hbar\omega}{1 - e^{-\beta_1\hbar\omega}}
\frac{1}{L^2} \times
\nonumber \\
& (\frac{\omega^2}{4\epsilon_0^2 c^4 |\gamma|^2}
  \Re \sigma_{2T}({\bf q},\omega) \Re \sigma_{1T}({\bf q},\omega) 
  |T_{12}^s|^2 +
\nonumber \\
& + \frac{|\gamma|^2}{4\epsilon_0^2 \omega^2} 
    \Re \sigma_{2L}({\bf q},\omega) \Re \sigma_{1L}({\bf q},\omega)
    |T_{12}^p|^2 ) .
\label{RjdotE}
\end{align}
This can be written in a more transparent form by using reflection and transmission coefficients (\ref{r_s}) - (\ref{t_p}). However, since $\gamma=\sqrt{\omega^2/c^2-q^2}$ we have to distinguish between the case of propagating waves in the far field ($\omega/c>q$) and evanescent waves in the near field ($\omega/c<q$). In the first case ($\omega/c>q$) one has
\begin{equation}
\frac{\omega}{2\epsilon_0 c^2 |\gamma|} \Re \sigma_{T} =
\frac{1-|r^s|^2-|t^s|^2}{2|t^s|^2} , \mbox{and}
\label{s_left_ll}
\end{equation}
\begin{equation}
\frac{|\gamma|}{2\epsilon_0 \omega} \Re \sigma_{L} =
\frac{1-|r^p|^2-|t^p|^2}{2|t^p|^2} .
\label{p_left_ll}
\end{equation}
It is convenient here to define the following quantities
\begin{align}
h^s_{ff}({\bf q},\omega) & \equiv 
\frac{\omega^2}{4\epsilon_0^2 c^4 |\gamma|^2}
\Re \sigma_{1T}({\bf q},\omega) \Re \sigma_{2T}({\bf q},\omega) |T_{12}^s|^2
\nonumber \\
& = 
\frac{(1-|r_1^s|^2-|t_1^s|^2)(1-|r_2^s|^2-|t_2^s|^2)}
{4|1 - r_1^s r_2^s e^{2i\gamma D}|^2} , \mbox{and}
\label{h_s_ff}
\end{align}
\begin{align}
h^p_{ff}({\bf q},\omega) & \equiv 
\frac{|\gamma|^2}{4\epsilon_0^2 \omega^2} 
\Re \sigma_{1L}({\bf q},\omega) \Re \sigma_{2L}({\bf q},\omega) |T_{12}^p|^2
\nonumber \\
& = 
\frac{(1-|r_1^p|^2-|t_1^p|^2)(1-|r_2^p|^2-|t_2^p|^2)}
{4|1 - r_1^p r_2^p e^{2i\gamma D}|^2} ,
\label{h_p_ff}
\end{align}
where we have used expression (\ref{transmission}) for the transmission coefficient $T_{12}$. In the second case ($\omega/c<q$) one has
\begin{equation}
\frac{\omega}{2\epsilon_0 c^2 |\gamma|} \Re \sigma_{T} =
\frac{\Im r^s}{|t^s|^2} ,
\label{s_right_ll}
\end{equation}
\begin{equation}
\frac{|\gamma|}{2\epsilon_0 \omega} \Re \sigma_{L} =
\frac{\Im r^p}{|t^p|^2} ,
\label{p_right_ll}
\end{equation}
\begin{equation}
h^s_{nf}({\bf q},\omega)  \equiv 
\frac{\omega^2}{4\epsilon_0^2 c^4 |\gamma|^2}
\Re \sigma_{1T}({\bf q},\omega) \Re \sigma_{2T}({\bf q},\omega) |T_{12}^s|^2
=
\frac{\Im r_1^s \Im r_2^s e^{-2|\gamma| D}}
{|1 - r_1^s r_2^s e^{-2|\gamma| D}|^2} , \mbox{and}
\label{h_s_nf}
\end{equation}
\begin{equation}
h^p_{nf}({\bf q},\omega)  \equiv 
\frac{|\gamma|^2}{4\epsilon_0^2 \omega^2} 
\Re \sigma_{1L}({\bf q},\omega) \Re \sigma_{2L}({\bf q},\omega) |T_{12}^p|^2
=
\frac{\Im r_1^p \Im r_2^p e^{-2|\gamma| D}}
{|1 - r_1^p r_2^p e^{-2|\gamma| D}|^2} .
\label{h_p_nf}
\end{equation}
At last we obtain for the heat transfer (equation (\ref{Ohmic_loss})) from the first graphene sheet to the second graphene sheet
\begin{equation}
H_{1\rightarrow 2}=H_{1\rightarrow 2,ff}+H_{1\rightarrow 2,nf} ,
\label{H12}
\end{equation}
where the far field ($\omega/c>q$) and near field ($\omega/c<q$) contributions are respectively given by
\begin{equation}
H_{1\rightarrow 2,ff}=\frac{1}{\pi}\int d\omega 
 \frac{\hbar\omega}{1-e^{-\beta_1\hbar\omega}}
\frac{1}{L^2}\sum_{{\bf q},\mu} 
h_{ff}^\mu (q,\omega) , \mbox{and}
\label{H_12ff}
\end{equation}
\begin{equation}
H_{1\rightarrow 2,nf}=\frac{1}{\pi}\int d\omega 
 \frac{\hbar\omega}{1-e^{-\beta_1\hbar\omega}}
\frac{1}{L^2}\sum_{{\bf q},\mu} 
h_{nf}^\mu (q,\omega) .
\label{H_12nf}
\end{equation}
In the same manner one can calculate heat transfer from the second graphene sheet to the first graphene sheet $H_{2\rightarrow 1}$, so the total heat transfer ($H=H_{1\rightarrow 2}-H_{2\rightarrow 1}$) between two graphene sheets can be written as
\begin{equation}
H=H_{ff}+H_{nf} ,
\label{H_tot}
\end{equation}
\begin{equation}
H_{ff}=\frac{2}{\pi} 
\int_{0}^{\infty} d\omega 
[ \Theta(\omega,T_1)-\Theta(\omega,T_2) ]
\frac{1}{(2\pi)^2} \int_0^{\omega/c} 2\pi q dq
\sum_{\mu} 
h_{ff}^\mu (q,\omega) ,
\label{H_ff0}
\end{equation}
\begin{equation}
H_{nf}=\frac{2}{\pi} 
\int_{0}^{\infty} d\omega 
[ \Theta(\omega,T_2)-\Theta(\omega,T_1) ]
\frac{1}{(2\pi)^2}\int_{\omega/c}^{\infty} 2\pi q dq
\sum_{\mu} 
h_{nf}^\mu (q,\omega) .
\label{H_nf0}
\end{equation}
Here we have introduced the Boltzman factor: 
$\Theta(\omega,T)=\hbar\omega/(e^{\beta\hbar\omega}-1)$, which comes about since the zero point energy cancels when taking the difference between emission and absorption. We write here again functions $h_{ff}^\mu$ and $h_{nf}^\mu$ for the sake of clearance
\begin{equation}
h^\mu_{ff}({\bf q},\omega)  \equiv 
\frac{(1-|r_1^\mu|^2-|t_1^\mu|^2)(1-|r_2^\mu|^2-|t_2^\mu|^2)}
{4|1 - r_1^\mu r_2^\mu e^{2i\gamma D}|^2}  ,
\label{h_ff0}
\end{equation}
\begin{equation}
h^\mu_{nf}({\bf q},\omega)  \equiv 
\frac{\Im r_1^\mu \Im r_2^\mu e^{-2|\gamma| D}}
{|1 - r_1^\mu r_2^\mu e^{-2|\gamma| D}|^2} .
\label{h_nf0}
\end{equation}
Note that for the case of black body which has perfect absorption $1=|a|^2=1-|r|^2-|t|^2$, i.e. zero reflection or transmission ($r=t=0$), equation (\ref{H_ff0}) simply gives the Stefan-Boltzman law:
\begin{equation}
H_{ff}=\frac{\pi^2 k^4}{60 c^2 \hbar^3}(T_1^4-T_2^4)
\label{Stefan-Boltzman}
\end{equation}

To summarize, in this section we have calculated the total heat transfer, that is, the transfer of heat energy per unit time per unit area between two graphene sheets at different temperatures. Total heat transfer $H=H_{ff}+H_{nf}$ has a contribution from the propagating waves in the far field ($H_{ff}$) and evanescent waves in the near field ($H_{nf}$), given by equations (\ref{H_ff0}) and (\ref{H_nf0}), respectively.

\chapter{Plasmonics in graphene} \label{chap:chap3}

In this chapter we investigate plasmons in doped graphene and demonstrate that they 
simultaneously enable low-losses and significant wave localization 
for frequencies of the light smaller than the optical phonon 
frequency $\hbar\omega_{Oph}\approx 0.2$~eV. 
Interband losses via emission of electron-hole pairs (1$^{\textrm{st}}$ order process) can be 
blocked by sufficiently increasing the doping level, which pushes the 
interband threshold frequency $\omega_{inter}$ toward higher values 
(already experimentally achieved doping levels can 
push it even up to near infrared frequencies). 
The plasmon decay channel via emission of an optical phonon 
together with an electron-hole pair (2$^{\textrm{nd}}$ order process) is inactive for 
$\omega<\omega_{Oph}$ (due to energy conservation), however, for frequencies larger than $\omega_{Oph}$ 
this decay channel is non-negligible. This is particularly important 
for large enough doping values when the interband threshold $\omega_{inter}$ is 
above $\omega_{Oph}$: in the interval $\omega_{Oph}<\omega<\omega_{inter}$ 
the 1$^{\textrm{st}}$ order process is suppressed, but the phonon decay channel is open. 
In this chapter, the calculation of losses is performed within the 
framework of a random-phase approximation (RPA) and number conserving 
relaxation-time approximation \cite{Mermin1970}; the measured 
DC relaxation-time from Ref. \cite{Novoselov2004} serves as an input parameter 
characterizing collisions with impurities, 
whereas the optical phonon relaxation times are estimated  
from the influence of the electron-phonon coupling \cite{Park2007} on the 
optical conductivity \cite{Stauber2008}.

In Sec. \ref{sec:sps}, we provide a brief review of conventional surface plasmons 
and their relevance for nanophotonics. In Sec. \ref{sec:res} we discuss the 
trade off between plasmon losses and wave localization in doped graphene, 
as well as the optical properties of these plasmons. 
We conclude and provide an outlook in Sec. \ref{sec:conc}.

\section{Surface plasmons}
\label{sec:sps}

\begin{figure*}[!ht]
\centerline{
\mbox{\includegraphics[width=0.85\textwidth]{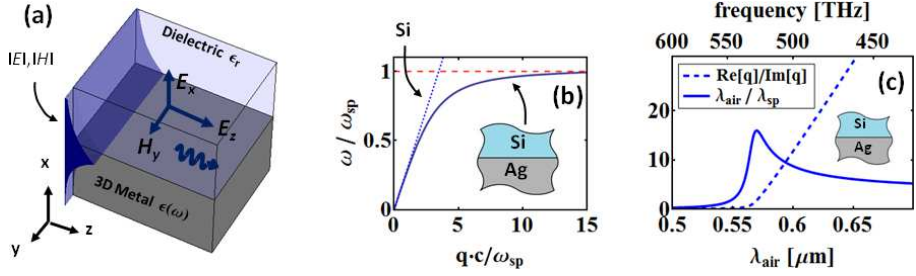}}
}
\caption{
(a) Schematic description of a surface plasmon (SP) on metal-dielectric 
interface. (b) SP dispersion curve (solid blue line) for Ag-Si interfaces; 
dotted blue is the light line in Si; dashed red line denotes the SP resonance. 
(c) Wave localization and propagation length for SPs at Ag-Si interface 
(experimental Ag losses are taken into account).
}
\label{Figure1_c3}
\end{figure*}

Surface plasmons (SPs) are electromagnetic (EM) waves that propagate along 
the boundary surface of a metal and a dielectric [see Fig. \ref{Figure1_c3}(a)]; these are 
transverse magnetic (TM) modes accompanied by collective oscillations of 
surface charges, which decay exponentially in the transverse directions 
(see, e.g., Refs. \cite{Barnes2003,Atwater2005} and Refs. therein). Their 
dispersion curve is given by:

\begin{equation}
q_{sp}=\frac{\omega}{c} 
\sqrt{\frac{\epsilon_r \epsilon(\omega)}{\epsilon_r+\epsilon(\omega)}}
\label{convP}
\end{equation}

[see Fig. \ref{Figure1_c3}(b)]; 
note that close to the SP resonance ($\omega=\omega_{SP}$), 
the SP wave vector [solid blue line in Fig. \ref{Figure1_c3}(b)] is much 
larger than the wave vector of the same frequency excitation in 
the bulk dielectric [dotted blue line in Fig. \ref{Figure1_c3}(b)]. 
As a result, a localized SP wave packet can be much smaller than a same 
frequency wave packet in a dielectric. Moreover, this “shrinkage” is accompanied 
by a large transverse localization of the plasmonic modes. These features are 
considered very promising for enabling nano-photonics 
\cite{Barnes2003,Atwater2005,Yablonovitch2005,Aristeidis2005}, 
as well as high field localization and enhancement. 
A necessary condition for the existence of SPs is $\epsilon(\omega)<-\epsilon_r$ 
(i.e., $\epsilon(\omega)$ is negative), which is why metals are usually used. 
However, SPs in metals are known to have small propagation lengths, which are 
conveniently quantified (in terms of the SP wavelength) with the ratio
$\Re q_{sp}/\Im q_{sp}$; this quantity is a measure of how many SP wavelengths 
can an SP propagate before it loses most of its energy. 
The wave localization (or wave "shrinkage") is quantified as 
$\lambda_{air}/\lambda_{sp}$, where $\lambda_{air}=2\pi c/\omega$ (the 
wavelength in air). These quantities are plotted in Fig. \ref{Figure1_c3}(c) for 
the case of Ag-Si interface, by using experimental data (see \cite{Yablonovitch2005} 
and references therein) to model silver (metal with the lowest losses for the frequencies 
of interest). Near the SP resonance, wave localization reaches its peak; 
however, losses are very high there resulting in a small propagation length 
$l\approx 0.1 \lambda_{sp}\approx 5$nm. At higher 
wavelengths one can achieve low losses but at the expense of poor wave 
localization.

\section{Plasmons and their losses in doped graphene}
\label{sec:res}

Graphene behaves as an essentially 2D electronic system. In the absence of 
doping, conduction and valence bands meet at a point (called Dirac point) 
which is also the position of the Fermi energy. The band structure, calculated in the tight 
binding approximation is shown in Fig. 2(b); for low energies the dispersion around the Dirac point 
can be expressed as $E_{n,{\bf k}}= n v_F \hbar |{\bf k}|$, where the 
Fermi velocity is $v_F=10^6$m/s, $n=1$ for conduction, and $n=-1$ for the 
valence band. Recent experiments \cite{Nair2008} have shown that this linear dispersion 
relation is still valid even up to the energies (frequencies) of visible 
light, which includes the regime we are interested in. 

\begin{figure*}[!ht]
\centerline{
\mbox{\includegraphics[width=0.85\textwidth]{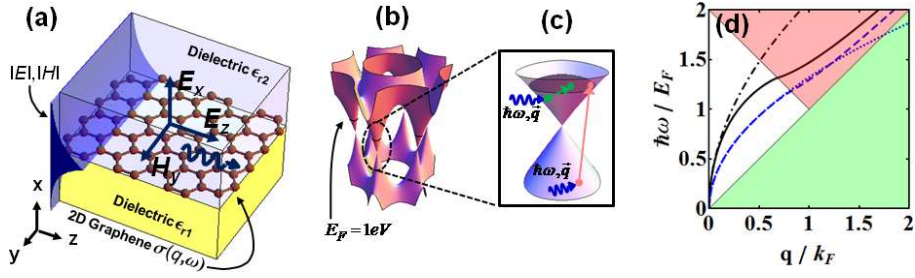}}
}
\caption{
(a) Schematic of the graphene system and TM plasmon modes. Note that the profile of 
the fields looks the same as the fields of an SP [Fig. \ref{Figure1_c3}(a)]. 
(b) Electronic band structure of graphene; to indicate the vertical scale 
we show the Fermi energy level for the case $E_F=1$~eV. 
(c) Sketch of the intraband (green arrows) and interband (red arrows) 
single particle excitations that can lead to large losses; these losses 
can be avoided by implementing a sufficiently high doping. 
(d) Plasmon RPA and semiclassical dispersion curves.
Black solid (RPA) and black dot-dashed (semiclassical) lines correspond to 
$\epsilon_{r1}=\epsilon_{r2}=1$; 
Blue dashed (RPA) and blue dotted (semiclassical) lines correspond to 
$\epsilon_{r1}=4$ and $\epsilon_{r2}=1$. The green (lower) and rose (upper) 
shaded areas represent regimes of intraband and interband excitations, 
respectively.
}
\label{Figure2_c3}
\end{figure*}

Here we consider TM modes in geometry depicted in figure \ref{Figure2_c3} (a), where graphene is 
surrounded with dielectrics of constants $\epsilon_{r1}$ and $\epsilon_{r2}$. 
Throughout the paper, for definiteness we use $\epsilon_{r1}=4$ corresponding 
to SiO$_2$ substrate, and $\epsilon_{r2}=1$ for air on top of graphene, 
which corresponds to a typical experimental setup. TM modes are found by 
assuming that the electric field has the form
\begin{align}
& E_z=Ae^{iqz-Q_1x},E_y=0,E_x=Be^{iqz-Q_1x},\ \mbox{for $x>0$}, \nonumber \\
& E_z=Ce^{iqz+Q_2x},E_y=0,E_x=De^{iqz+Q_2x},\ \mbox{for $x<0$}.
\label{ansatz}
\end{align}
After inserting this ansatz into Maxwell’s equations and matching the boundary 
conditions [which include the conductance of the 2D graphene layer, $\sigma(\omega,q)$], 
we obtain the dispersion relation for TM modes:
\begin{equation}
\frac{\epsilon_{r1}}{\sqrt{q^2-\frac{\epsilon_{r1} \omega^2}{c^2}}}+
\frac{\epsilon_{r2}}{\sqrt{q^2-\frac{\epsilon_{r2} \omega^2}{c^2}}}=
-\frac{\sigma(\omega,q)i}{\omega\epsilon_0}
\label{disp1}
\end{equation}
By explicitly writing the dependence of the conductivity on the wave vector $q$ 
we allow for the possibility of nonlocal effects, where the mean free path of 
electrons can be smaller than $q^{-1}$ \cite{AM}. Throughout this work we 
consider the nonretarded regime ($q\gg \omega/c$), so equation (\ref{disp1}) simplifies to
\begin{equation}
q\approx Q_1\approx Q_2\approx 
\epsilon_0 \frac{\epsilon_{r1}+\epsilon_{r2}}{2}
\frac{2i\omega}{\sigma(\omega,q)}.
\label{disp2}
\end{equation}
Note that a small wavelength (large $q$) leads to a high transversal localization 
of the modes, which are also accompanied by a collective surface charge 
oscillation, similar to SPs in metals; however, it should be understood that, 
in contrast to SPs, here we deal with 2D collective excitations, i.e. plasmons. 
We note that even though field profiles 
of plasmons in graphene and SPs in metals look the same, these two systems are
qualitatively different since electrons in graphene are essentially frozen 
in the transverse dimension \cite{Backes1992}. This fact and the differences 
in electronic dispersions (linear Dirac cones vs. usual parabolic) lead 
to qualitatively different dispersions of TM modes in these two systems 
[see Fig. \ref{Figure1_c3}(b) and Fig. \ref{Figure2_c3}(d)]. 
To find dispersion of plasmons in graphene we need 
the conductivity of graphene $\sigma(\omega,q)$, which we now proceed to 
analyze by employing the semiclassical model \cite{AM} (in subsection \ref{subsemi}), 
RPA and number conserving relaxation-time approximation \cite{Mermin1970} (in subsection 
\ref{subRPA}), and by estimating the relaxation-time due to the influence of 
electron-phonon coupling \cite{Park2007} on the optical 
conductivity \cite{Stauber2008} (in subsection \ref{subeph}).

\subsection{Semiclassical model}
\label{subsemi}

For the sake of the clarity of the presentation, we first note that 
by employing a simple semi-classical model for the conductivity 
(see Ref. \cite{AM}), one obtains a Drude-like expression: 
\begin{equation}
\sigma(\omega)=\frac{e^2 E_F}{\pi \hbar^2} \frac{i}{\omega+i \tau^{-1}}
\label{sigDrude}
\end{equation}
(the semiclassical conductivity does not depend on $q$). 
Here $\tau$ denotes the relaxation-time (RT), which in a phenomenological 
way takes into account losses due to electron-impurity, electron-defect, 
and electron-phonon scattering. 
Equation (\ref{sigDrude}) is obtained by assuming zero temperature 
$T\approx 0$, which is a good approximation for highly doped 
graphene considered here, since $E_F\gg k_B T$. 
From Eqs. (\ref{disp2}) and (\ref{sigDrude}) it is straightforward to obtain 
plasmon dispersion relation: 
\begin{equation}
q(\omega)=
\frac{\pi \hbar^2 \epsilon_0 (\epsilon_{r1}+\epsilon_{r2})}
{e^2 E_F}(1+\frac{i}{\tau \omega})\omega^2, 
\label{qomDrude}
\end{equation}
as well as losses,
\begin{equation}
\frac{\Re q}{\Im q}=
\omega\tau=
\frac{2\pi c \tau}{\lambda_{air}}. 
\end{equation}
In order to quantify losses one should estimate the relaxation time $\tau$. 
If the frequency $\omega$ is below the interband threshold frequency $\omega_{inter}$, 
and if $\omega<\omega_{Oph}$, then both interband damping and 
plasmon decay via excitation of optical phonons together with an 
electron-hole pair are inactive. 
In this case, the relaxation time can be estimated from DC measurements \cite{Novoselov2004}, 
i.e., it can be identified with DC relaxation 
time which arises mainly from impurities (see Refs. \cite{Novoselov2004}). 
It is reasonable to expect that 
impurity related relaxation time will not display large frequency dependence. 
In order to gain insight into the losses by using this line of reasoning let us 
assume that the doping level is given by $E_F=0.64$~eV (corresponding to 
electron concentration of $n=3\times 10^{13}$~cm$^{-2}$); 
the relaxation time corresponds to DC mobility $\mu=10000$~cm$^2$/Vs 
measured in Ref. \cite{Novoselov2004}: 
$\tau_{DC}=\mu\hbar\sqrt{n\pi}/e v_F=6.4 \times 10^{-13}$s. 
As an example, for the frequency $\hbar\omega=0.155$~eV 
($\lambda_{air}=8\, \mu$m), the semiclassical model yields $\Re q/\Im q \approx 151$ 
for losses and $\lambda_{air}/\lambda_{p}\approx 42$ for wave localization. 
Note that both of these numbers are quite favorable compared to conventional SPs
[e.g., see Fig. \ref{Figure1_c3}(c)].
It will be shown in the sequel that for the doping value $E_F=0.64$~eV this
frequency is below the interband loss threshold, and it is evidently also smaller
than the optical phonon loss threshold $\hbar\omega_{Oph}\approx 0.2$~eV, so both of 
these loss mechanisms can indeed be neglected.

\subsection{RPA and relaxation-time approximation}
\label{subRPA}

In order to take the interband losses into account, we use the self-consistent 
linear response theory, also known as the random-phase approximation (RPA) \cite{AM}, 
together with the relaxation-time (finite $\tau$) approximation introduced by Mermin 
\cite{Mermin1970}. Both of these approaches, 
that is, the collisionless RPA ($\tau\rightarrow \infty$) 
\cite{Wunsch2006,Hwang2007}, and the RPA-RT approximation (finite $\tau$) \cite{Rana2008}, 
have been applied to study graphene. 
In the $\tau\rightarrow \infty$ case, the RPA 2D polarizability of graphene 
is given by \cite{Hwang2007}:
\begin{equation}
\bar{\chi}(q,\omega) = \frac{e^2}{q^2}\Pi(q,\omega),
\label{chiRPA}
\end{equation}
where
\begin{align}
\Pi(q,\omega) = \frac{4}{\Omega} \sum_{{\bf k},n_1,n_2} &
  \frac{f(E_{n_2,{\bf k}+{\bf q}})-f(E_{n_1,{\bf k}})}
       {\hbar \omega+E_{n_1,{\bf k}}-E_{n_2,{\bf k}+{\bf q}}+i\eta} \nonumber \\
 & \times |\langle n_1,{\bf k}|e^{-i{\bf q}\cdot {\bf r}}|n_2,{\bf k}+{\bf q} \rangle|^2.
\label{PiRPA}
\end{align}
Here $f(E)=(e^{(E-E_F)/k_BT }+1)^{-1}$ is the Fermi distribution function,
$E_F$ is the Fermi energy and factor 4 stands for 2 spin and 2 valley degeneracies. Note that polarizability $\bar{\chi}(q,\omega)$ is simply related to the density-density response function $\chi(q,\omega)$, introduced in chapter \ref{chap:chap2}, since $\Pi(q,\omega)=-\chi(q,\omega)$.

Now, in Eq. (\ref{chiRPA}) $\omega$ is given an infinitesimally small 
imaginary part which leads to the famous Landau damping; that is, 
plasmons can decay by exciting an electron-hole pair 
(interband and intraband scattering) as illustrated in Fig. \ref{Figure2_c3}(c). 
The effects of other types of scattering (impurities, phonons) can be 
accounted for by using the relaxation-time $\tau$ as a parameter 
within the RPA-RT approach \cite{Mermin1970}, which takes into account conservation of local electron number. Within this approximation the 2D polarizability is 

\begin{equation}
\bar{\chi}_{\tau}(q,\omega) =\frac{(1+i/\omega\tau)\bar{\chi}(q,\omega+i/\tau)}
{1+(i/\omega\tau) \bar{\chi}(q,\omega+i/\tau)/\bar{\chi}(q,0)}. 
\label{chitau}
\end{equation}
The 2D dielectric function and conductivity are respectively given by (see 
\cite{Stern1967}):
and 
\begin{equation}
\sigma_{RPA}(q,\omega)=-i\omega \bar{\chi}_{\tau}(q,\omega).
\label{sigmaRPA}
\end{equation}
We note here that throughout the text only $\pi$–--bands are taken into 
consideration; it is known that in graphite, higher $\sigma$–--bands give 
rise to a small background dielectric constant \cite{Taft1965} at low energies, 
which is straightforward to implement in the formalism. Using Eqs. (\ref{disp2}) 
and (\ref{sigmaRPA}) we obtain that the properties of plasmons (i.e., dispersion, 
wave localization and losses) can be calculated by solving 
\begin{equation}
\epsilon_{RPA}(q,\omega)=0,
\label{eps=0}
\end{equation}
with complex wave vector $q=q_1+iq_2$. The calculation is simplified by 
linearizing Eq. (\ref{eps=0}) in terms of small $q_2/q_1$, to obtain,
\begin{equation}
\frac{\epsilon_{r1}+\epsilon_{r2}}{2}+
\frac{e^2}{2\epsilon_0 q_1} \Re [\Pi(q_1,\omega)]=0,
\label{dispfin}
\end{equation}
for the plasmon dispersion, and 
\begin{equation}
q_2=\frac{\Im [\Pi(q_1,\omega)]+\frac{1}{\tau}
\frac{\partial}{\partial\omega} \Re [\Pi(q_1,\omega)]
+\frac{1}{\omega\tau} \Re [\Pi(q_1,\omega) (1-\Pi(q_1,\omega))/\Pi(q_1,0)]}
{\frac{1}{q_1}\Re [\Pi(q_1,\omega)]-\frac{\partial}{\partial q_1}\Re [\Pi(q_1,\omega)]}
\label{q2}
\end{equation}
yielding losses. 
Note that in the lowest order the dispersion relation 
(and consequently $\lambda_{air}/\lambda_p$ and the group velocity 
$v_g$) does not depend on $\tau$. 
This linearization is valid when $q_2 \ll q_1$; as the plasmon 
losses increase, e.g., after entering the interband regime [the rose area in Fig. 
\ref{Figure2_c3}(d)], results from Eqs. (\ref{dispfin}) and (\ref{q2}) should be 
regarded as only qualitative. The characteristic shape of the plasmon 
dispersion is shown in Fig. \ref{Figure2_c3}(d). 
Note that the semi-classical model and the RPA model agree well 
if the system is sufficiently below the interband threshold
[for small $q$, $\omega(q)\sim \sqrt{q}$ as in Eq. (\ref{qomDrude})]. 
By comparing Figs. \ref{Figure2_c3}(d) and \ref{Figure1_c3}(b) we see that the dispersion 
for SPs on silver-dielectric surface qualitatively differs from the plasmon 
dispersion in graphene \cite{Backes1992}.
While SPs' dispersion relation approaches an asymptote ($\omega\rightarrow\omega_{SP}$) 
for large $q$ values [Eq. (\ref{convP})], graphene plasmon relation gives 
$\omega(q)$ which continuously increases [Fig. \ref{Figure2_c3}(d)].

\begin{figure*}[!ht]
\centerline{
\mbox{\includegraphics[width=0.85\textwidth]{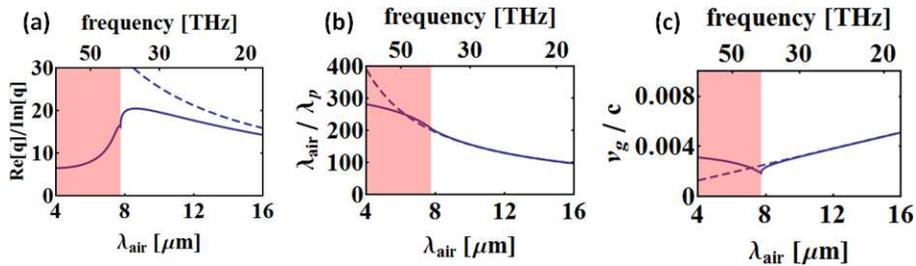}}
}
\caption{
Properties of plasmons in doped graphene. Solid-lines are obtained with the 
number-conserving RPA calculation, and the dashed lines with the semiclassical 
approach. Losses (a), field localization (wave "shrinkage") (b), and group 
velocity (c) for doping $E_F=0.135$~eV, and relaxation time 
$\tau=1.35\times 10^{-13}$~s, which corresponds to the mobility of $10000$~cm$^2/$Vs.
The upper scale in all figures is frequency $\nu=\omega/2\pi$, 
whereas the rose shaded areas denote the region of high interband losses. 
}
\label{Figure3_c3}
\end{figure*}

Theoretically predicted plasmon losses $\Re q/\Im q$ and wave 
localization $\lambda_{air}/\lambda_{p}$ are illustrated in Fig. 
\ref{Figure3_c3} for doping level $E_F=0.135$~eV and relaxation time 
$\tau=1.35\times 10^{-13}$~s. 
We observe that for this particular doping level, 
for wavelengths smaller than $\lambda_{inter}\approx 7.7\, \mu$m, 
the system is in the regime of high interband losses (rose shaded region). 
Below the interband threshold, both losses and wave localization obtained by 
employing RPA-RT approach are quite well described by the previously 
obtained semiclassical formulae. 
Since the frequencies below the interband threshold are (for the assumed 
doping level) also below the optical phonon frequency, the relaxation time 
can be estimated from DC measurements. 

At this point we also note that in all our calculations we have 
neglected the finite temperature effects, i.e., $T\approx 0$. 
To justify this, we note that for doping values utilized in this paper 
the Fermi energies are $0.135$~eV$\approx 5.2k_BT_{r}$ ($n=1.35\times 10^{12}$~cm$^{-2}$) 
and $0.64$~eV$\approx 25k_BT_{r}$ ($n=3\times 10^{13}$~cm$^{-2}$) for room 
temperature $T_{r}=300$~K.
The effect of finite temperature is to slightly smear the sharpness of the 
interband threshold, but only in the vicinity ($\sim k_BT_{r}$) of the threshold.

By increasing the doping, $E_F$ increases, and the region of interband 
plasmonic losses moves towards higher frequencies (smaller wavelengths). 
However, by increasing the doping, 
the interband threshold frequency will eventually become larger than graphene's optical 
phonon frequency $\omega_{Oph}$: there will exist an interval of 
frequencies, $\omega_{Oph}<\omega<\omega_{inter}$, where it is kinematically 
possible for the photon of frequency $\omega$ to excite an 
electron-hole pair together with emission of an optical phonon. 
This second order process can reduce the relaxation time estimated
from DC measurements and should be taken into account, as we show in the 
following subsection.

\subsection{Losses due to optical phonons}
\label{subeph}

In what follows, we estimate and discuss the relaxation time due to the 
electron-phonon coupling. 
This can be done by using the Kubo formula which has been utilized in 
Ref. \cite{Stauber2008} to calculate the real part of the 
optical conductivity, $\Re\sigma(\omega,q=0)$. The calculation of conductivity 
$\Re\sigma(\omega,0)$ involves the electron self-energy 
$\Sigma(E)$, whose imaginary part expresses the width of a 
state with energy $E$, whereas the real part corresponds to the energy shift. 
Let us assume that the electron self-energy stems from the 
electron-phonon coupling and impurities, 
\begin{equation}
\Sigma(E)=\Sigma_{e-ph}(E)+\Sigma_{imp}(E). 
\end{equation}
For $\Sigma_{e-ph}$ we utilize a simple yet fairly accurate model 
derived in Ref. \cite{Park2007}: If $|E-E_F|>\hbar\omega_{Oph}$, 
then 
\begin{equation}
\Im\Sigma_{e-ph}(E)=\gamma |E-\mbox{sgn}(E-E_F)\hbar \omega_{Oph}|,
\end{equation}
while elsewhere $\Im\Sigma_{e-ph}(E)=0$; the dimensionless constant 
$\gamma=18.3\times 10^{-3}$ \cite{Park2007} is proportional to the 
square of the electron-phonon matrix element \cite{Park2007}, i.e., 
the electron-phonon coupling coefficient.
In order to mimic impurities, we will assume that $\Im\Sigma_{imp}(E)$ 
is a constant (whose value can be estimated from DC measurements). 
The real parts of the self-energies are calculated 
by employing the Kramers-Kr\" onig relations. In all our calculations 
the cut-off energy is taken to be $8.4$~eV, which corresponds to the 
cut-off wavevector $k_{c}=\pi/a$, where $a=2.46$~\AA. 
By employing these self-energies we calculate the conductivity 
$\Re\sigma(\omega,q=0)$, from which we estimate the relaxation time 
by using Eq. (\ref{sigDrude}), i.e., 
\begin{equation}
\tau(\omega)\approx 
\frac{e^2 E_F}{\pi \hbar^2 \omega^2}
\frac{1}{\Re \sigma (\omega,0)}
\label{taufin}
\end{equation}
for the region below the interband threshold; in deriving (\ref{taufin}) 
we have assumed $\tau\omega\gg 1$. 

\begin{figure*}[!ht]
\centerline{
\mbox{\includegraphics[width=0.75\textwidth]{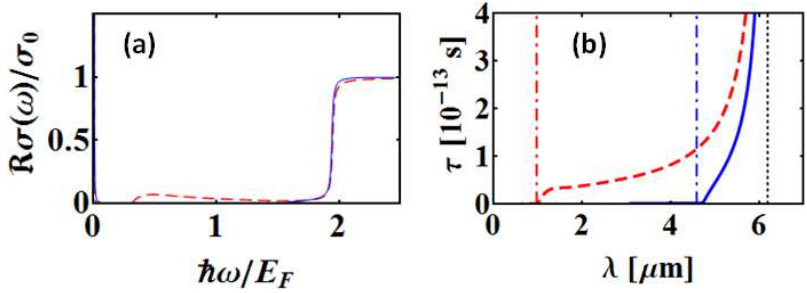}}
}
\caption{
(a) The real part of the conductivity in units of $\sigma_0=\pi e^2/2h$ 
in dependence of frequency $\hbar\omega/E_F$, and (b) the corresponding relaxation time 
as a function of wavelength. The contribution to $\Re\sigma(\omega)$ from impurities 
is chosen to be negligible. The displayed graphs correspond to two different values 
of doping which yield $E_F=0.135$ eV (solid blue line), and 
$E_F=0.640$ eV (dashed red line). The position of the optical phonon frequency 
$\hbar\omega_{Oph}\approx0.2$ eV is depicted by the dotted vertical line in (b); 
dot-dashed lines depict the values of wavelengths corresponding to $2E_F$, that is, 
the interband threshold value (for $q=0$) for the two doping concentrations. 
}
\label{Figure4_c3}
\end{figure*}

Figure \ref{Figure4_c3} plots the real part of the conductivity and the relaxation time 
for two values of doping: $E_F=0.135$~eV ($n=1.35\times 10^{12}$~cm$^{-2}$, 
solid line) and $E_F=0.64$~eV ($n=3\times 10^{13}$~cm$^{-2}$, dashed line). 
In order to isolate the influence of the electron-phonon coupling 
on the conductivity and plasmon losses, the contribution from impurities 
is assumed to be very small: $\Im\Sigma_{imp}(E)=10^{-6}$~eV. 
The real part of the conductivity has a universal value $\sigma_0=\pi e^2/2h$ 
above the interband threshold value $\hbar\omega=2E_F$ (for $q=0$), e.g., see 
\cite{Nair2008,Mak2008}. 
We clearly see that the relaxation time is not affected by the 
electron-phonon coupling for frequencies below $\omega_{Oph}$, that is, 
we conclude that scattering from impurities and defects is a dominant decay 
mechanism for $\omega<\omega_{Oph}$ (assuming we operate below the interband threshold). 
However, for $\omega>\omega_{Oph}$, the relaxation times in Fig. \ref{Figure4_c3} 
are on the order of $10^{-14}-10^{-13}$~s, 
indicating that optical phonons are an important decay mechanism. 

It should be emphasized that the exact calculated values should be 
taken with some reservation for the following reason: strictly speaking, one should 
calculate the relaxation times $\tau(\omega,q)$ along the plasmon 
dispersion curve given by Eq. (\ref{dispfin}); namely the matrix elements
which enter the calculation depend on $q$, whereas the phase space 
available for the excitations also differ for $q=0$ and $q>0$. 
Moreover, the exact value of the matrix element for electron phonon coupling 
is still a matter of debate in the community. 
Therefore, the actual values for plasmon losses could be somewhat different 
for $\omega>\omega_{Oph}$. Nevertheless, fairly small values of relaxation times presented in Fig. \ref{Figure4_c3} for $\omega>\omega_{Oph}$ indicate that emission of an optical phonon together with an electron-hole pair is an important decay mechanism in this regime. Precise calculations for $q>0$ and $\omega>\omega_{Oph}$ are a topic for a future paper. 

\begin{figure*}[!ht]
\centerline{
\mbox{\includegraphics[width=0.85\textwidth]{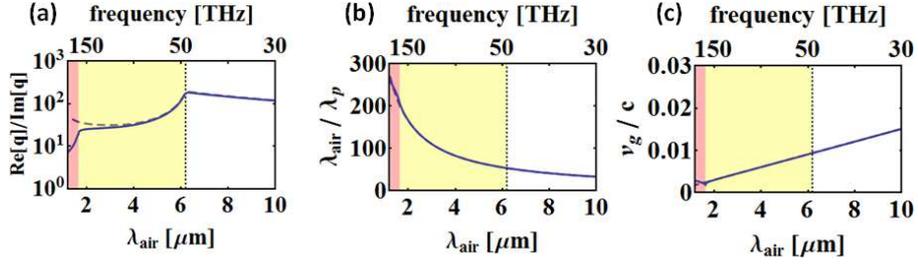}}
}
\caption{
Properties of plasmons in doped graphene. Solid-lines are obtained with the 
number-conserving RPA calculation, and the dashed lines with the semiclassical 
approach. Losses (a), field localization (wave "shrinkage") (b), and group 
velocity (c) for doping $E_F=0.64$~eV; losses are calculated by 
using the relaxation time $\tau^{-1}=\tau^{-1}_{DC}+\tau^{-1}_{e-ph}$, where 
$\tau_{DC}=6.4\times 10^{-13}$~s, and $\tau_{e-ph}$ is the relaxation time 
from the electron-phonon coupling for the given parameters. 
In the white regions (right regions in all panels), losses are 
determined by $\tau_{DC}$. 
In the yellow shaded regions (central regions in all panels), losses are 
determined by the optical phonon emission, i.e., $\tau_{e-ph}$. 
The rose shaded areas (left region in all panels) denote the region 
of high interband losses.
Dotted vertical lines correspond to the optical phonon frequency 
$\omega_{Oph}\approx 0.2$~eV. The upper scale in all figures is 
frequency $\nu=\omega/2\pi$. 
 See text for details. 
}
\label{Figure5_c3}
\end{figure*}

Plasmonic losses and wave localization calculated from the RPA-RT approximation 
are illustrated in Fig. \ref{Figure5_c3} for doping level $E_F=0.64$~eV
and the relaxation time $\tau$ given by $\tau^{-1}=\tau^{-1}_{DC}+\tau^{-1}_{e-ph}$, where 
$\tau_{DC}=6.4\times 10^{-13}$~s (mobility $10000$~cm$^2/$Vs), 
whereas $\tau_{e-ph}$ is frequency dependent and corresponds to electron-phonon 
coupling assuming very clean samples 
[see dashed line in Fig. \ref{Figure4_c3}(b)]. 
Interband losses [left (rose shaded) regions in all panels] are active for wavelengths smaller 
than $\lambda_{inter}\approx 1.7\,\mu$m. In the frequency interval 
$\omega_{inter}>\omega>\omega_{Oph}$ [central (yellow shaded) regions in all panels], 
the decay mechanism via electron phonon coupling determines the loss rate, 
i.e., $\tau\approx \tau_{e-ph}$. 
For $\omega<\omega_{Oph}$ [right (white) regions in all panels], the DC 
relaxation time $\tau_{DC}$ can be used to estimate plasmon losses. 

It should be noted that the mobility of $10000$~cm$^2/$Vs could be improved, 
likely even up to mobility $100000$~cm$^2/$Vs \cite{Geim2007}, thereby further improving 
plasmon propagation lengths for frequencies below the optical phonon frequency. 
However, for these larger mobilities the calculation of losses should also include 
in more details the frequency dependent contribution to the relaxation time from 
acoustic phonons (this decay channel is open at all frequencies); 
such a calculation would not affect losses for $\omega>\omega_{Oph}$ where 
optical phonons are dominant.

\section{Conclusion and Outlook}
\label{sec:conc}

In conclusion, we have used RPA and number-conserving relaxation-time approximation 
with experimentally available input parameters, and theoretical estimates for the 
relaxation-time utilizing electron-phonon coupling, to study plasmons and their 
losses in doped graphene. 
We have shown that for sufficiently large doping values
high wave localization and low losses are simultaneously 
possible for frequencies below that of the optical phonon branch 
$\omega<\omega_{Oph}$ (i.e., $E_{plasmon}<0.2$~eV). 
For sufficiently large doping values, 
there is an interval of frequencies above $\omega_{Oph}$ and below 
interband threshold, where an important decay mechanism for plasmons 
is excitation of an electron-hole pair together with an optical 
phonon (for $\omega<\omega_{Oph}$ this decay channel is inactive);
the relaxation times for this channel were estimated and discussed. 
We point out that further more precise calculations of plasmon relaxation 
times should include coupling to the substrate (e.g., coupling to surface-plasmon 
polaritons of the substrate), a more precise shape of the phonon dispersion 
curves, and dependence of the relaxation time via 
electron-phonon coupling on $q>0$ (see subsection \ref{subeph}). 

The main results, shown in Figures \ref{Figure3_c3} and \ref{Figure5_c3} point out 
some intriguing opportunities offered by plasmons in graphene for the field of 
nano-photonics and metamaterials in infrared (i.e. for $\omega<\omega_{Oph}$). 
For example, we can see in those figures that high field localization and enhancement  
$\lambda_{air}/\lambda_p\sim 200$ [see Figure \ref{Figure3_c3}(b)] are possible 
(resulting in $\lambda_p<50$~nm), while plasmons of this kind could have propagation 
loss-lengths as long as $\sim 10\lambda_p$ [see Fig. \ref{Figure5_c3}(a)];
these values (albeit at different frequencies) are substantially 
more favorable than the corresponding values for 
conventional SPs, for example, for SPs at the Ag/Si interface 
$\lambda_{air}/\lambda_p\sim 20$, whereas propagation lengths are only 
$\sim 0.1\lambda_{sp}$ [see Fig. \ref{Figure1_c3}(c)]. 
Another interesting feature of plasmons in graphene is that, similar to usual 
SP-systems \cite{Aristeidis2005}, wave localization is followed by a group 
velocity decrease; the group velocities can be of the order 
$v_g=10^{-3}-10^{-2}$c, and the group velocity can be low over a wide frequency 
range, as depicted in Figs. \ref{Figure3_c3}(c) and \ref{Figure5_c3}(c). This is of interest 
for possible implementation of novel nonlinear optical devices in graphene, 
since it is known that small group velocities can lead to savings in both the 
device length and the operational power \cite{Soljacic2004}; the latter would also be reduced 
because of the large transversal field localization of the plasmon modes.

\chapter{Transverse-electric plasmons} 
\label{chap:chap4}

In Chapter \ref{chap:chap3} we were studying longitudinal charge density oscillations i.e. longitudinal plasmons or TM modes. However, due to unusual electron dispersion, graphene can also support transverse plasmons or TE modes \cite{Mikhailov2007}. These excitation are possible only if the imaginary part of the conductivity of a thin sheet of material is negative \cite{Mikhailov2007}. 
On the other hand, such a conductivity requires some complexity of the 
band structure of the material involved. For example, TE plasmons 
cannot occur if the 2D material possesses a single parabolic 
electron band. From this perspective, bilayer graphene, with its rich band 
structure and optical conductivity (e.g., see \cite{Nicol2008} and references therein), 
seems as a promising material for exploring the possibility of existence of TE plasmons. 
Here we predict the existence of TE plasmons in bilayer graphene.
We find that their plasmonic properties are much more pronounced in 
bilayer than in monolayer graphene, in a sense that the
wavelength of TE plasmons in bilayer can be smaller than in monolayer graphene 
at the same frequency.

Throughout this work we consider bilayer graphene as an infinitely 
thin sheet of material with conductivity $\sigma ({\bf q},\omega)$. 
We assume that air with $\epsilon_r=1$ is above and below bilayer graphene. 
Given the conductivity, by employing classical electrodynamics, one 
finds that self-sustained oscillations of the charge occur when 
(see \cite{Mikhailov2007} and references therein)
\begin{equation}
1+\frac{i\sigma({\bf q},\omega) \sqrt{q^2-\omega^2/c^2}}
{2\epsilon_0\omega}=0
\end{equation}
for TM modes, and 
\begin{equation}
1-
\frac{\mu_0 \omega i\sigma({\bf q},\omega)}
{2\sqrt{q^2-\omega^2/c^2}}=0
\label{TE}
\end{equation}
for TE modes. The TM plasmons can considerably depart from the light 
line, that is, their wavelength can be considerably smaller than that 
of light at the same frequency. For this reason, when calculating 
TM plasmons it is desirable to know the conductivity as a function of 
both frequency $\omega$ and wavevector ${\bf q}$. However, it turns out 
that the TE plasmons (both in monolayer \cite{Mikhailov2007} and 
bilayer graphene, as will be shown below) are quite close to the light 
line $q=\omega /c$, and therefore it is a good approximation 
to use $\sigma (\omega)=\sigma ({\bf q}=0,\omega)$. 
Moreover, these plasmons are expected to show strong polariton character, 
i.e., creation of hybrid plasmon-photon excitations. 
At this point it is worthy to note that if the relative permittivity of
dielectrics above and below graphene are sufficiently different, so that
light lines differ substantially, then TE plasmon will not exist 
(perhaps they could exist as leaky modes).

\section{Optical conductivity of bilayer graphene}

The conductivity $\sigma (\omega)=\Re \sigma (\omega)+i\Im \sigma (\omega)$
is complex, and plasmon dispersion is characterized by the imaginary part 
$\Im \sigma (\omega)$, whereas $\Re \sigma (\omega)$ determines 
plasmon losses, or more generally absorption of the sheet. 
From Eq. (\ref{TE}) it follows that the TE plasmons exist only if 
$\Im \sigma (\omega)<0$ \cite{Mikhailov2007}. 

\begin{figure}[htbp]
\centerline{
\mbox{\includegraphics[width=0.44\textwidth]{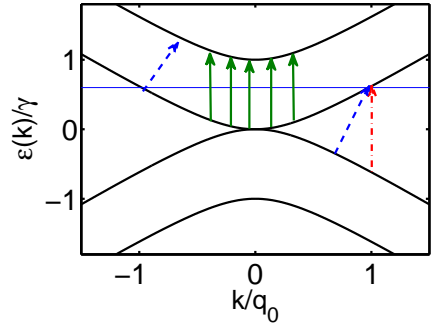}}
}
\caption{The band-structure of bilayer graphene. 
The two upper bands (as well as the two lower bands) are perfectly nested 
and separated by $\gamma\sim 0.4$~eV; $q_0=\gamma/\hbar v_F$. 
Horizontal line depicts one possible value of the Fermi level, and 
arrows denote some of the possible interband electronic transitions. 
See text for details. 
}
\label{Figure1_c4}
\end{figure}

In order to calculate the imaginary part of the conductivity, we 
employ Kramers-Kronig relations and the calculation of absorption 
by Nicol and Carbotte \cite{Nicol2008}, where $\Re\sigma(\omega)$ 
[see Eqs. (19)-(21) in Ref. \cite{Nicol2008}]
was calculated by using the Kubo formula. 
The optical conductivity has rich structure due to the 
fact that the single-particle spectrum of graphene is organized 
in four bands given by \cite{Nicol2008}, 
\begin{equation}
\frac{\epsilon({\bf k})}{\gamma} =
\pm\sqrt{\frac{1}{4}+\left(\frac{\hbar v_F k}{\gamma}\right)^2}
\pm \frac{1}{2},
\label{bandstructure}
\end{equation}
where $v_F=10^6$~m/s, the parameter $\gamma\approx 0.4$~eV 
is equal to the separation between the two conduction bands 
(which is equal to the separation between the valence bands).
The band structure (\ref{bandstructure}) is calculated from the tight binding 
approach, where $v_F$ is connected to the nearest-neighbour hopping terms 
for electrons to move in each of the two graphene planes, and the distance 
between Carbon atoms in one monolayer (see Ref. \cite{Nicol2008}), whereas 
$\gamma$ is the hopping parameter corresponding to electrons hoping from one 
layer to the other and vice versa \cite{Nicol2008}. 
The two graphene layers are stacked one above the other according to the 
so-called Bernal-type stacking (e.g., see Ref. \cite{CastroNeto}). 
We emphasize that the perfect nesting of bands gives rise to the stronger 
plasmon like features of TE plasmons in bilayer than in monolayer graphene. 
The four bands are illustrated in Fig. \ref{Figure1_c4} along 
with some of the electronic transitions which result in absorption. 
Absorption depends on $\gamma$ and the Fermi level $\mu$; the latter can be 
changed by applying external bias voltage. 

The imaginary part of the conductivity can be calculated from 
$\Re\sigma(\omega)$ by using the Kramers-Kronig relations 
\begin{equation}
\Im\sigma(\omega)=-\frac{2\omega}{\pi}
{\cal P}\int_{0}^{\infty}
\frac{\Re\sigma(\omega')}{\omega'^2-\omega^2}d\omega',
\end{equation}
which yields
\begin{eqnarray}
\frac{\Im\sigma(\omega)}{\sigma_0} 
& = & f(\Omega,2\mu)+g(\Omega,\mu,\gamma) \nonumber \\
& + & [f(\Omega,2\gamma)+g(\Omega,\gamma,-\gamma)]\Theta(\gamma-\mu) \nonumber \\
& + & [f(\Omega,2\mu)+g(\Omega,\mu,-\gamma)]\Theta(\mu-\gamma) \nonumber \\
& + & \frac{\gamma^2}{\Omega^2}
\left[ \frac{\Omega}{\pi(2\mu+\gamma)}+f(\Omega,2\mu+\gamma) \right] \nonumber \\
& + & \frac{\gamma^2}{\Omega^2}
\left[ \frac{\Omega}{\pi\gamma}+f(\Omega,\gamma) \right]\Theta(\gamma-\mu) \nonumber \\
& + & \frac{\gamma^2}{\Omega^2}
\left[ \frac{\Omega}{\pi(2\mu-\gamma)}+f(\Omega,2\mu-\gamma) \right]\Theta(\mu-\gamma) \nonumber \\
& + & \frac{a(\mu)}{\pi\Omega}+\frac{2 \Omega b(\mu)}{\pi(\Omega^2-\gamma^2)},
\label{ImS}
\end{eqnarray}
where
\begin{eqnarray}
f(x,y) & = & \frac{1}{2\pi} \log\left| \frac{x-y}{x+y} \right|,  \nonumber \\
g(x,y,z) & = & \frac{z}{2\pi} 
\frac{(x-z)\log|x-2y|+(x+z)\log|x+2y|-2x \log|2y+z|}
{x^2-z^2} , \nonumber \\
a(\mu) & = & \frac{4\mu(\mu+\gamma)}{2\mu+\gamma}+
\frac{4\mu(\mu-\gamma)}{2\mu-\gamma}\Theta(\mu-\gamma) , \nonumber \\
b(\mu) & = & \frac{\gamma}{2} \left[\log\frac{2\mu+\gamma}{\gamma} -
\log\frac{2\mu-\gamma}{\gamma}\Theta(\mu-\gamma) \right],
\label{ImS1}
\end{eqnarray}
$\sigma_0=e^2/2\hbar$, $\Theta(x)=1$ if $x\geq 0$ and zero otherwise, and $\Omega=\hbar\omega$. 
Here we assume zero temperature $T\approx 0$, which is a good approximation for sufficiently 
doped bilayer graphene where $\mu \gg k_B T$.
Formulae (\ref{ImS}) and (\ref{ImS1}) are used to describe the properties of 
TE plasmons. 

\begin{figure}[htbp]
\centerline{
\mbox{\includegraphics[width=0.44\textwidth]{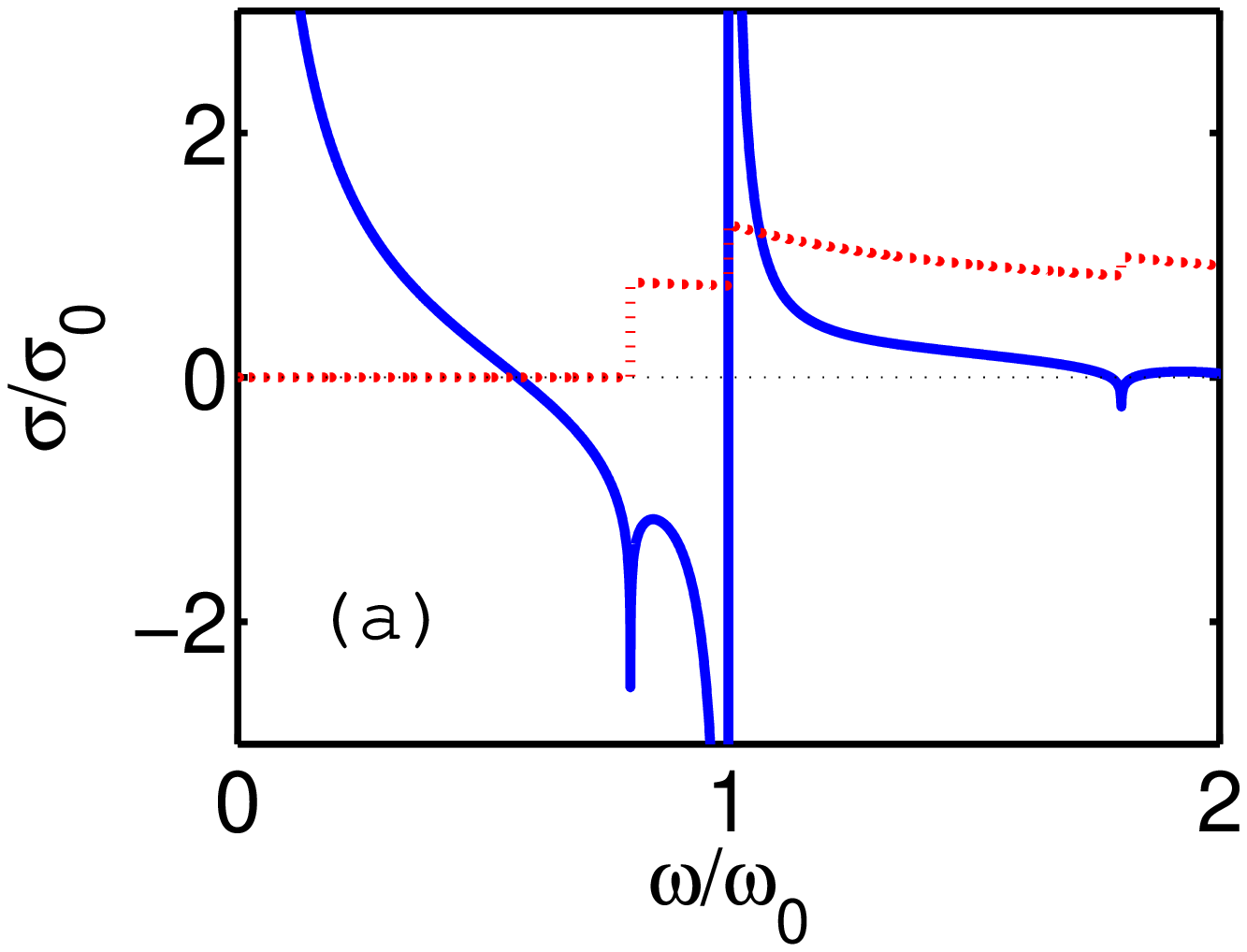}}
\mbox{\includegraphics[width=0.44\textwidth]{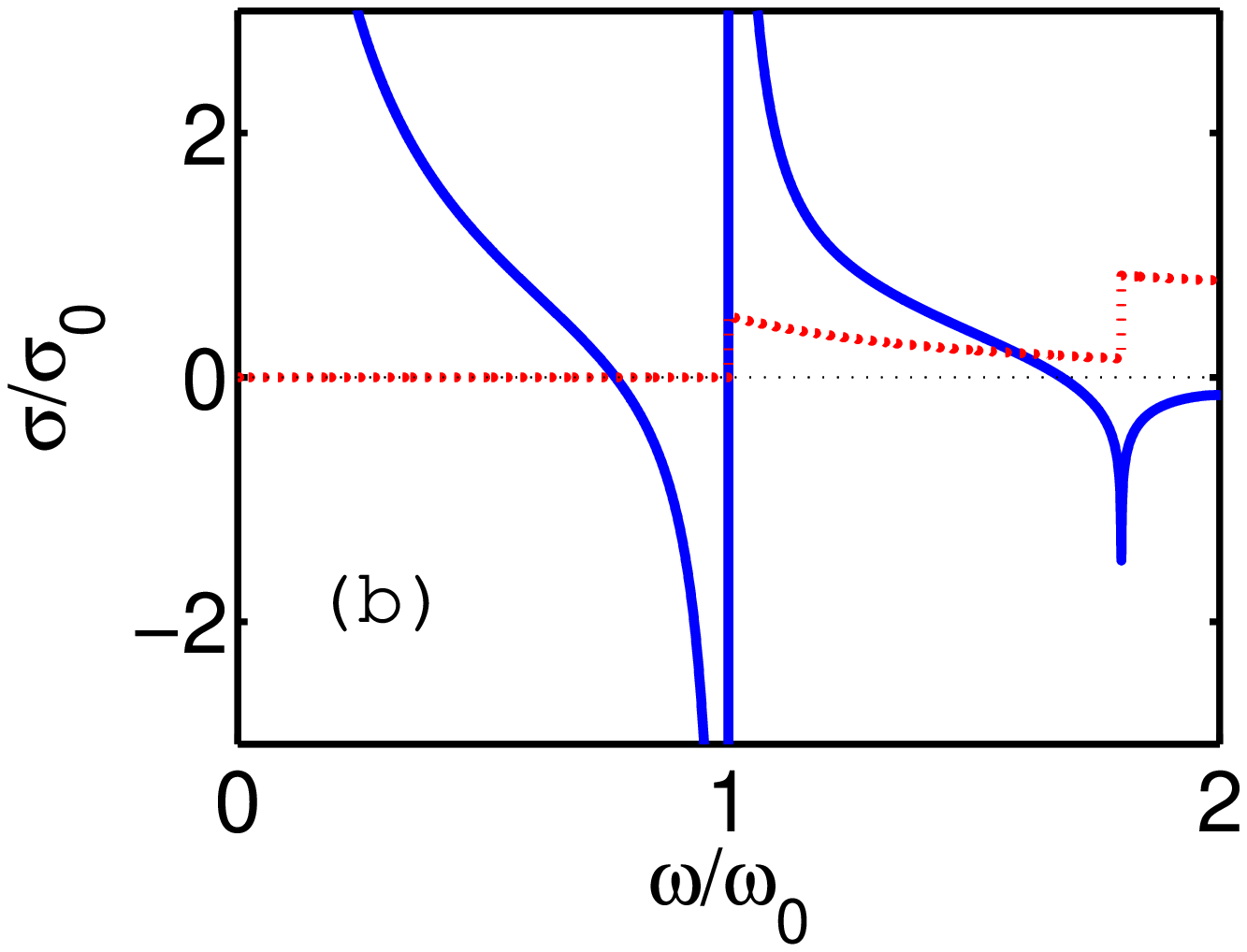}}
}
\caption{
The real (red dotted lines) and imaginary (blue solid lines) part of the 
conductivity of bilayer graphene for two values of doping: 
$\mu=0.4\gamma$ (a), and $\mu=0.9\gamma$ (b). 
The conductivity is in units of $\sigma_0=e^2/2\hbar$, and the 
frequency is in units of $\omega_{0}=\gamma/\hbar$.
The $\delta$-functions in $\Re\sigma(\omega)$ at $\omega=0$ (intraband transitions)
and $\omega=\gamma/\hbar$ (transitions from the lower to the upper conduction band
depicted as green solid arrows in Fig. \ref{Figure1_c4}) are not shown
(see \cite{Nicol2008}).
}
\label{Figure2_c4}
\end{figure}

In Figure \ref{Figure2_c4} we show the real and imaginary part of the 
conductivity for two different values of the Fermi level:
$\mu=0.4\gamma$ and $\mu=0.9\gamma$ (we focus on the electron doped system $\mu>0$). 
Because plasmons are strongly damped by interband transitions, 
it is instructive at this point to discuss the kinematical requirements 
for the excitation of electron-hole pairs. 
If the doping is such that $\mu<\gamma/2$,
a quantum of energy $\hbar\omega$ (plasmon or photon) with in-plane momentum 
$q=0$ can excite an electron-hole pair only if $\hbar\omega>2\mu$ (excitations from the 
upper valence to the lower conduction band shown as red dot-dashed line in 
Fig. \ref{Figure1_c4}). 
If $\mu>\gamma/2$, the $(q=0,\omega)$-quantum can excite an electron-hole pair only for 
$\hbar\omega\geq \gamma$ (excitations from the lower to the upper conduction band
shown as green solid lines in Fig. \ref{Figure1_c4} occur at $\hbar\omega=\gamma$). 
If the plasmon/photon has in-plane momentum $q$ larger than zero, then 
interband transitions are possible for smaller frequencies
(see blue dashed lines in Fig. \ref{Figure1_c4}). 
There is a region in the $(q,\omega)$-plane where electron-hole excitations 
are forbidden due to the Pauli principle. 
Because plasmons are strongly damped by these interband transitions 
(this is Landau damping), in our search for the 
TE plasmons, we focus on their dispersion curve in the regime where 
electron-hole pair formation is inadmissible (via first-order transition). 

\section{Transverse-electric plasmon dispersion in bilayer graphene}

\begin{figure}[htbp]
\centerline{
\mbox{\includegraphics[width=0.44\textwidth]{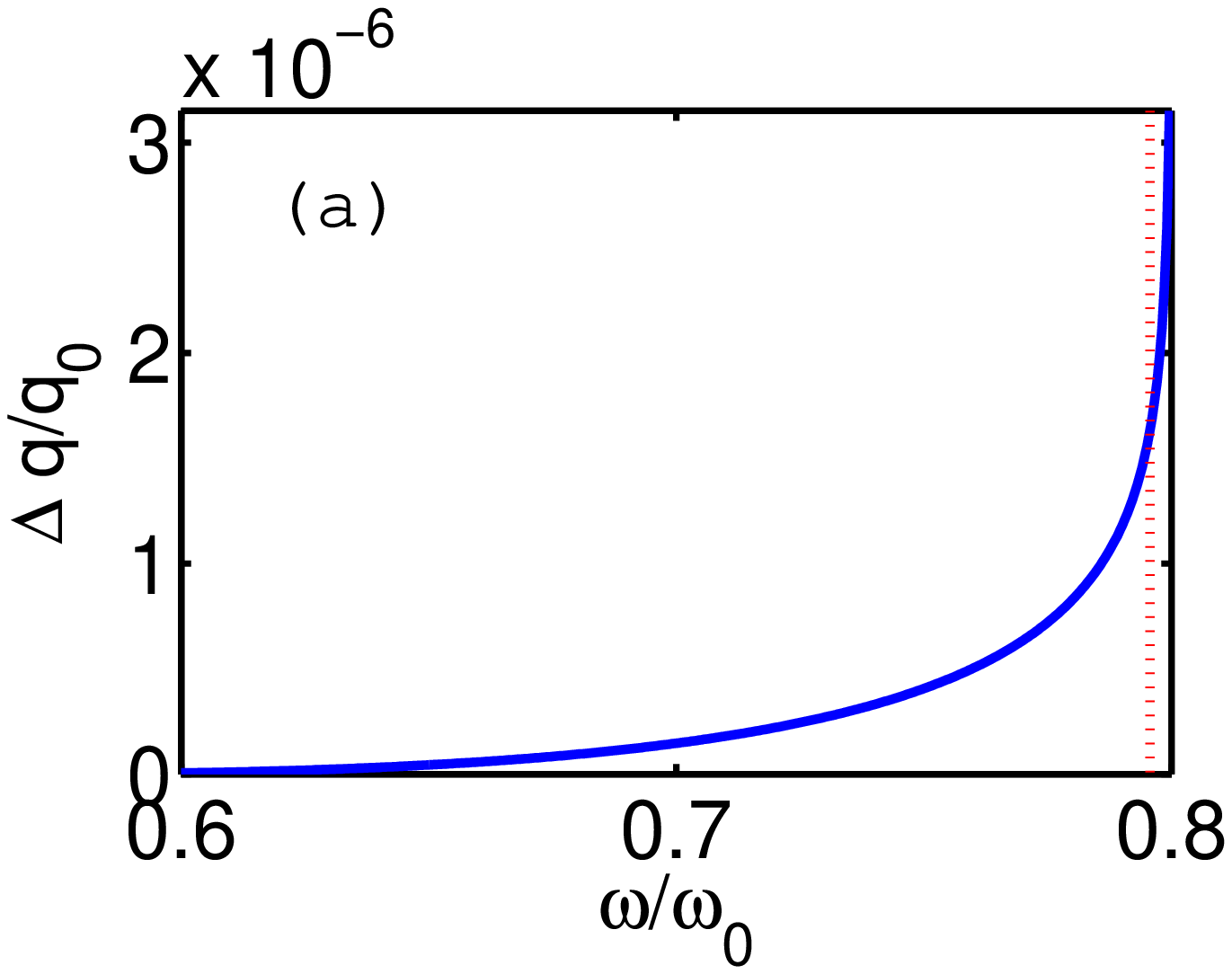}}
\mbox{\includegraphics[width=0.44\textwidth]{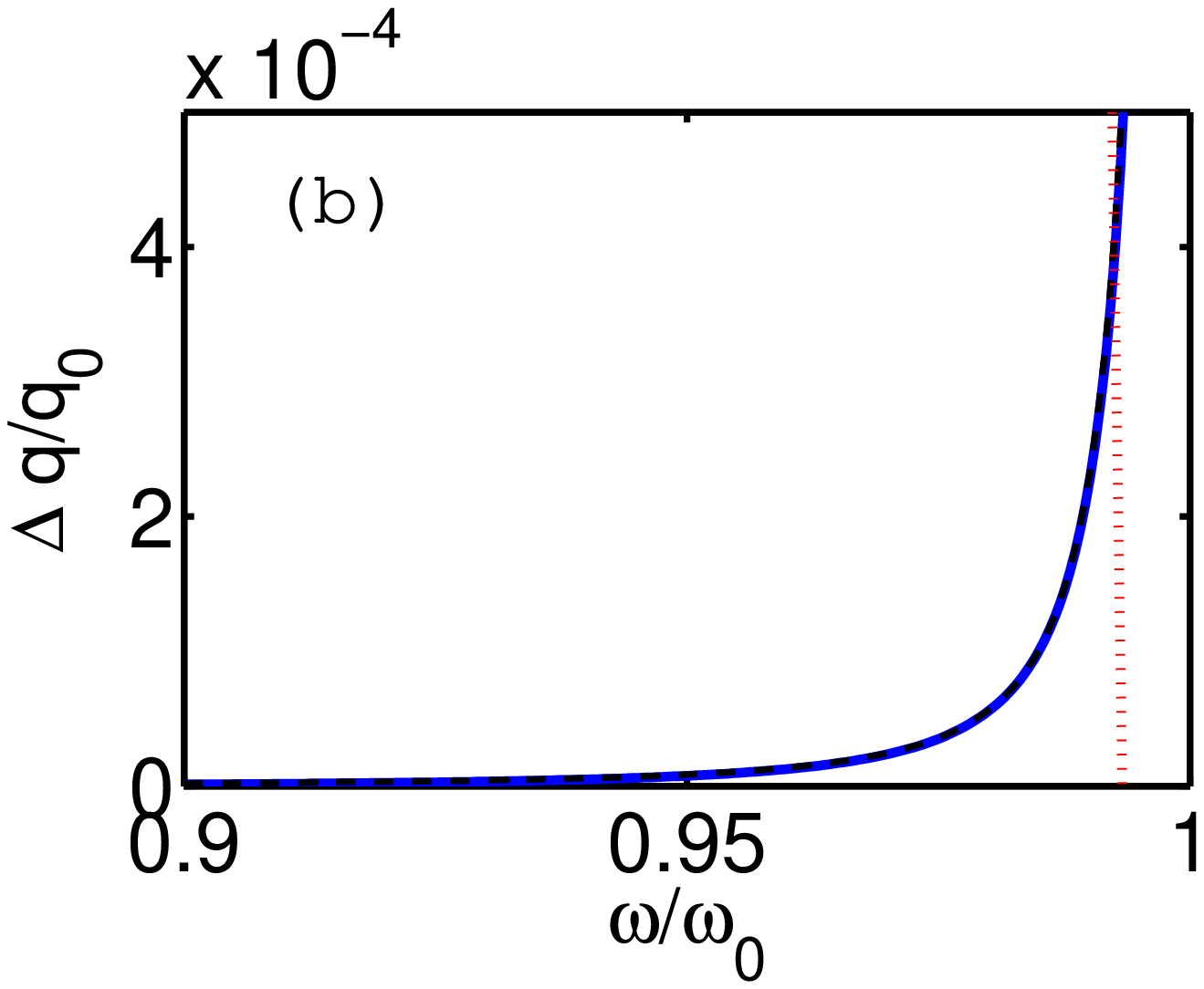}}
}
\caption{
The plasmon dispersion curve $\Delta q=q-\omega/c$ vs. $\omega$ 
for $\mu=0.4\gamma$ (a), and $\mu=0.9\gamma$ (b) is shown as blue solid 
line. To the right of the vertical red dotted lines plasmons can 
be damped via excitation of electron-hole pairs, whereas to the left of 
this line these excitations are forbidden due to the Pauli principle. 
Black dashed line in (b) (which closely follows the blue line) 
corresponds to Eq. (\ref{approxgamma}). 
The wave vector is in units of $q_0=\gamma/\hbar v_F$, and the 
frequency is in units of $\omega_{0}=\gamma/\hbar$. 
}
\label{Figure3_c4}
\end{figure}

In Figure \ref{Figure3_c4} we show the plasmon dispersion curves for 
$\mu=0.4\gamma$ and $\mu=0.9\gamma$; in the spirit of Ref. 
\cite{Mikhailov2007}, we show $\Delta q=q-\omega/c$ as a function 
of frequency $\omega$. 
Plasmons are very close to the light line and thus one can 
to a very good approximation write the dispersion curve as 
\begin{equation}
\Delta q\approx \frac{\omega}{8\epsilon_0^2c^3}\Im\sigma(\omega)^2.
\end{equation}
To the left (right) of the vertical red dotted line in Fig. \ref{Figure3_c4}, 
plasmon damping via excitation of electron-hole pairs is (is not) forbidden.  

For $\mu=0.4\gamma$, $\Im\sigma(\omega)$ is smaller than zero 
for $\omega$ in an interval of frequencies just below $2\mu$. 
From the leading term in $\Im \sigma (\omega)$ we find that 
departure of the dispersion curve from the light line is logarithmically slow: 
$\Delta q_{0<\mu<\gamma/2}\propto [\log|\hbar\omega-2\mu|]^2$. 
The same type of behavior occurs in monolayer graphene \cite{Mikhailov2007}. 

However, for $\mu=0.9\gamma$, one can see the advantage of bilayer over
monolayer graphene in the context of TE plasmons. 
The conductivity $\Im\sigma(\omega)$ is smaller than zero in an interval 
of frequencies below $\gamma$. In this interval, 
the most dominant term to the conductivity is the last one from 
Eq. (\ref{ImS}), that is,
\begin{equation}
\Delta q_{\gamma/2<\mu<\gamma}\approx
\frac{\omega\sigma_0^2 }{2\pi^2\epsilon_0^2c^3}
\left[\frac{\hbar\omega b(\mu)}{\gamma^2-(\hbar\omega)^2}\right]^2.
\label{approxgamma}
\end{equation}
This approximation is illustrated with black dashed line in Fig. \ref{Figure3_c4},
and it almost perfectly matches the dispersion curve. 
Note that the singularity in $\Im\sigma(\omega)$ at $\hbar\omega=\gamma$ is  
of the form $1/(\gamma-\hbar\omega)$, whereas the singularity at $\hbar\omega=2\mu$
is logarithmic (as in monolayer graphene \cite{Mikhailov2007}). 
As a consequence, the departure of the dispersion curve from the 
light line in bilayer graphene is much faster for $\mu>\gamma/2$ than 
for $\mu<\gamma/2$, and it is faster than in monolayer graphene as well 
[note the two orders of magnitude difference between the abscissa scales 
in Figs. \ref{Figure3_c4}(a) and (b)]. 
Thus, we conclude that more pronounced plasmonic features of TE plasmons (shrinking of wave length 
which is measured as departure of $q$ from the light line) can be obtained 
in bilayer graphene. The term in $\Im\sigma(\omega)$ which is responsible 
for TE plasmons for $\mu>\gamma/2$ corresponds (via Kramers-Kronig relations) 
to the absorption term $b(\mu)\delta(\hbar\omega-\gamma)$ \cite{Nicol2008}, which arises from the 
transitions from the first to the second valence band
(shown as green solid arrows in Fig. \ref{Figure1_c4}), which are perfectly nested 
and separated by $\gamma$. Thus, this unique feature of bilayer graphene 
gives rise to TE plasmons with more pronounced plasmon like features 
than in monolayer graphene. 

Before closing this chapter, let us discuss some properties and possible observation of TE plasmons. 
First, note that since the electric field oscillations are 
both perpendicular to the propagation vector ${\bf q}$, and 
lie in the bilayer graphene plane, the electric current ${\bf j}=\sigma(\omega) {\bf E}$
is also perpendicular to ${\bf q}$. Thus, ${\bf j}\cdot {\bf q}=0$, and the 
equation of continuity yields that the charge density is zero 
(i.e., one has self-sustained oscillations of the current).
In order to excite plasmons of frequency $\omega$ with light of the same 
frequency, one has to somehow account for the conservation of the momentum 
which is larger for plasmons. Since the momentum mismatch is relatively 
small, the standard plasmon excitation schemes such as the 
prism or grating coupling methods (e.g., see \cite{Barnes2003} and 
references therein) could be used for the excitation of these plasmons. 

To conclude this chapter, we have predicted the existence of transverse electric (TE) 
plasmons in bilayer graphene. Since they exist very close to 
the light line, these plasmons are expected to show strong polariton 
character, i.e., mixing with photon modes. 
However, due to the perfectly nested valence bands of bilayer graphene, 
their dispersion departs much more from the light line than in monolayer 
graphene.

\chapter{Plasmon-phonon coupling} \label{chap:chap5}

In this chapter we analyze the coupling of plasmons with intrinsic optical phonons
in graphene by using the self-consistent linear response formalism. 
We find that longitudinal plasmons (LP) 
couple only to transverse optical (TO) phonons, while transverse 
plasmons (TP) couple only to longitudinal optical (LO) phonons. 
The LP-TO coupling is stronger for larger concentration of carriers, 
in contrast to the TP-LO coupling (which is fairly weak). 
The former could be measured via current experimental techniques. 
Thus, plasmon-phonon resonance could serve as a magnifier for exploring 
the electron-phonon interaction, and for novel electronic 
control (by externally applied voltage) over crystal lattice 
vibrations in graphene. 

To analyze plasmon-phonon coupling let us start with the Hamiltonian for the Dirac electrons in graphene
\begin{equation}
H_e=\hbar v_F \mbox{\boldmath $\sigma$} \cdot {\bf k} ,
\label{Dirac}
\end{equation}
where $v_F=10^6$~m/s, ${\bf k} = (k_x,k_y)=-i\mbox{\boldmath $\nabla$}$ is the 
wave-vector operator, $\mbox{\boldmath $\sigma$}=(\sigma_x,\sigma_y)$, and $\sigma_{x,y}$ are the Pauli spin matrices. We label the eigenstates of Hamiltonian $H_e$ by $| s, {\bf k} \rangle $ and the appropriate eigenvalues by $E_{s,\bf k}=s \hbar v_F |\bf k|$, where $s=1$ for the conduction band and $s=-1$ for the valence band.

\begin{figure}
\centerline{
\mbox{\includegraphics[width=0.60\textwidth]{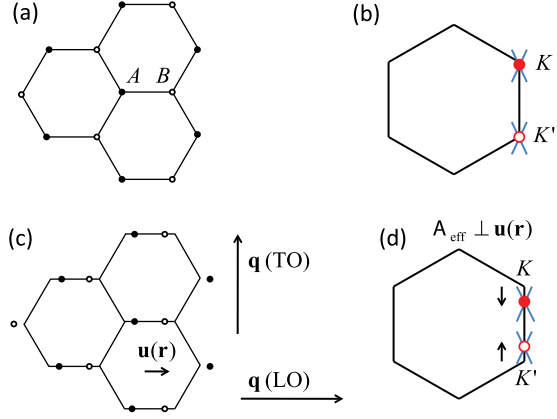}}
}
\caption{
(a) Schematic illustration of the lattice structure with two sublattices (A and B).
(b) The two degenerate Dirac cones are centered at K and K' points at the edge 
of the Brillouin zone. 
(c) A displacement of lattice atoms ${\bf{u}}({\bf r})$ is parallel (perpendicular) 
to the propagation wave vector ${\bf q}$ of a LO (TO) phonon.
(d) The displacement ${\bf{u}}({\bf r})$ creates an effective vector potential 
${\bf A}_{\textrm{eff}}$ {\em perpendicular} to ${\bf{u}}({\bf r})$ 
(the sign of ${\bf A}_{\textrm{eff}}$ for the K' point is opposite to 
that for the K point). 
}
\label{Figure1_c5}
\end{figure}

The long-wavelength in-plane optical phonon branch in graphene consists of two modes (LO and TO) which are effectively dispersionless and degenerate at energy $\hbar\omega_0=0.196eV$. 
Let ${\bf{u}}({\bf R})=[{\bf u}_A({\bf R})-{\bf u}_B({\bf R})]/\sqrt{2}$ denote the relative displacements of the sub-lattice atoms $A$ and $B$ of a unit cell specified by a coordinate ${\bf R}$ [see Fig. \ref{Figure1_c5}(c)]. 
Then, in the long-wavelength limit ${\bf R}$ can be replaced 
by a continuous coordinate {\bf r} and we have
\begin{equation}
{\bf{u}}({\bf r})=\sum_{\mu {\bf q}} \frac{1}{\sqrt{NM}}Q_{\mu {\bf q}}{\bf e}_{\mu {\bf q}}e^{i{\bf q}{\bf r}},
\label{ph-displace}
\end{equation}
where $N$ is the number of unit cells, $M$ is the carbon atom mass, 
${\bf q}=q(\cos\phi_{\bf q},\sin\phi_{\bf q})$ is the phonon wave vector, 
$\mu=L,T$ stands for the polarization, and the polarization unit vectors are 
${\bf e}_{L{\bf q}}=i(\cos\phi_{\bf q},\sin\phi_{\bf q})$, and 
${\bf e}_{T{\bf q}}=i(-\sin\phi_{\bf q},\cos\phi_{\bf q})$. 
The displacement vector ${\bf{u}}({\bf r})$ is parallel (perpendicular) to the 
phonon propagation wave vector ${\bf q}$ for LO (TO, respectively) phonons 
[see Fig. \ref{Figure1_c5}(c)]. The phonon Hamiltonian is given by
\begin{equation}
H_{ph}=\frac{1}{2} \sum_{\mu {\bf q}} (P_{\mu \bf q}^{\dag} P_{\mu \bf q}+
\omega_0^2 Q_{\mu \bf q}^{\dag} Q_{\mu \bf q}),
\label{ph-Hamilt}
\end{equation}
where $Q_{\mu \bf q}$ and $P_{\mu \bf q}$ denote phonon coordinate and momentum. 
The electron-phonon interaction takes a peculiar form in graphene (see chapter \ref{chap:chap2}):
\begin{equation}
H_{e-ph} = -\sqrt{2} \frac{\beta \hbar v_F}{b^2} \mbox{\boldmath $\sigma$} \times {\bf{u(r)}},
\label{e-ph1}
\end{equation}
where $\mbox{\boldmath $\sigma$} \times {\bf{u}}=\sigma_x u_y-\sigma_y u_x$, $b=0.142$~nm is 
the nearest carbon atoms distance, and $\beta=2$. 
We find it convenient to write Eq. (\ref{e-ph1}) as
\begin{equation}
H_{e-ph} = L^2 F \sum_{\mu {\bf q}} {\bf j}_{\bf q}^{\dag} \times {\bf e}_{\mu {\bf q}} Q_{\mu {\bf q}}
\label{e-ph2}
\end{equation}
where ${\bf j}_{\bf q}=-e v_F L^{-2} \mbox{\boldmath $\sigma$} e^{-i{\bf q}{\bf r}}$ is the 
single-particle current-density 
operator, $L^2$ is the area of the system, $e$ is charge of the electron, and 
$F=\frac{\sqrt{2} \beta \hbar}{e b^2 \sqrt{NM}}$.

The electromagnetic field in the plane of graphene is completely described by 
the vector potential ${\bf A}=\sum_{\mu {\bf q}} {\bf e}_{\mu {\bf q}} A_{\mu {\bf q}} e^{i{\bf q}{\bf r}}$ 
(scalar potential is gauged to zero, time dependence is implicitly assumed, and 
$\mu=L,T$ denote polarizations). 
The interaction with Dirac electrons is obtained by substitution 
$\hbar {\bf k}\rightarrow \hbar {\bf k}+ e {\bf A}$ in Eq. (\ref{Dirac}),
which leads to 
\begin{equation}
H_{e-em}=e v_F \mbox{\boldmath $\sigma$} \cdot {\bf A} =
-L^2 \sum_{\mu {\bf q}} {\bf j}_{\bf q}^{\dag} \cdot {\bf e}_{\mu {\bf q}} A_{\mu {\bf q}}.
\label{e-em}
\end{equation}
By comparing Eqs. (\ref{e-ph1}) and (\ref{e-em}) 
it follows that electron-phonon interaction can be regarded 
as a presence of an effective vector potential 
\begin{equation}
{\bf A}_{\textrm{eff}}=F \sum_{\bf q} 
({\bf e}_{T {\bf q}} Q_{L {\bf q}} -
{\bf e}_{L {\bf q}} Q_{T {\bf q}})
e^{i{\bf q}{\bf r}},
\label{Aeff}
\end{equation}
that is, $H_{e-ph}=e v_F \mbox{\boldmath $\sigma$} \cdot {\bf A}_{\textrm{eff}}$. 
It is evident that ${\bf A}_{\textrm{eff}} \cdot {\bf{u}}({\bf r})=0$ that is 
the effective vector potential ${\bf A}_{\textrm{eff}}$ is perpendicular to 
${\bf{u}}({\bf r})$ as illustrated in Figs. \ref{Figure1_c5}(c) and (d) (see also 
Ref. \cite{CastroNeto}), which is responsible for the mixing of polarizations in 
plasmon-phonon coupling. 

As a first pass, let us ignore the phonons and focus on the Hamiltonian $H=H_e+H_{e-em}$.
Without an external perturbation, the electrons in graphene fill the Fermi sea according 
to the Fermi distribution function $f_{s {\bf k}}$. 
A field $A_{\mu {\bf q}} (\omega)$ oscillating at frequency $\omega$ 
will induce an average current density (up to a linear order in the vector 
potential) 
\begin{equation}
\langle { J}_{\mu}({\bf q},\omega) \rangle = - \chi_{\mu}({\bf q},\omega)  A_{\mu {\bf q}} (\omega),
\label{current}
\end{equation} 
where the current-current response function (including 2-spin and 2-valley degeneracy) 
is given by \cite{PinesBook}
\begin{align}
\chi_{\mu}({\bf q},\omega)=4 L^2 \sum_{s_1s_2{\bf k}} & \frac{f_{s_1 {\bf k}}-f_{s_2 {\bf k}+{\bf q}}}{\hbar\omega+\hbar\omega_{s_1 {\bf k}}-\hbar\omega_{s_2 {\bf k}+{\bf q}}+i\eta} 
\nonumber \\
& \times |\langle s_1{\bf k}|{\bf j}_{\bf q} \cdot {\bf e}_{\mu {\bf q}}^*|s_2 {\bf k}+{\bf q}\rangle|^2.
\label{chi}
\end{align}
For the response function  $\chi_{\mu}({\bf q},\omega)$ we utilize the 
analytical expression from Ref. \cite{Principi2009}. 
The subtlety involved with the divergence in Eq. (\ref{chi}) is 
solved by subtracting from $\chi_{L}({\bf q},\omega)$ 
[$\chi_{T}({\bf q},\omega)$] the value $\chi_{L}({\bf q},\omega=0)$ 
[$\chi_{T}({\bf q} \to 0,\omega=0)$]
to take into account that there is no current response 
to the longitudinal [transverse] time [time and space] independent vector potential, 
see \cite{Principi2009, Falkovsky2007} for details. 
We would like to note that when working with the current-current response 
function, rather than with the density-density response function, the 
nature of the plasmon-phonon interaction (especially the mixing of 
polarizations as shown below) is far more transparent. 

\begin{figure}
\centerline{
\mbox{\includegraphics[width=0.44\textwidth]{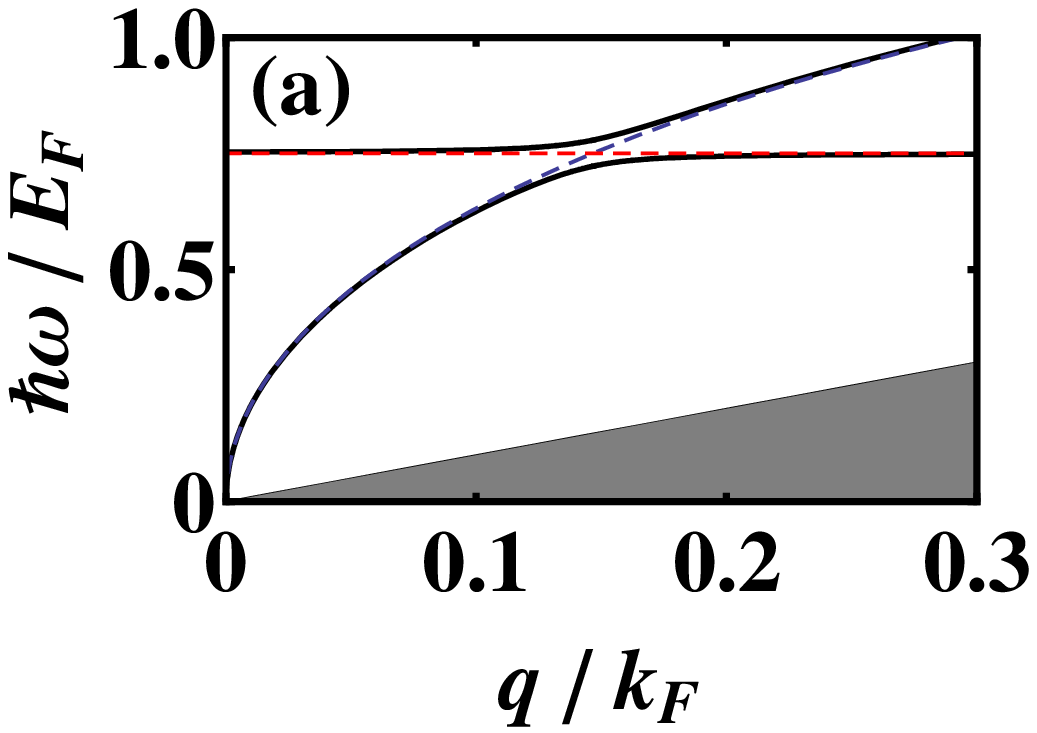}}
\mbox{\includegraphics[width=0.44\textwidth]{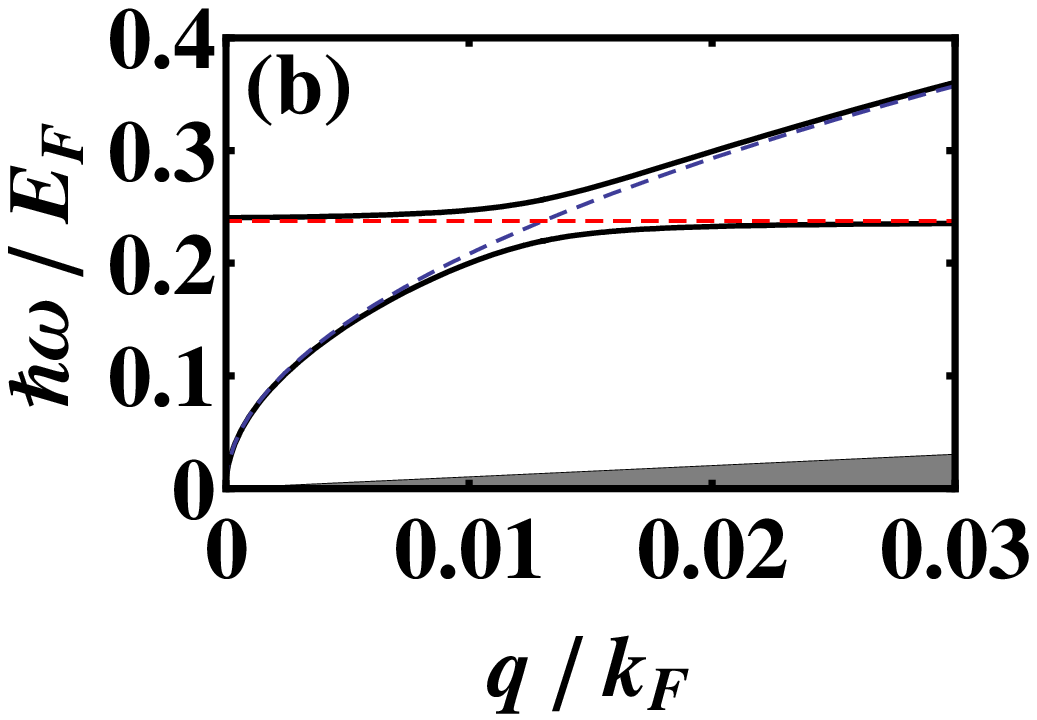}}
}
\caption{
Dispersion lines of hybrid LP-TO plasmon-phonon modes (solid lines) 
and of the uncoupled modes (dashed lines) for two values of doping:
(a) $n=5\times 10^{12}$~cm$^{-2}$, and 
(b) $n=5\times 10^{13}$~cm$^{-2}$. 
The hybridization is stronger for larger doping values. 
Grey areas denote the region of single-particle damping.
}
\label{Figure2_c5}
\end{figure}

\begin{figure}
\centerline{
\mbox{\includegraphics[width=0.44\textwidth]{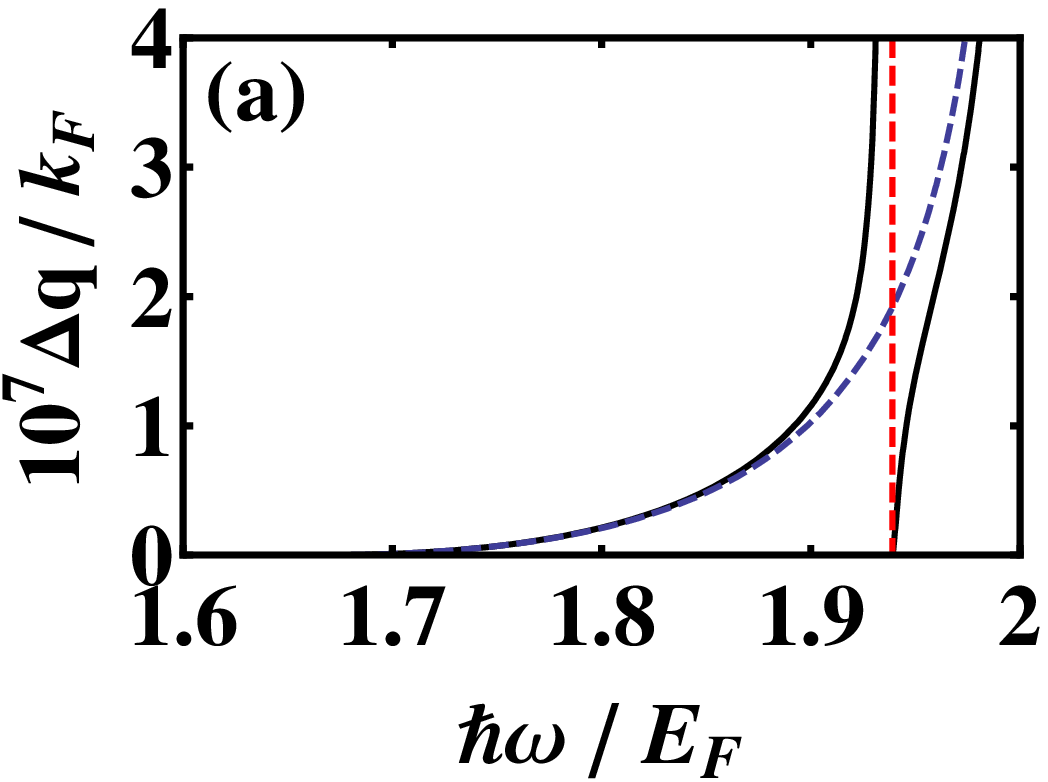}}
\mbox{\includegraphics[width=0.44\textwidth]{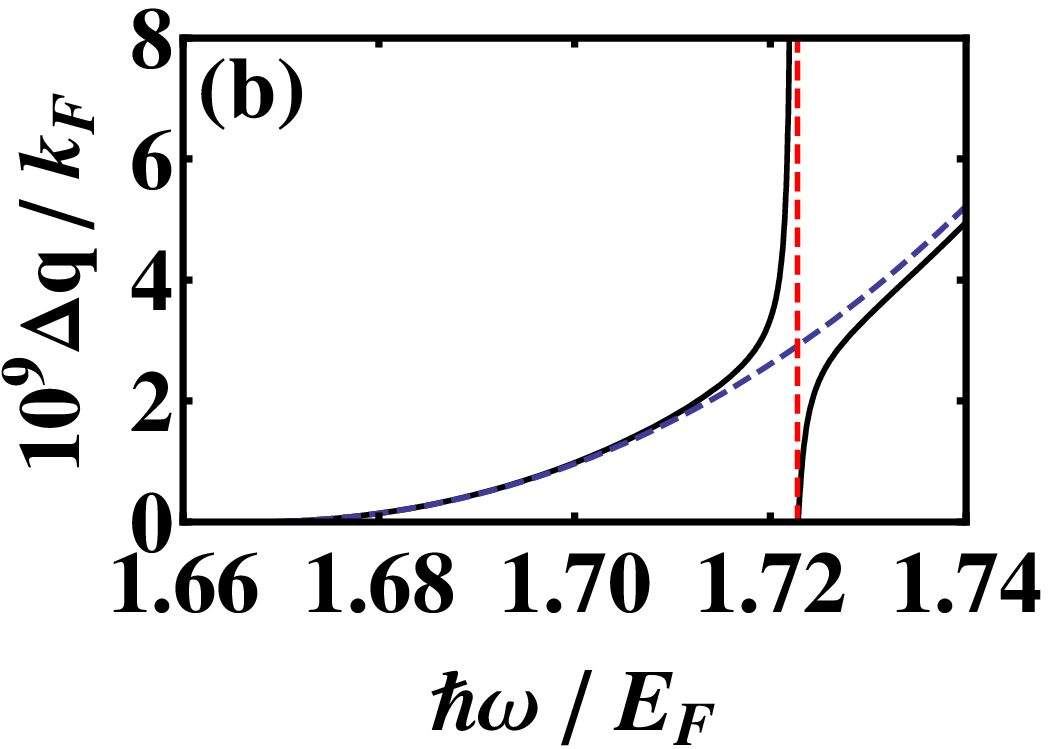}}
}
\caption{
Dispersion lines of hybrid TP-LO plasmon-phonon modes (solid lines) 
and of the uncoupled modes (dashed lines) for two values of doping:
(a) $n=7.5\times 10^{11}$~cm$^{-2}$, and 
(b) $n=9.5\times 10^{11}$~cm$^{-2}$. The plasmon-like dispersion 
is very close to the light line $q=\omega/c$; therefore, the ordinate shows 
$\Delta q=q-\omega/c$. 
}
\label{Figure3_c5}
\end{figure}

Next, it is straightforward to show from the Maxwell equations that an electric current 
oscillating in a two-dimensional plane will induce a vector potential
\begin{equation}
\langle A_{L {\bf q}}(\omega) \rangle = \langle { J}_L({\bf q},\omega) \rangle 
\frac{\sqrt{q^2-\omega^2/c^2}}{-2\omega^2\epsilon_0},
\label{Maxwell-long}
\end{equation} 
and
\begin{equation}
\langle A_{T {\bf q}}(\omega) \rangle = \langle { J}_T({\bf q},\omega) \rangle 
\frac{\mu_0}{2\sqrt{q^2-\omega^2/c^2}},
\label{Maxwell-trans}
\end{equation} 
where we have assumed that graphene is suspended in air and 
that there are no other sources present in space.  
This induced vector potential in turn acts on electrons in graphene through the interaction 
Hamiltonian $H_{e-em}$ which can result in plasmons - self-sustained collective oscillations of 
electrons. From Eqs. (\ref{current}) and (\ref{Maxwell-long}) we get the dispersion relation 
for longitudinal plasmons \cite{Wunsch2006,Hwang2007}
\begin{equation}
1-\frac{\sqrt{q^2-\omega^2/c^2}}{2\omega^2\epsilon_0}\chi_L({\bf q},\omega)=0.
\label{plasmon-long}
\end{equation} 
From Eqs. (\ref{current}) and (\ref{Maxwell-trans}) we get the dispersion relation 
for transverse plasmons \cite{Mikhailov2007}
\begin{equation}
1+\frac{\mu_0}{2\sqrt{q^2-\omega^2/c^2}}\chi_T({\bf q},\omega)=0.
\label{plasmon-trans}
\end{equation} 
Longitudinal plasmons are also referred to as transverse magnetic modes since they 
are accompanied by a longitudinal electric ($E$) and a transverse magnetic field 
($B$) in the plane of graphene. Likewise transverse plasmons or transverse electric 
modes are accompanied by a transverse electric and a longitudinal magnetic 
field \cite{Mikhailov2007}. 
Dispersion relation of LP (TP) modes is shown by the blue dashed line in 
Fig. \ref{Figure2_c5}. (Fig. \ref{Figure3_c5}, respectively). 
Finally we note that we are primarily interested in non-radiative modes ($q>\omega/c$) in 
which case fields are localized near the graphene plane ($z=0$) and decay 
exponentially: $E(z),B(z) \propto e^{-|z|\sqrt{q^2-\omega^2/c^2}}$.

In order to find the plasmon-phonon coupled excitations we consider the 
complete Hamiltonian $H=H_e+H_{e-em}+H_{e-ph}+H_{ph}$. We assume that the 
hybrid plasmon phonon mode oscillates at some frequency $\omega$ 
with wavevector $q$ (which are to be found). From the equation of motion for 
the phonon amplitudes $Q_{\mu {\bf q}}$ one finds \cite{PinesBook}
\begin{equation}
(\omega^2 - \omega_0^2)\langle Q_{T {\bf q}} \rangle = 
L^2 F \langle { J}_{L}({\bf q},\omega) \rangle ,
\label{ph-resp-trans}
\end{equation} 
and 
\begin{equation}
(\omega^2 - \omega_0^2 ) \langle Q_{L {\bf q}} \rangle = 
-L^2 F \langle { J}_{T}({\bf q},\omega) \rangle .
\label{ph-resp-long}
\end{equation} 
The electron phonon interaction (\ref{e-ph2}) is included as 
the effective vector potential (\ref{Aeff}) in Eq. (\ref{e-em}), 
which from Eq. (\ref{current}) immediately yields 
\begin{equation}
\langle { J}_L({\bf q},\omega) \rangle = \chi_L({\bf q},\omega) 
( - \langle A_{L {\bf q}} (\omega) \rangle + F \langle Q_{T {\bf q}} \rangle),
\label{e-resp-long}
\end{equation}
and
\begin{equation}
\langle { J}_T({\bf q},\omega) \rangle = \chi_T({\bf q},\omega) 
( - \langle A_{T {\bf q}} (\omega) \rangle - F \langle Q_{L {\bf q}} \rangle).
\label{e-resp-trans}
\end{equation}
From Eqs. (\ref{ph-resp-trans}) - (\ref{e-resp-trans}) it is clear that transverse 
(longitudinal) phonons couple only to longitudinal (transverse) plasmons. 
Apparently, this follows from the fact that LO (TO, respectively) phonons are 
equivalent to oscillations of an effective vector potential ${\bf A}_{\textrm{eff}}$
[see Eq. (\ref{Aeff})], and therefore an effective electric field, perpendicular 
(parallel, respectively) to ${\bf q}$. 

Finally using Eqs. (\ref{Maxwell-long}), (\ref{ph-resp-trans}), and 
(\ref{e-resp-long}) we get the dispersion relation for the 
LP-TO coupled mode
\begin{equation}
\omega^2 - \omega_0^2 = \frac{L^2F^2\chi_L({\bf q},\omega)}
{1-\frac{\sqrt{q^2-\omega^2/c^2}}{2\omega^2\epsilon_0}\chi_L({\bf q},\omega)},
\label{LP-TO}
\end{equation}
and from Eqs. (\ref{Maxwell-trans}), (\ref{ph-resp-long}), and (\ref{e-resp-trans}) 
dispersion relation for the TP-LO coupled mode
\begin{equation}
\omega^2 - \omega_0^2 = \frac{L^2F^2\chi_T({\bf q},\omega)}
{1+\frac{\mu_0}{2\sqrt{q^2-\omega^2/c^2}}\chi_T({\bf q},\omega)}.
\label{TP-LO}
\end{equation}
The plasmon dispersions relations (\ref{plasmon-long}) and (\ref{plasmon-trans}) appear 
as poles in the Eqs. (\ref{LP-TO}) and (\ref{TP-LO}) for the coupled modes, which means 
that the coupling is greatest at the resonance point where plasmon momentum and energy 
match that of the appropriate phonon mode. We denote this point (where the uncoupled plasmon 
and phonon dispersion cross) by $(q_c,\omega_0)$. One can quantify the strength of the 
coupling effect by calculating the frequency difference between the hybrid modes 
at the wavevector $q_c$ in units of the uncoupled frequency value: $\Delta\omega/\omega_0$. 
Finally by doping one can change plasmon dispersion which in turn changes $q_c$ and 
the strength of the plasmon-phonon coupling.

The dispersion lines for the hybrid LP-TO modes are shown in Fig. \ref{Figure2_c5} for two values of doping, (a) $n=5\times 10^{12}$~cm$^{-2}$, $E_F=0.261$~eV, $k_F=3.96\times 10^{8}$~m$^{-1}$, and 
(b) $n=5\times 10^{13}$~cm$^{-2}$, $E_F=0.825$~eV, $k_F=1.25\times 10^{9}$~m$^{-1}$. 
The strength of the coupling increases with increasing values of doping, and one has for the case
(a) $\Delta\omega/\omega_0=7.5 \% $, and (b) $\Delta\omega/\omega_0=15.5 \% $. 
To describe graphene sitting on a substrate (say SiC, which is a polar material), 
one only needs to include the dielectric function of the substrate into our calculation. 
In that case plasmons can also couple to surface phonon modes of the polar substrate \cite{Liu2010}. 
However, since these surface phonons have sufficiently smaller energies than 
optical phonons in graphene out results are qualitatively unchanged in that case. 
LP-TO hybrid modes could be measured by observing the change in the phonon 
dispersion with the Neutron Spectroscopy or Inelastic X-ray Scattering.
Alternatively, one could use grating coupler or Electron Energy Loss 
Spectroscopy to measure the shift in the plasmon energy.
Our results imply that plasmon-phonon coupling could serve to explore 
the electron-phonon interaction (the frequency shifts at resonance are much 
larger then the G peak shift recently measured by Raman Spectroscopy \cite{BOA}), and that by 
externally appling voltage one can influence the properties of lattice vibrations.

In spite of the fact that the formal derivation of hybrid TP-LO coupled 
modes is equivalent to the derivation of the LP-TO modes, their 
properties qualitatively differ. First, we note that the dispersion of 
transverse plasmons is extremely close to the light line, and we 
plot $\Delta q=q-\omega/c$ vs. frequency $\omega$ following Ref. \cite{Mikhailov2007}. 
For this reason, transverse plasmons are expected to have strong polariton character 
and they will be hard to distinguish from free photons (also, even a small plasmon 
linewidth will obscure the distinction). 
Moreover, they do not exist in graphene between two dielectrics with sufficiently 
different relative permittivity, where the light lines for the dielectrics are 
separated.
Next, transverse plasmons exist only in the frequency interval 
$2E_F > \hbar\omega > 1.667 E_F$ \cite{Mikhailov2007}, which means that 
the LO phonon energy must be in the same interval  
for the hybridization to occur. 
Figure \ref{Figure3_c5} shows the dispersion curves of the hybrid TP-LO modes 
for two values of doping, 
(a) $n=7.5\times 10^{11}$~cm$^{-2}$, $E_F=0.101$~eV, $k_F=1.53\times 10^{8}$~m$^{-1}$, and 
(b) $n=9.5\times 10^{11}$~cm$^{-2}$, $E_F=0.114$~eV, $k_F=1.73\times 10^{8}$~m$^{-1}$. 
We observe that the trend here is opposite to that of the LP-TO coupling, as the strength of the coupling 
decreases with increasing doping; specifically, one has for the case 
(a) $\Delta\omega/\omega_0=0.17 \% $, and (b) $\Delta\omega/\omega_0=0.02 \%$.
The maximal coupling occurs when $2E_F$ is just above $\hbar \omega_0$, 
and it is zero when $\hbar \omega_0=1.667E_F$. 
We emphasize that the strength of the coupling for TP-LO modes is in general 
much weaker than in LP-TO modes.

Before closing this chapter, we note another interesting result which is captured by our calculations.
Equations (\ref{LP-TO}) and (\ref{TP-LO}) for shifts in 
the energies of TO and LO modes at $q=0$ reduce to 
\begin{equation}
\omega^2 - \omega_0^2 = \frac{L^2F^2\chi_{L,T}(0,\omega)}
{1+\frac{i}{2\omega\epsilon_0 c}\chi_{L,T}(0,\omega)},
\label{q0}
\end{equation}
which is identical to the result of Ref. \cite{Ando_anomaly}, where the 
coupling of optical phonons to single-particle excitations was studied, 
appart from the imaginary term in the denominator which is zero in \cite{Ando_anomaly}. 
This small but qualitative difference is consequence of phonon coupling to the 
radiative electromagnetic modes, which increases the phonon linewidth. 
For example, for the doping values of $n=5\times 10^{12}$~cm$^{-2}$, 
$5\times 10^{13}$~cm$^{-2}$, and $5\times 10^{14}$~cm$^{-2}$, Eq. (\ref{q0}) yields 
$0.005\%$, $0.07\%$, and $0.7\%$, respectively, for the 
linewidths, while there is no linewidth from single-particle damping 
at these doping values. This effect is qualitatively unchanged for graphene 
sitting on a substrate and could be measured by Raman spectroscopy. 
Finally, we note an interesting solution of Eq. (\ref{TP-LO}) (valid for suspended graphene): 
when the hybrid TP-LO mode dispersion crosses the light line it has the same 
energy as the uncoupled phonon mode, i.e.,  $\omega=\omega_0$. In other words, LO phonon at 
a wavevector $q=\omega_0/c$ decouples from all (single particle and collective) electron 
excitations, while no such effect exists for the TO phonons.

In conclusion, we have predicted hybridization of plasmons and intrinsic optical 
phonons in graphene using self-consistent linear response theory. 
To the best of our knowledge, this is the first study of such resonance 
in an isolated 2D material. 
We found that graphene's unique electron-phonon interaction leads to unconventional 
mixing of plasmon and optical phonon polarizations: 
longitudinal plasmons couple exclusively to transverse optical 
phonons, whereas graphene's transverse plasmons couple to longitudinal optical phonons;
this contrasts plasmon-phonon coupling in all previously studied systems. 
The strength of the hybridization increases with doping 
in LP-TO coupled modes, while the trend is opposite for TP-LO modes. 
The LP-TO coupling is much stronger than TP-LO coupling, 
and it could be measured by current experiments, which would 
act as a magnifier for exploring the electron-phonon interaction in graphene. 
This coupling is an even more striking example of a breakdown of Born-Oppenheimer 
approximation in graphene than the recently measured stiffening of the Raman G peak \cite{BOA}. 
Moreover, plasmon-phonon interaction can serve to electronically control 
the frequencies of lattice vibrations in graphene, which could have 
interesting technological implications.

\chapter{Near field heat transfer} \label{chap:chap6}

\section{Near field heat transfer between two graphene sheets}

\begin{figure}[htb] 
\centerline{\mbox{\includegraphics[width=1.00\textwidth]{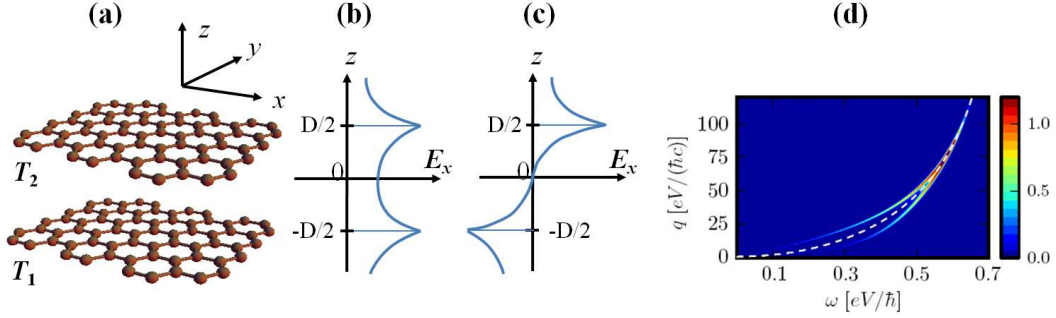}}}
\caption{
(a) Schematic diagram of the radiation transfer problem: a free standing sheet of graphene at temperature $T_1$ is radiating to another free standing graphene sheet at temperature $T_2$ and distance $D$ away. (b) Schematic diagram of the field profile for even mode. (c) Odd mode. (d) Contour plot of the transfer function $h_{nf}^p$ for the case of two graphene sheets at the same chemical potentials $\mu_{1,2}=0.5eV$ and same temperatures $T_{1,2}=300K$, separated by a distance of $D=10$ nm. Dashed line denotes the plasmon dispersion relation for a single isolated graphene sheet, while poles of transfer function $h_{nf}^p$ show dispersion relations of the coupled (even and odd) modes of the two sheets.} 
\label{Figure1_c6}
\end{figure}

In this chapter we analyze the near field heat transfer between two graphene sheets mediated by thermally excited plasmon modes and demonstrate that there is a large enhancement of heat transfer compared to the far field black body radiation. The system we analyze, shown in figure \ref{Figure1_c6}, consists of a suspended graphene sheet at temperature $T_1$ emitting to another
suspended graphene sheet held at room temperature $T_2=300K$, and a
distance $D$ away. 
 
In chapter \ref{chap:chap2} we calculated the expression for the radiative heat exchange between two graphene sheets. Total heat transfer $H=H_{ff}+H_{nf}$ can be conveniently separated into the contribution from the propagating waves in the far field
\begin{equation}
H_{ff}=\frac{1}{\pi^2} 
\int_{0}^{\infty} d\omega 
[ \Theta(\omega,T_1)-\Theta(\omega,T_2) ]
\int_0^{\omega/c} q dq
\sum_{\mu} 
h_{ff}^\mu (q,\omega) ,
\label{H_ff}
\end{equation}
and evanescent waves in the near field
\begin{equation}
H_{nf}=\frac{1}{\pi^2} 
\int_{0}^{\infty} d\omega 
[ \Theta(\omega,T_2)-\Theta(\omega,T_1) ]
\int_{\omega/c}^{\infty} q dq
\sum_{\mu} 
h_{nf}^\mu (q,\omega) .
\label{H_nf}
\end{equation}
Here $\Theta(\omega,T)=\hbar\omega/(e^{\beta\hbar\omega}-1)$ is the Boltzman factor, $\mu$ stands for $s$ or $p$ polarization and functions $h_{ff}^\mu$ and $h_{nf}^\mu$ are given by:
\begin{equation}
h^\mu_{ff}({\bf q},\omega)  \equiv 
\frac{(1-|r_1^\mu|^2-|t_1^\mu|^2)(1-|r_2^\mu|^2-|t_2^\mu|^2)}
{4|1 - r_1^\mu r_2^\mu e^{2i\gamma D}|^2} , \mbox{and}
\label{h_ff}
\end{equation}
\begin{equation}
h^\mu_{nf}({\bf q},\omega)  \equiv 
\frac{\Im r_1^\mu \Im r_2^\mu e^{-2|\gamma| D}}
{|1 - r_1^\mu r_2^\mu e^{-2|\gamma| D}|^2} .
\label{h_nf}
\end{equation}

Let us first note that graphene is a poor absorber in the far field, since it is only one atom thick. Indeed, it was experimentally demonstrated \cite{Nair2008} that graphene absorbs only $|a|^2\approx 2\%$ of the incident light (see equation (\ref{absorption_coefficient})). Since one can also write $|a|^2=1-|r|^2-|t|^2$ we can simply neglect the far field transfer (see equations (\ref{H_ff}) and (\ref{h_ff})), at least compared to the black body case which is characterized by $|a_{BB}|^2=100\%$. On the other hand, we will see that near field heat transfer can be significantly greater than the black body case, if graphene sheets are sufficiently close to allow the tunneling of surface modes (plasmons). 
 
To analyze the near field heat transfer between two graphene sheets let us write the $p$ polarization reflection coefficient (\ref{r_p}) for a single sheet as $r^p=(1-\epsilon)/\epsilon$ where
$\epsilon=1+\gamma\sigma/(2\epsilon_0\omega)$ is the dielectric function
of graphene \cite{Falkovsky2008}. We immediately see that poles of $r^p$ are located at the plasmon dispersion $\epsilon=0$ which was derived in chapter \ref{chap:chap3} but we write it here again for the sake of clearance:
\begin{equation}
  q = \epsilon_0 \frac{2i\omega}{\sigma(\omega,T)} .
\label{eq:disp-vac}
\end{equation}
We assumed here $\gamma=\sqrt{\omega^2/c^2-q^2}\approx iq$ since we have shown that plasmon dispersion is mainly located in the non-retarded regime $q>>\omega/c$. We have already pointed out that strong near field heat transfer requires graphene sheets to be very close, which in turn allows coupling of two plasmon modes (see figure \ref{Figure1_c6}). In the case when two graphene sheets have identical parameters ($r_1^p=r_2^p$) this coupling results in two new modes: even mode described by an equation $r^p=e^{qD}$, and an odd mode described by an equation $r^p=-e^{qD}$. Naturally, when sheets are sufficiently far apart ($D>>$) or the wave vector is sufficiently large ($q>>$) so that the coupling becomes irrelevant, these two modes become degenerate again, described by a pole in $r^p$. Further on, note that these two equations can be joined in a single one: $1-(r^p)^2e^{-2qD}=0$, which is precisely a denominator in equation (\ref{h_nf}) which determines the poles of the function $h_{nf}^p$. In other words, these two coupled surface modes, strongly enhance and dominate the near field heat transfer. Note however from equation (\ref{h_nf}) that $h_{nf}^p\propto e^{-2qD}$ so that graphene sheets have to be very close to have a significant near field heat transfer. In other words plasmon surface modes can act as an excellent heat conductors, only the graphene sheets have to be very close to allow the coupling of exponentially decaying ($E\propto e^{-qD}$) plasmon field. 

Finally note that $h_{nf}^p\propto \Im r_1^p \Im r_2^p$, while $\Im r^p$ has a pole at the bare plasmon dispersion so there will be a competition between this factor and a pole at a dispersion of a coupled mode. Therefore the function $h_{nf}^p$ will increase with increasing wave vector since the coupling between the modes will decrease and even/odd mode dispersion (and a corresponding pole) will join that of a bare plasmon (and a corresponding pole). However, when these dispersions meet ($q\approx 1/D$) the function $h_{nf}^p$ will start to decrease with the wave vector due to exponentially decaying factor $h_{nf}^p\propto e^{-2qD}$. At last note that function $h_{nf}^p$ is multiplied with a Boltzman factor $\Theta(\omega,T)$ which shifts everything to lower frequencies so there are several competing effects in action which will be hard to disentangle in the end when everything gets integrated over all $q,\omega$ values.

The same analysis applies to the $s$ polarization however it is easy to see that it will have a minor contribution to the total near field heat transfer. The reason for this is a large difference in the character of the plasmon dispersion relations (compare figure \ref{Figure2_c5} and figure \ref{Figure3_c5}). On one hand longitudinal plasmons, described by a pole in $r^p$, are located in the non-retarded regime ($q>>\omega/c$) with a large density of states, while transverse plasmons, describe by a pole in $r^s$, are located in the strongly retarded regime ($q\approx \omega/c$) with a tiny density of states. Since each $q$ value can be thought of as a separate heat channel, and if graphene sheets are close enough so that all relevant $q$ modes are active, the $p$ polarization will have many more heat channels and dominate over the $s$ polarization.

To model graphene we shall use $q$-independent conductivity which simplifies the mathematical calculations and gives a good order of magnitude on the heat transfer (see discussion below). In chapter \ref{chap:chap2} we showed that the total conductivity $\sigma(\omega)=\sigma_D(\omega)+\sigma_I(\omega)$, can be separated into Drude (intraband) and interband part, expressed
respectively as (see also \cite{Falkovsky2008}):
\begin{align} 
  \sigma_{D}(\omega) &= \frac{i}{\omega+i/\tau}\frac{e^2 2k_{b}T}{\pi\hbar^2}
    \textrm{ln}\left[2\textrm{cosh}\frac{\mu}{2k_bT}\right]
  \label{eq:cond}
  \\
  \sigma_{I}(\omega) &= \frac{e^2}{4\hbar}
    \left[G\left(\frac{\hbar\omega}{2}\right) + i
    \frac{4\hbar\omega}{\pi}\int_0^{\infty}
    \frac{G(\epsilon)-G(\hbar\omega/2)}
      {(\hbar\omega)^2-4\epsilon^2}d\epsilon
    \right] \nonumber .
\end{align} 
where $G(\epsilon) =\textrm{sinh}(\epsilon/k_bT)/(\textrm{cosh}
(\mu/k_bT)+\textrm{cosh}(\epsilon/k_bT))$, and $\mu$ is the chemical
potential. Various electron scattering processes are taken into account
through the relaxation time $\tau$. From DC mobility measurements in
graphene, one obtains an order-of-magnitude value of $\tau\approx
10^{-13}s$. Now, due to fluctuation-dissipation theorem $h_{nf}^p\propto \Im r^p$ and $\Im r^p\propto\Re\sigma$ (see equation (\ref{p_right_ll})), so we have to take particular attention to the origin of dissipation ($\Re\sigma$) in our system. At zero temperature the situation is very simple since Drude term (intraband contribution) and relaxation time $\tau$ determines the losses for low frequencies, while interband contribution is dominant for frequencies above the interband threshold ($\hbar\omega=2\mu$). However, at finite temperature, interband processes can play a leading role even below the absorption threshold $\omega\approx 2\mu$, particularly for small chemical potential where thermal broadening of interband threshold (on the order of few $k_bT$) becomes more significant. While the use of $q$-independent expression
for graphene conductivity (\ref{eq:cond}) for intraband processes
is a good approximation, one must take care when
applying (\ref{eq:cond}) to interband transitions. Here, the contribution
from the finite wave-vector becomes important since it broadens the
interband threshold from $2\mu$ to $2\mu-\hbar qv_F$. On the other hand,
this is similar to finite temperature effects which also broaden the
interband threshold, so we do not expect a qualitatively different result
with $q$-dependent conductivity.

\begin{figure}[htb] 
\centerline{\includegraphics[width=0.50\textwidth]{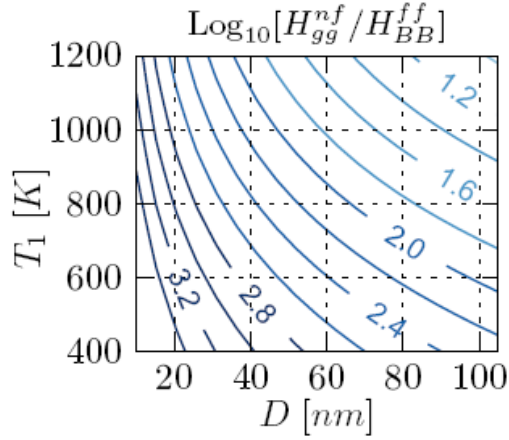}}
\caption{
Contour plot of the near field heat transfer between two graphene sheets $H_{gg}^{nf}$ normalized to the far field heat transfer between two black bodies $H_{BB}^{ff}$ of the same temperatures, in the log scale. Here $T_2=300K$, $\mu_{1,2}=0.1eV$ and $\tau_{1,2}=10^{-13}s$.}
\label{Figure2_c6}
\end{figure}

To quantify the heat exchange in the near field we plot in figure \ref{Figure2_c6} the total transfer $H_{nf}$ (\ref{H_nf})
normalized to the transfer between two black bodies in the far field.
Since the exponentially decaying Boltzman factor shifts all the contributions to the lower frequencies we will focus on the small values of chemical potential. For $\mu_{1,2}=0.1eV$, we observe orders-of-magnitude increase in
heat exchange particularly at small separations ($\times 1000$ for
$D=20nm,T_1=800K$), but also at separations as large as $0.1 \mu m$. In
general, dependence of transfer on separation $D$ is non-uniform and does not seem to yield a simple functional dependence on the emitter and absorber temperatures (as is the case for two black bodies). This efficient
heat exchange between two graphene sheets in the near field, together
with recently reported advances in hot carrier extraction from
graphene \cite{Gabor2011}, may offer a potential for a novel, hybrid
thermophotovoltaic/thermoelectric solid-state heat-to-electricity
conversion device.


\section{Near field thermo-photo-voltaics using graphene as a thermal emitter}

Here we show that graphene can be used as a thermal emitter in the near field thermo-photo-voltaic (TPV) system resulting in high efficiencies and power densities. The near field heat transfer is mediated by thermally excited plasmon modes in graphene similarly to the situation in the last section. 

The system we analyze consists of a hot graphene emitter at a temperature $T_1$ and a photo-voltaic (PV) cell held at room temperature $T_2=300$K and distance $D$ away from graphene. It is interesting to note that the expression for the near field heat transfer between graphene and a PV cell is given by the same expression (\ref{H_nf}) for the graphene to graphene heat transfer
\begin{equation}
H_{nf}=\frac{1}{\pi^2} 
\int_{0}^{\infty} d\omega 
\left[ 
\frac{\hbar\omega}{e^{\beta_1\hbar\omega}-1} -
\frac{\hbar\omega}{e^{\beta_2\hbar\omega}-1}
\right]
\int_{\omega/c}^{\infty} q dq
h_{nf}^p (q,\omega) , 
\label{H_nf2}
\end{equation}
where
\begin{equation}
h^p_{nf}({\bf q},\omega)  \equiv 
\frac{\Im r_1^p \Im r_2^p e^{-2q D}}
{|1 - r_1^p r_2^p e^{-2q D}|^2} .
\label{h_nf2}
\end{equation}
Here we have neglected the contribution from $s$ polarization, and assumed $\gamma=\sqrt{\omega^2/c^2-q^2}\approx iq$ since the near field heat transfer is mediated by graphene plasmon modes in the non-retarded regime ($q>>\omega/c$). To model graphene we use $q$-independent conductivity $\sigma(\omega)$ from equation (\ref{eq:cond}) as before, and the reflection coefficient is given by equation (\ref{r_p}) which we write again for the sake of clearance:
\begin{equation}
r_1^p(q,\omega)=\frac{ \frac{i q \sigma(\omega)}{2\epsilon_0\omega} }
         { 1+ \frac{i q \sigma(\omega)}{2\epsilon_0\omega}} .
\label{r_graphene}
\end{equation}
Above hot graphene emitter we now have PV cell which we model as a simple direct band-gap semiconductor with parameters:
\begin{align}
    \epsilon_2(\omega) = \left(n+i\frac{\alpha}{2k_0}\right)^2
    \quad \mathrm{where} \quad \alpha(\omega) = \left\{ \begin{array}{ll} 0
    &, \omega < \omega_g \\ \alpha_0 \sqrt{\frac{\omega-\omega_g}{\omega_g}}
    &, \omega > \omega_g \end{array} \right . 
\label{eq:eps2}
\end{align}
Here $n$ is the refractive index, $k_0=2\pi/\lambda=c/\omega$ is the photon wavelength in vacuum, and $\omega_g$ is the bandgap frequency.
Specifically we will discuss the case of indium antimonide (InSb)     
with parameters $\omega_g=0.17eV$ and $\alpha_0 \approx 0.7 \times 10^4 cm^{-1}$ 
(at room temperature \cite{Gobeli1960}). Finally the reflection coefficient of the PV cell, in the non-retarded regime ($q>>\omega/c$), is simply given by \cite{Jackson}
\begin{equation}
r_2^p(\omega)=\frac{\epsilon_2(\omega)-1}{\epsilon_2(\omega)+1}.
\label{r_PV}
\end{equation}
When the PV cell is biased at a voltage \footnote{In
general, the optimal voltage $V_o$ depends on other parameters in the
system. We avoid the full optimization procedure, and, motivated by the
observed dependence of efficiency on $V_o$, choose a voltage slightly
below the limit $V_o^{max}=\omega_g(1-T_2/T_1)$.} $V_o$, we can express the total radiative power exchange as
\cite{Celanovic2004}
\begin{equation}
P_{rad}=\frac{1}{\pi^2} 
\int_{0}^{\infty} d\omega 
\left[ 
\frac{\hbar\omega}{e^{\beta_1\hbar\omega}-1} -
\frac{\hbar\omega}{e^{\beta_2(\hbar\omega-V_o)}-1}
\right]
\int_{\omega/c}^{\infty} q dq
h_{nf}^p (q,\omega) .
\label{P_rad}
\end{equation}
On the other hand we can also write the total photon flux into the PV cell as
\begin{equation}
j_{ph}=\frac{1}{\pi^2} 
\int_{0}^{\infty} d\omega 
\left[ 
\frac{1}{e^{\beta_1\hbar\omega}-1} -
\frac{1}{e^{\beta_2(\hbar\omega-V_o)}-1}
\right]
\int_{\omega/c}^{\infty} q dq
h_{nf}^p (q,\omega).
\label{j_ph}
\end{equation}
In the Shockley-Quiesser limit \cite{Shockley} of ideal PV cell, the only recombination of the charge carriers happens through the radiative processes so the electron current is simply: $j_e=e j_{ph}$. Then the electrical power generated in the PV cell is $P_{PV}=j_e V_o$, and efficiency of  device is
\begin{equation}
\eta_{TPV} = 
\frac{P_{PV}}{P_{rad}} = 
\frac{e j_{ph}V_o}{P_{rad}}.
\label{efficiency}
\end{equation}
Let us now choose the graphene's chemical potential to be $\mu=0.25$ eV, and that the PV cell is held at room temperature $T_2=300$K and distance $D=10$ nm away from the graphene sheet. Then for the case of graphene's temperature $T_1=600$ K and biased voltage $V_o=0.08$ V, the output power density of our TPV device is $P_{PV}/A=6$ W/cm$^2$ with an efficiency of $\eta=35\%$. We note that these are remarkably high power densities considering that our thermal emitter is only one atom thick. To get a better sense of the scales involved we can compare the far field radiative power exchange $P_{rad}^{BB}$ between two black bodies held at temperatures $T_1$ and $T_2$ but involving only the photons of energies above the given PV band gap, and the near field radiative power exchange $P_{rad}^{gPV}$ between graphene and a PV cell held at these temperatures. For the temperatures $T_1=600$ K and $T_2=300$ K one obtains $P_{rad}^{gPV}/P_{rad}^{BB}=62$ times increase over the black body case. 

Further on, note that the near field heat transfer is particularly convenient since the energy is transfered by the evanescent modes and the photons with energy below the band gap, that are not absorbed by the PV cell, simply return to the graphene emitter as heat, unlike the far field case where they are lost in the form of propagating waves. This results in the high efficiencies $\eta=35\%$, however note that these numbers are still below the Carnot limit $\eta=50\%$  for temperatures $T_1=600$ K and $T_2=300$ K. The reason for this is the broad band plasmon spectrum contributing to the heat transfer with the high energy photons ($\omega>\omega_g$) wasting the energy difference ($\Delta E=\hbar\omega-\hbar\omega_g$) on the thermalization losses heating up the system. 

To achieve even higher efficiencies one would need to tailor the emitter properties so that it selectively radiates only in the small interval around the band gap of the PV cell. One way to do this would be to use surface plasmons at metal-dielectric surface, since they have very large density of states around the surface plasmon resonance; see figure \ref{Figure1_c3} b, and compare it to broad band spectrum of graphene plasmon mode from figure \ref{Figure2_c3} d. However, the problem with metals is that surface plasmon resonance usually falls in the visible/ultraviolet regime which is impossible to excite thermally. Alternatively one could use highly doped semiconductors like Indium-Tin-Oxide \cite{ITO} which has a resonance in the infra-red \cite{ITO_plasmonic}, however due to high doping level there is a lot of electron-impurity scattering and high losses result in reduced efficiencies. In that regards graphene TPV system shows large promise for a new temperature range ($600-1200$K) solid state energy conversion, where conventional thermoelectrics can not operate due to high temperatures and far field TPV schemes suffer from low efficiency and power density.

\chapter{Summary} \label{chap:summary}

We have explored light-matter interaction in graphene in the context of plasmonics and other technological applications but also used graphene as a platform to explore many body physics phenomena like the interaction between plasmons, phonons and other elementary excitations. Plasmons and plasmon-phonon interaction were analyzed within self-consistent linear response approximation. We demonstrated that electron-phonon interaction leads to large plasmon damping when plasmon energy exceeds that of the optical phonon but also a peculiar mixing of plasmon and optical phonon polarizations. Plasmon-phonon coupling is strongest when these two excitations have similar energy and momentum. We also analyzed properties of transverse electric plasmons in bilayer graphene. Finally we have showed that thermally excited plasmons strongly mediate and enhance the near field radiation transfer between two closely separated graphene sheets. We also demonstrated that graphene can be used as a thermal emitter in the near field thermophotovoltaics leading to large efficiencies and power densities. Near field heat transfer was analyzed withing the framework of fluctuational electrodynamics.

In Chapter \ref{chap:chap2} we presented analytical methods that were used throughout the text. We have derived electron band structure and electron-phonon interaction using the tight binding approximation. After that we derived the linear response functions (density-density and current-current) and used the fluctuation dissipation theorem to calculate the current-current correlation function induced by the thermal fluctuations in the system. Finally we employed these results to calculate radiative heat transfer between two graphene sheets. In Chapter \ref{chap:chap3} we have investigated plasmons in doped graphene and demonstrated that they simultaneously enable low-losses and significant wave localization for frequencies of the light smaller than the optical phonon 
frequency $\hbar\omega_{Oph}\approx 0.2$~eV. Interband losses via emission of electron-hole pairs (1$^{\textrm{st}}$ order process) were shown to be 
blocked by sufficiently increasing the doping level, which pushes the 
interband threshold frequency $\omega_{inter}$ toward higher values 
(already experimentally achieved doping levels can push it even up to near infrared frequencies). The plasmon decay channel via emission of an optical phonon together with an electron-hole pair (2$^{\textrm{nd}}$ order process) is inactive for $\omega<\omega_{Oph}$ (due to energy conservation), however, for frequencies larger than $\omega_{Oph}$ this decay channel is non-negligible. This is particularly important for large enough doping values when the interband threshold $\omega_{inter}$ is above $\omega_{Oph}$: in the interval $\omega_{Oph}<\omega<\omega_{inter}$ the 1$^{\textrm{st}}$ order process is suppressed, but the phonon decay channel is open. 

In Chapter \ref{chap:chap4} we showed that graphene can also support unusual transverse electric plasmons and we predicted the existence of TE plasmons in bilayer graphene. We found that their plasmonic properties are much more pronounced in bilayer than in monolayer graphene, in a sense that the
wavelength of TE plasmons in bilayer can be smaller than in monolayer graphene 
at the same frequency.  

In Chapter \ref{chap:chap5} we analyzed the coupling of plasmons with intrinsic optical phonons in graphene by using the self-consistent linear response formalism. We found that longitudinal plasmons (LP) 
couple only to transverse optical (TO) phonons, while transverse 
plasmons (TP) couple only to longitudinal optical (LO) phonons. 
The LP-TO coupling is stronger for larger concentration of carriers, 
in contrast to the TP-LO coupling (which is fairly weak). 
The former could be measured via current experimental techniques. 
Thus, plasmon-phonon resonance could serve as a magnifier for exploring 
the electron-phonon interaction in graphene.

In Chapter \ref{chap:chap6} we analyzed the near field heat transfer between two graphene sheets mediated by thermally excited plasmon modes, and we demonstrated that there is a large enhancement of heat transfer compared to the far field black body radiation. Finally we showed that graphene can be used as a thermal emitter in the thermo-photo-voltaic system resulting in high device efficiencies and power densities.

\appendix

\chapter{Plasmon-phonon coupling in the context of Feynman diagrams} \label{app:Bubble_pl_ph}

\begin{figure}
\centerline{
\mbox{\includegraphics[width=0.8\textwidth]{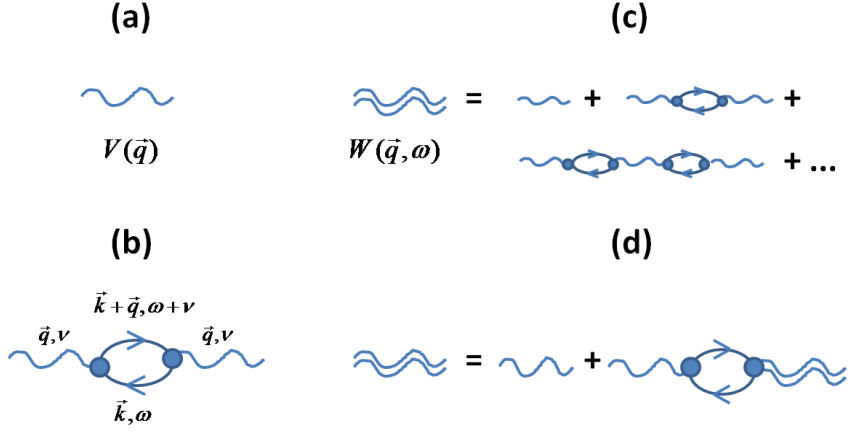}}
}
\caption{
(a) Feynman diagram for bare Coulomb interaction $V({\bf q})$. 
(b) Polarizability $\Pi({\bf q},\omega)$. (c) and (d) Screened Coulomb interaction $W({\bf q},\omega)$ in the Random Phase Approximation.
}
\label{Figure_App_pl_ph1}
\end{figure}

In this appendix we give alternative derivation of plasmon-phonon coupling in the context of Feynman diagrams. In that respect let us start by writing Coulomb potential 
\begin{equation}
V({\bf r})=\frac{e^2}{4\pi\epsilon_0 r} ,
\label{Coulomb-real}
\end{equation}
and its Fourier transform in two dimensions
\begin{equation}
V({\bf q})=\frac{e^2}{2\epsilon_0 q} .
\label{Coulomb-fourier}
\end{equation}
Bare Coulomb interaction $V({\bf q})$ can polarize electron gas by creating electron-hole pair which in turn screens the bare interaction resulting with an effective interaction $W({\bf q},\omega)$. This process can happen several times in a row (see figure \ref{Figure_App_pl_ph1}) so we can write self-consistent equation for the effective interaction  \cite{Sunjic}
\begin{equation}
\left[ -i W({\bf q},\omega) \right] =
\left[ -i V({\bf q}) \right] + 
\left[ -i V({\bf q}) \right] \left[ -i \Pi({\bf q},\omega) \right] 
\left[ -i W({\bf q},\omega) \right]  .
\label{zasjenjenje}
\end{equation}
If we now use the Random Phase Approximation which neglects higher order scattering of the created electron-hole pair, then the polarizability $\Pi({\bf q},\omega)$, depicted with a Feynman diagram in figure \ref{Figure_App_pl_ph1} (b), can be written as
\begin{align}
-i\Pi({\bf q},\omega) = 
-4 & \int \frac{d{\bf k}d\nu}{(2\pi)^3} \sum_{n,n'} 
iG_0(n',{\bf k}+{\bf q},\omega+\nu) iG_0(n,{\bf k},\nu) 
\nonumber \\
& \times
\langle n' {\bf k}+{\bf q} | e^{i{\bf q}{\bf r}} | n {\bf k} \rangle
\langle n {\bf k} | e^{-i{\bf q}{\bf r}} | n' {\bf k}+{\bf q} \rangle .
\label{polarizabilnost}
\end{align}
Here $| n {\bf k} \rangle$, i.e. the wave function 
$\psi_{n {\bf k}}({\bf r})=\langle {\bf r} | n {\bf k} \rangle$,
denotes single particle free Dirac electron states (see relation (\ref{wavef-Dirac})) 
and the Green function $G_0(n,{\bf k},\nu)$ is given by expression \cite{Sunjic} 
\begin{equation}
G_0(n,{\bf k},\nu)=
\frac{1-f_{n {\bf k}}}{\hbar\nu-E_{n {\bf k}}+i\eta} +
\frac{f_{n {\bf k}}}{\hbar\nu-E_{n {\bf k}}-i\eta} ,
\label{Green}
\end{equation}
where $f_{n {\bf k}}$ denotes the Fermi-Dirac distribution. After integration over energy $\nu$ we obtain
\begin{align}
\Pi({\bf q},\omega) =
4 & \int \frac{d{\bf k}}{(2\pi)^2} \sum_{n,n'}
\frac{f_{n {\bf k}}-f_{n' {\bf k}+{\bf q}}}
{\hbar\omega+E_{n {\bf k}}-E_{n' {\bf k}+{\bf q}}}
\nonumber \\
& \times
\langle n' {\bf k}+{\bf q} | e^{i{\bf q}{\bf r}} | n {\bf k} \rangle
\langle n {\bf k} | e^{-i{\bf q}{\bf r}} | n' {\bf k}+{\bf q} \rangle ,
\label{polarizabilnost1}
\end{align}
and then by using the exact wave function for Dirac electrons 
$\psi_{n {\bf k}}({\bf r})=\langle {\bf r} | n {\bf k} \rangle$, 
given in equation (\ref{wavef-Dirac}), we obtain the polarizability
\begin{align}
\Pi({\bf q},\omega) =
4 & \int \frac{d{\bf k}}{(2\pi)^2} \sum_{n,n'}
\frac{f_{n {\bf k}}-f_{n' {\bf k}+{\bf q}}}
{\hbar\omega+E_{n {\bf k}}-E_{n' {\bf k}+{\bf q}}}
\nonumber \\
& \times
\frac{1}{2}(1+nn'\cos[\varphi({\bf k}+{\bf q})-\varphi({\bf k})]) .
\label{polarizabilnost2}
\end{align}
Further on, by using relation (\ref{zasjenjenje}) we can write the screened interaction
\begin{equation}
W({\bf q},\omega)=\frac{V({\bf q})}{1-\Pi({\bf q},\omega)V({\bf q})} ,
\label{zasjenjenje2}
\end{equation}
where we recognize the expression for dielectric function of electron gas
\begin{equation}
\epsilon({\bf q},\omega)=1-\Pi({\bf q},\omega)V({\bf q})=
1-\frac{e^2}{2\epsilon_0 q}\Pi({\bf q},\omega) .
\label{epsilon}
\end{equation}
Finally, we note that plasmons are simply defined as zeros of the dielectric function: $\epsilon({\bf q},\omega)=0$.

\begin{figure}
\centerline{
\mbox{\includegraphics[width=0.8\textwidth]{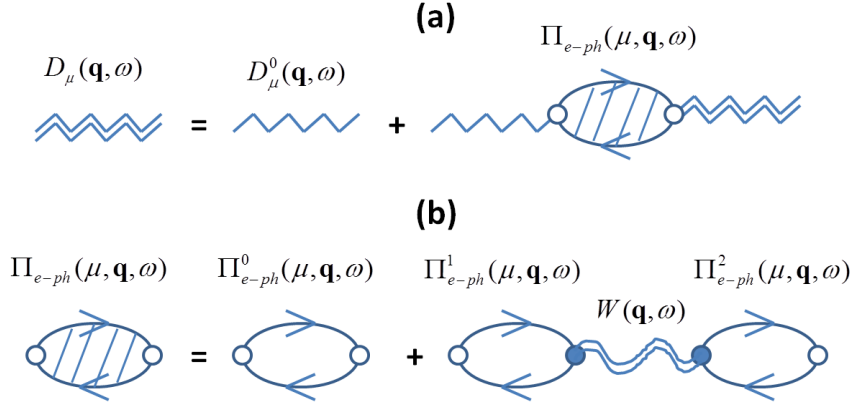}}
}
\caption{
(a) Feynman diagrams for phonon Green function renormalization. 
(b) Feynman diagrams for polarizability function. Note that the electron-electron interaction vertex is different from the electron-phonon vertex.
}
\label{Figure_App_pl_ph2}
\end{figure}

Let us find now the phonon Green function for free phonons at zero temperature. Since longitudinal and transverse optical phonons are degenerate at energy $\hbar\omega_0=0.196$ eV, then the Green function for both branches is given by
\begin{equation}
D_{\mu}^0({\bf q},\omega)=
\frac{2\hbar\omega_0}{\hbar\omega(\hbar\omega+i\eta)-(\hbar\omega_0)^2} .
\label{GreenFon}
\end{equation}
Now, the electron-phonon interaction was given in equation (\ref{Hamilton-e-ph-density})
\begin{equation}
H_{e-ph}=
L^2 \sum_{{\bf q},\mu} g M_{{\bf q}\mu} \rho_{\bf q}^+ Q_{{\bf q}\mu} .
\label{Hamilton-e-ph-final2}
\end{equation}
The phonon motion can in turn polarize the electron gas which is described by a self-consistent equation for the phonon Green function renormalization (see figure \ref{Figure_App_pl_ph2})
\begin{equation}
\left[ -i D_{\mu}({\bf q},\omega) \right] =
\left[ -i D_{\mu}^0({\bf q},\omega) \right] + 
\left[ -i D_{\mu}^0({\bf q},\omega) \right] \left[ -i \Pi_{e-ph}(\mu,{\bf q},\omega) \right] 
\left[ -i D_{\mu}({\bf q},\omega) \right]  ,
\label{FonRenormalization}
\end{equation}
so the renormalized Green function is given by an expression:
\begin{align}
D_{\mu}({\bf q},\omega) & =
\frac{D_{\mu}^0({\bf q},\omega)}
{1-D_{\mu}^0({\bf q},\omega)\Pi_{e-ph}(\mu,{\bf q},\omega)}
\nonumber \\
& =
\frac{2\hbar\omega_0}
{(\hbar\omega)^2-(\hbar\omega_0)^2-2\hbar\omega_0\Pi_{e-ph}(\mu,{\bf q},\omega)}.
\label{GreenRen}
\end{align}
Finally, the renormalized phonon frequency is defined by a pole of the Green function
\begin{equation}
\omega^2-\omega_0^2=2\frac{\omega_0}{\hbar}\Pi_{e-ph}(\mu,{\bf q},\omega) .
\label{FonFreqRen}
\end{equation}
Up to the lowest order, the interaction will create virtual electron-hole pair (see figure \ref{Figure_App_pl_ph2}) which can be described with a polarization function
\begin{align}
-i\Pi_{e-ph}^0(\mu,{\bf q},\omega)=
-4 g^2 & \int \frac{d{\bf k}d\nu}{(2\pi)^3} \sum_{n,n'}
iG_0(n',{\bf k}+{\bf q},\omega+\nu) iG_0(n,{\bf k},\nu)
\nonumber \\
& \times
\langle n' {\bf k}+{\bf q} | M_{{\bf q}\mu} e^{i{\bf q}{\bf r}} | n {\bf k} \rangle
\langle n {\bf k} | M_{{\bf q}\mu}^* e^{-i{\bf q}{\bf r}} | n' {\bf k}+{\bf q} \rangle .
\label{polarizabilnost-eph}
\end{align}
We note here that $M_{{\bf q}\mu}$ given by equation (\ref{M}) is two by two  matrix so that polarizability $\Pi_{e-ph}^0(\mu,{\bf q},\omega)$ isn't simply proportional to the function $\Pi({\bf q},\omega)$ which was obtained in relation to the screened Coulomb interaction (see relation (\ref{polarizabilnost})). In the context of Feynman diagrams we can say that diagram vertices are different for the case of Coulomb (electron-electron) interaction from the case of electron-phonon interaction. Further on, we obtain
\begin{align}
\Pi_{e-ph}^0(\mu,{\bf q},\omega)=
4 g^2 &\int \frac{d{\bf k}}{(2\pi)^2} \sum_{n,n'}
\frac{f_{n {\bf k}}-f_{n' {\bf k}+{\bf q}}}
{\hbar\omega+E_{n {\bf k}}-E_{n' {\bf k}+{\bf q}}}
\nonumber \\
& \times
\langle n' {\bf k}+{\bf q} | M_{{\bf q}\mu} e^{i{\bf q}{\bf r}} | n {\bf k} \rangle
\langle n {\bf k} | M_{{\bf q}\mu}^* e^{-i{\bf q}{\bf r}} | n' {\bf k}+{\bf q} \rangle .
\label{polarizabilnost-eph2}
\end{align}
Let us now take the exact form of the wave function $\psi_{n {\bf k}}$ and the matrix element $M_{{\bf q}\mu}$ according to relations (\ref{wavef-Dirac}) and 
(\ref{M}). We can see that the polarizability depends on the phonon polarization and we obtain
\begin{align}
\Pi_{e-ph}^0(L,{\bf q},\omega)=
4 g^2 & \int \frac{d{\bf k}}{(2\pi)^2} \sum_{n,n'}
\frac{f_{n {\bf k}}-f_{n' {\bf k}+{\bf q}}}
{\hbar\omega+E_{n {\bf k}}-E_{n' {\bf k}+{\bf q}}}
\nonumber \\
& \times
\frac{1}{2}
(1-nn'\cos[2\varphi({\bf q})-\varphi({\bf k})-\varphi({\bf k}+{\bf q})]) 
, \mbox{and}
\label{polarizabilnost-eph-L}
\end{align}
\begin{align}
\Pi_{e-ph}^0(T,{\bf q},\omega)=
4 g^2 & \int \frac{d{\bf k}}{(2\pi)^2} \sum_{n,n'}
\frac{f_{n {\bf k}}-f_{n' {\bf k}+{\bf q}}}
{\hbar\omega+E_{n {\bf k}}-E_{n' {\bf k}+{\bf q}}}
\nonumber \\
& \times
\frac{1}{2}
(1+nn'\cos[2\varphi({\bf q})-\varphi({\bf k})-\varphi({\bf k}+{\bf q})]) .
\label{polarizabilnost-eph-T}
\end{align}

If we now imagine that phonon energy and momentum matches plasmon energy and momentum, then the electron-phonon interaction will be amplified through the collective electron response. In that case it won't be sufficient to calculate only the polarization of single electron hole pair and we will have to take into consideration contribution from the infinite sequence of bubble diagrams (which are in fact necessary to describe plasmon excitation). The easiest way to do this is to take the electron-phonon interaction which polarizes a single electron hole pair and include the possibility that Coulomb interaction can in turn create another electron hole pair. The infinite sequence of diagrams can be included if we work from the start with screened Coulomb interaction instead of bare interaction and one should take special account of the nature of diagram vertices considering if the electron hole pair was created by Coulomb or electron-phonon interaction (see figure \ref{Figure_App_pl_ph2}). In that way we obtain the screened electron-phonon polarizability in the Random Phase Approximation
\begin{equation}
\Pi_{e-ph}(\mu,{\bf q},\omega)=\Pi_{e-ph}^0(\mu,{\bf q},\omega)+
\Pi_{e-ph}^1(\mu,{\bf q},\omega)W({\bf q},\omega)\Pi_{e-ph}^2(\mu,{\bf q},\omega) .
\label{zasj-polariz}
\end{equation}
Here $W({\bf q},\omega)=\frac{V({\bf q})}{1-\Pi({\bf q},\omega)V({\bf q})}$
so we immediately see that if phonon dispersion crosses the plasmon dispersion then the electron-phonon interaction will be amplified due the collective electron response where we have  
$\epsilon({\bf q},\omega)=1-\Pi({\bf q},\omega)V({\bf q})=0$.
That part is in fact responsible for the plasmon phonon coupling. Finally the polarizability describing the bubble with different vertices is given by
\begin{align}
\Pi_{e-ph}^1(\mu,{\bf q},\omega)={\Pi_{e-ph}^2}(\mu,{\bf q},\omega)^*=
4 g & \int \frac{d{\bf k}}{(2\pi)^2} \sum_{n,n'}
\frac{f_{n {\bf k}}-f_{n' {\bf k}+{\bf q}}}
{\hbar\omega+E_{n {\bf k}}-E_{n' {\bf k}+{\bf q}}}
\nonumber \\
& \times
\langle n' {\bf k}+{\bf q} | M_{{\bf q}\mu} e^{i{\bf q}{\bf r}} | n {\bf k} \rangle
\langle n {\bf k} | e^{-i{\bf q}{\bf r}} | n' {\bf k}+{\bf q} \rangle .
\label{polarizabilnost-eph3}
\end{align}
If we include here the exact wave function $\psi_{n {\bf k}}$ and matrix elements $M_{{\bf q}\mu}$, we obtain different expressions depending on the phonon polarization:
\begin{align}
\Pi_{e-ph}^1(L,{\bf q},\omega)=
4 g & \int \frac{d{\bf k}}{(2\pi)^2} \sum_{n,n'}
\frac{f_{n {\bf k}}-f_{n' {\bf k}+{\bf q}}}
{\hbar\omega+E_{n {\bf k}}-E_{n' {\bf k}+{\bf q}}}
\nonumber \\
& \times
\frac{i}{2}
(n\sin[\varphi({\bf q})-\varphi({\bf k})]+
 n'\sin[\varphi({\bf q})-\varphi({\bf k}+{\bf q})]) , \mbox{and}
\label{polarizabilnost-eph3-L}
\end{align}
\begin{align}
\Pi_{e-ph}^1(T,{\bf q},\omega)=
4 g & \int \frac{d{\bf k}}{(2\pi)^2} \sum_{n,n'}
\frac{f_{n {\bf k}}-f_{n' {\bf k}+{\bf q}}}
{\hbar\omega+E_{n {\bf k}}-E_{n' {\bf k}+{\bf q}}}
\nonumber \\
& \times
\frac{i}{2}
(n\cos[\varphi({\bf q})-\varphi({\bf k})]+
 n'\cos[\varphi({\bf q})-\varphi({\bf k}+{\bf q})]) .
\label{polarizabilnost-eph3-T}
\end{align}
Let us first analyze interaction with the longitudinal optical phonons. In that respect let us take expression (\ref{polarizabilnost-eph3-L}) and assume, without the loss of generality, that $\varphi({\bf q})=0$ i.e. vector $\bf q$ is along $\hat{x}$ direction. We than obtain
\begin{align}
\Pi_{e-ph}^1(L,{\bf q},\omega)=
4 g & \int \frac{d{\bf k}}{(2\pi)^2} \sum_{n,n'}
\frac{f_{n {\bf k}}-f_{n' {\bf k}+{\bf q}}}
{\hbar\omega+E_{n {\bf k}}-E_{n' {\bf k}+{\bf q}}}
\nonumber \\
& \times
\frac{i}{2}(-n\sin[\varphi({\bf k})]- n'\sin[\varphi({\bf k}+{\bf q})]).
\label{polarizabilnost-eph3-L2}
\end{align}
But the function under the integral sign is odd with respect to reflection across the x axis, meaning that the entire integral vanishes i.e. 
$\Pi_{e-ph}^1(L,{\bf q},\omega)=0$.
In other words we have shown analytically that there is no whatsoever coupling of plasmons and longitudinal optical phonons! Finally, to find the coupling of plasmons with transverse optical phonons one only needs to solve self-consistent set of equations (\ref{FonFreqRen}) and (\ref{zasj-polariz}) which was done numerically and demonstrated to agree with the results of chapter \ref{chap:chap5}, where we used different gauge to obtain the same result.

\listoffigures




\end{onehalfspace}
\end{document}